\documentclass{aa}

\usepackage{twoopt}
\usepackage[breaklinks=true]{hyperref} 
\bibpunct{(}{)}{;}{a}{}{,} 
\makeatletter
\newcommandtwoopt{\citeads}[3][][]{\href{http://adsabs.harvard.edu/abs/#3}%
{\def\hyper@linkstart##1##2{}%
\let\hyper@linkend\@empty\citealp[#1][#2]{#3}}}
\newcommandtwoopt{\citepads}[3][][]{\href{http://adsabs.harvard.edu/abs/#3}%
{\def\hyper@linkstart##1##2{}%
\let\hyper@linkend\@empty\citep[#1][#2]{#3}}}
\newcommandtwoopt{\citetads}[3][][]{\href{http://adsabs.harvard.edu/abs/#3}%
{\def\hyper@linkstart##1##2{}%
\let\hyper@linkend\@empty\citet[#1][#2]{#3}}}
\newcommandtwoopt{\citeyearads}[3][][]%
{\href{http://adsabs.harvard.edu/abs/#3}
{\def\hyper@linkstart##1##2{}%
\let\hyper@linkend\@empty\citeyear[#1][#2]{#3}}}
\makeatother
\usepackage{graphicx}
\usepackage{siunitx}
\usepackage{subcaption}
\usepackage{adjustbox}
\usepackage{makecell}
\usepackage[usenames,dvipsnames]{xcolor}
\usepackage[normalem]{ulem}
\usepackage{multirow}
\usepackage{amsmath}
\usepackage{float}
\usepackage{longtable}
\usepackage{xcolor}
\usepackage{rotating}
\usepackage{enumitem}

\usepackage{txfonts}
\newcommand{\cosf}{\mbox{CO\,6--5}}

\newcommand{\coet}{\mbox{CO\,11--10}}
\newcommand{\cott}{\mbox{CO\,13--12}}
\newcommand{\coftft}{\mbox{CO\,15--14}}
\newcommand{\costft}{\mbox{CO\,16--15}}
\newcommand{\cottto}{\mbox{CO\,22--21}}
\newcommand{\cstott}{\mbox{C$^{17}$O\,3--2}}
\newcommand{\cetosf}{\mbox{C$^{18}$O\,6--5}}
\newcommand{\cetone}{\mbox{C$^{18}$O\,9--8}}
\newcommand{\ttcosf}{\mbox{$^{13}$CO\,6--5}}
\newcommand{\ttcotn}{\mbox{$^{13}$CO\,10--9}}

\begin{document}

   \title{Velocity-resolved high-$J$ CO emission from \\ massive star-forming clumps}

   \author{Hoang Thanh Dat \inst{1}\fnmsep\thanks{tdhoang@mpifr-bonn.mpg.de},
          Agata Karska \inst{1,2,3}
          \and
          Min Young Lee \inst{4}
          \and
          Friedrich Wyrowski \inst{1}
          \and
          Le Ngoc Tram \inst{1}
          \and
          Aiyuan Y. Yang \inst{5,6}
          \and
          Karl M. Menten \inst{1}
          }
   \institute{Max Planck Institute for Radio Astronomy, Auf dem H\"ugel 69, 53121 Bonn, Germany
 \and
  Argelander-Institut für Astronomie, Universität Bonn, Auf dem Hügel 71, 53121 Bonn, Germany
\and
  Institute of Astronomy, Faculty of Physics,
    Astronomy and Informatics, Nicolaus Copernicus University, ul. Grudziądzka 5, 87-100 Toruń, Poland
\and
   Korea Astronomy and Space Science Institute, 776 Daedeok-daero, Yuseong-gu, Daejeon 34055, Republic of Korea
 \and
    National Astronomical Observatories, Chinese Academy of Sciences, A20 Datun Road, Chaoyang District, Beijing 100101,\\ P.R. China
\and
    Key Laboratory of Radio Astronomy and Technology, Chinese Academy of Sciences, A20 Datun Road, Chaoyang District, Beijing 100101, P.R. China}

\date{Received June 12, 2023; accepted September 21, 2023}
\titlerunning{high-$J$ CO emission in high-mass clumps}
\authorrunning{H.~T.~Dat et al. 2023}

  \abstract
   {Massive star formation is associated with energetic processes, which result in significant gas cooling via far-infrared (IR) lines. Velocity-resolved observations can constrain the kinematics of the gas, allowing the identification of the physical mechanisms responsible for gas heating.}
   {Our aim is to quantify far-infrared CO line emission toward high-mass star-forming regions, identify the high-velocity gas component associated with outflows, and estimate the physical conditions required for the excitation of the observed lines.}
   {Velocity-resolved SOFIA/GREAT spectra of 13 high-mass star forming clumps of various luminosities and evolutionary stages are studied in highly-excited rotational lines of CO. For most targets, the spectra are from frequency intervals covering the \coet{} and 16--15 lines. Toward two sources, also the \cott{} line was observed with SOFIA/4GREAT. Angular resolutions at the line frequencies range from $14\arcsec$ to $20\arcsec$, corresponding to spatial scales of $\sim$ 0.1--0.8 pc. Radiative transfer models are used to determine the physical conditions giving rise to the emission in the line wings.}
   {All targets in our sample show strong high-$J$ CO emission in the far-IR, characterized by broad line wings associated with outflows, thereby significantly increasing the sample of high-mass objects with velocity-resolved high-$J$ CO spectra. Twelve sources show emission in the line wings of the \coet{} line ($E_\mathrm{u}/k_\mathrm{B}$=365\,K), and 8 sources in the \costft{} line ($E_\mathrm{u}/k_\mathrm{B}$=752\,K). The contribution of the emission in the line wings to the total emission ranges from $\sim$28\% to 76\%, and does not correlate with the envelope mass or evolutionary stage. Gas excitation temperatures cover a narrow range of 120--220\,K for the line wings, and 110--200\,K for the velocity-integrated line emission, assuming Local Thermodynamics Equilibrium (LTE). For the two additional sources with the \cott{} line ($E_\mathrm{u}/k_\mathrm{B}$=503\,K) data, wing emission rotational temperatures of \mbox{$\sim$130\,K} and \mbox{165\,K} are obtained using Boltzmann diagrams. The corresponding non-LTE radiative transfer models indicate gas densities of \mbox{$10^{5}$--$10^{7}$\,cm$^{-3}$} and CO column densities of $10^{17}$--$10^{18}$\,cm$^{-2}$ in the line wings, similar to physical conditions in deeply-embedded low- and high-mass protostars. The velocity-integrated CO line fluxes correlate with the bolometric luminosity over 7 orders of magnitude including data on the low-mass protostars from the literature. This suggests that similar processes are responsible for the high-$J$ CO excitation over a significant range of physical scales.
}
   {Velocity-resolved line profiles allow the detection of outflows toward massive star-forming clumps spanning a broad range of evolutionary stages. The lack of clear evolutionary trends suggest that mass accretion and ejection prevail during the entire lifetime of star-forming clumps.}


   \maketitle
%

\section{Introduction}

High-mass stars have a significant impact on their environments and on galaxy evolution globally through their ionising radiation, stellar winds, and their deaths in supernova explosions \citep{zin07}. Already during the earliest stages of their formation, massive protostars might inject significant amounts of energy and momentum into the interstellar medium (ISM) in the form of outflows, capable of disrupting clumps and cores \citep{beuther02outflow,bally16}.

Outflows, a ubiquitous phenomenon in both low and high mass star-forming regions, play an essential role in transporting angular-momentum and regulating the star-forming process across multiple spatial scales \citep{bally83,evans99}. Both, the dissipation of the envelope material and mass loss via the outflows lower the core-to-star formation efficiency \citep{Kru14,off17}. At cluster/clump scales, outflows drive turbulence that provides additional support against gravitational collapse \citep{frank14}.

Outflows are typically detected using low-$J$ ($J\lesssim$5) velocity-resolved rotational lines of carbon monoxide (CO), which is the second most abundant molecule in the interstellar medium \citep[CO/H$_{2}=1.2\times10^{-4}$,][]{fre82}. The low-lying rotational levels of CO are easily collisionally-excited even at low densities and can readily be observed at millimeter wavelengths. These lines constitute  a useful diagnostic of the gas kinetic temperature of outflows \citep{bally83,yil15}. An extensive search for outflows traced by such low-$J$ CO lines toward a total of 2052 massive star-forming clumps that were identified in the APEX Telescope Large Area Survey of the Galaxy \citep[ATLASGAL,][]{schuller09}, provided an overall outflow detection rate of $58\%$ \citep{yang18,yang22}.

Observations of high-$J$ CO ($J\gtrsim10$) lines provide an opportunity to study denser and warmer parts of star-forming clumps and the ouflows that arise in them. Recent surveys with the \textit{Herschel} Space Observatory \citep{pil10} found that CO lines account for the bulk far-infrared (IR) gas cooling in both low- and high-mass star-forming regions \citep{kar13,kar14,kar18,vd21}. The velocity-resolved profiles of high-$J$ CO toward low-mass protostars revealed a significant contribution of high-velocity ($v\sim$ 20--30\,km\,s$^{-1}$) gas to the total far-IR line emission \citep{irene13,yil13}, and similarity to the H$_2$O emission likely arising from the same gas \citep{irene16,kri17}. Single-pointing observations toward high-mass sources have also revealed broad, outflow wings in high-$J$ CO line profiles, but have been limited to just a few sources: \mbox{W3 IRS5} \citep{irene13}, AFGL 2591 \citep{maja}, Orion KL, Orion S, Sgr B$^{*}$, and W49N \citep{indriolo17}. Complementary observations have been obtained with the German REceiver for Astronomy at Terahertz frequencies\footnote{GREAT is a development by the MPI f\"{u}r Radioastronomie and the KOSMA/Universit\"{a}t zu K\"{o}ln, in cooperation with the MPI f\"{u}r Sonnensystemforschung and the DLR Institut für Planetenforschung.} \citep[GREAT, ][]{risacher2018} onboard the Stratospheric Observatory For Infrared Astronomy \citep[SOFIA, ][]{young2012early}. High-resolution spectroscopy of far-IR CO lines from an intermediate-mass protostar Cep E revealed an extremely high-velocity gas (EHV; $\varv$ up to $\sim140$ km s$^{-1}$) tracing shocks associated with the jet and intermediate-to-high velocity gas ($\varv$ from 50 to 100 km s$^{-1}$) associated with outflow cavities and a bow shock \citep{ruiz12,lefloch15,gusdorf17}. The line profiles of CO\,$16-15$ toward two high-mass sources, however, lacked the EHV component and revealed broad line wings extending up to $\varv$ $\sim50$ km s$^{-1}$ \citep{leu15,gusdorf16}. Other surveys, conducted with the PACS and SPIRE instruments aboard the \textit{Herschel} space telescope \citep{pil10}, lacked the high spectral resolution necessary to disentangle the envelope and outflow emission in the spectra \citep{kar14,goi13,goi15}.

In this paper, we use SOFIA/GREAT to quantify high-$J$ CO emission toward 13 high-mass star forming clumps with the aim to isolate the contribution from the outflows and estimate excitation conditions associated with the line wing emission. We also examine how the high-$J$ CO emission varies as a function of clump properties and evolutionary stages.

The paper is organized as follows: Section \ref{sec:obs} describes the source sample, observations with SOFIA, and the complementary CO observations with the APEX telescope and \textit{Herschel}. In Section \ref{sec:results_n_analysis}, we present line profiles of high-$J$ CO transitions (Section \ref{sec:results}) and decompose the emission that belongs to the line wings (Section \ref{sec:profile-decomposition}). In addition, we study the correlations of velocity-integrated emission with source properties, and those of the fraction of wing emission with source evolutionary stages (Section \ref{sec:ratio}). Subsequently, we analyse the excitation of high-$J$ CO lines using LTE and non-LTE approaches (Sections \ref{sec:lte} and \ref{sec:nonlte}). Section \ref{sec:dis} consists of the discussion of our results in the context of previous studies and Section \ref{sec:conclusions} presents a summary and our conclusions.

\section{Observations} \label{sec:obs}
\subsection{Sample}
 \begin{table*}[h!]
\centering
\small
\caption{Catalog of source properties}
\label{tab:catalog}
\begin{tabular}{llccccccccccc}
 \hline
 \hline
No. & Source & ATLASGAL name\tablefootmark{a} & RA & Dec & $V_{\text{lsr}}$\tablefootmark{b} & $D$\tablefootmark{c} & $L_{\text{bol}}$\tablefootmark{d} & $M_{\text{clump}}$\tablefootmark{e} & $D_{\text{GC}}$ & Type\tablefootmark{f}  \\
     &    & & (J2000) & (J2000) & (km s$^{-1}$) & (kpc) & ($L_{\odot}$) & ($M_{\odot}$) & (kpc)\\
 \hline
 \hline
1 & G351.16+0.7 & AGAL351.161+00.697 & 17:19:56.69 & -35:57:53.0 & -6.0 & 1.3 & $8.8\times 10^3$  & $1.2\times 10^3$ & 6.7 & IRb  \\
2 & G351.25+0.7 & AGAL351.244+00.669 & 17:20:18.86 & -35:54:42.5 & -2.8 & 1.3 & $4.9 \times 10^4$ & $3.7 \times 10^2$ & 6.7 & IRb  \\
3 & G351.44+0.7 & AGAL351.444+00.659 & 17:20:55.20 & -35:45:08.0 & -3.8 & 1.3 & $2.0 \times 10^4$ & $1.0 \times 10^3$ & 6.7 & 24d   \\
4 & G351.58$-$0.4 & AGAL351.581$-$00.352 & 17:25:25.03 & -36:12:45.4 & -95.6 & 8.0 & $4.6 \times 10^5$ & $2.3 \times 10^3$ & 2.0 & IRb   \\
5 & G351.77$-$0.5 & AGAL351.774$-$00.537 & 17:26:42.54 & -36:09:20.1 & -2.8 & 1.3 & $ 3.7\times 10^4$ & $3.3 \times 10^2$ & 7.8 & IRb  \\
\hline
6 & G12.81$-$0.2 & AGAL012.804$-$00.199 & 18:14:13.54 & -17:55:32.0 & 34.6 & 2.6 & $2.5 \times 10^5$  & $1.9 \times 10^3$  & 6.2 & HII   \\
7 & G14.19$-$0.2 & AGAL014.194$-$00.194 & 18:16:58.63 & -16:42:16.4 & 39.7 & 3.1 & $3.7 \times 10^3$ & $5.1 \times 10^2$ & 4.8 & 24d   \\
8 & G13.66$-$0.6 & AGAL013.658$-$00.599 & 18:17:24.09 & -17:22:10.3 & 48.5 & 4.5 & $2.4 \times 10^4$ & $2.7 \times 10^2$ & 4.3 & IRb   \\
9 & G14.63$-$0.6 & AGAL014.632$-$00.577 & 18:19:14.65 & -16:30:02.7 & 18.5 & 1.5 & $1.6 \times 10^3$ & $1.6 \times 10^2$ & 6.3 & 24d  \\
\hline
10 & G34.41+0.2 & AGAL034.411+00.234 & 18:53:18.13 & +01:25:23.7 & 57.9 & 2.9 & $3.1 \times 10^3$ & $4.4 \times 10^2$ & 7.2 & IRb  \\
11 & G34.26+0.15 & AGAL034.258+00.154 & 18:53:18.51 & +01:14:57.6 & 58.0 & 2.9 & $6.1 \times 10^4$ & $1.7 \times 10^3$ & 6.9 & HII \\
12 & G34.40$-$0.2 & AGAL034.401+00.226 & 18:53:18.63 & +01:24:40.4 & 57.1 & 2.9 & $3.2 \times 10^3$ & $7.9 \times 10^2$ & 7.2 & HII \\
13 & G35.20$-$0.7 & AGAL035.197$-$00.742 & 18:58:12.94 & +01:40:40.6 & 33.5 & 2.2 & $2.4 \times 10^4$  & $4.6 \times 10^2$  & 6.8 & IRb  \\
 \hline
  \hline
\end{tabular}
\begin{flushleft}
\tablefoot{
\tablefoottext{a}{Source names from the ATLASGAL Compact Source Catalogue \citep{contreras2013atlasgal,urquhart2014atlasgal}.}
\tablefoottext{b}{Source velocities ($V_{\text{lsr}}$) estimated from proxy lines (see Appendix\,\ref{app:decomposition-method}).}
\tablefoottext{c}{Heliocentric distances estimated in \citet{urqu22}}
\tablefoottext{d}{Source bolometric luminosities, $L_{\text{bol}}$, estimated from greybody fit to dust emission in \citet{atlasgal_james18}. $L_{\text{bol}}$ of source 1, 6, and 13 are estimated in \citet{konig2017atlasgal} }
\tablefoottext{e}{Clump masses, $M_{\text{clump}}$, estimated from cold dust emission at 870\,$\mu$m in \cite{urqu22}. $M_{\text{clump}}$ of source 1, 6, and 13 are obtained from \citet{konig2017atlasgal}  }
\tablefoottext{f}{ Source classification using the criteria from \citet{konig2017atlasgal}, and refers to IR-bright sources (IRb), IR-weak sources (24d), and HII regions (HII).}
}
\end{flushleft}
\end{table*}

\begin{table*}
\begin{center}
\caption{Overview of the observations}
\label{tab:observations}
\renewcommand{\footnoterule}{}
\begin{tabular}{c c c c c c c c c}
\hline\hline
Molecule & Trans. & Freq. & $E_\mathrm{u}$/$k_\mathrm{B}$ & $A_\mathrm{u}$ & $g_\mathrm{u}$ & Receiver & Beam  & $\Delta v$\\

~ & $J_{u}-J_{l}$ & (GHz) & (K) & (s$^{-1}$) & ~ & ~ & ('') & (km s$^{-1}$)\\
\hline
CO & 6--5   & 691.5 & 116.16 & 2.1(-5) & 13 & APEX/CHAMP$^+$ & 9  & 0.318\\
CO & 11--10 & 1267.0 & 364.97 & 1.3(-4) & 23  & SOFIA/GREAT & 20 & (0.361, 0.578)\\
CO & 13--12 & 1496.9 & 503.13 & 2.2(-4) & 27 & SOFIA/GREAT & 20 & 0.978\\
CO & 16--15 & 1841.4  & 751.72 & 4.1(-4) & 33  & SOFIA/GREAT & 14 & (0.248, 0.795)\\
\hline
$^{13}$CO & 6--5 & 661.1 & 111.05 & 1.9(-5) & 13 & APEX/CHAMP$^+$ & 9 & 0.332\\
$^{13}$CO & 10--9 & 1101.4 & 290.79 & 8.8(-5) & 21 & Herschel/HIFI & 19 & 0.136\\
\hline
C$^{18}$O & 6--5 & 658.6 & 110.63 & 1.9(-5) & 13 & APEX/CHAMP$^+$ & 9 & 0.334\\
C$^{18}$O & 9--8 & 987.6 & 237.03 & 6.4(-5) & 19 & Herschel/HIFI & 22 & 0.152\\
\hline
\end{tabular}
\begin{flushleft}
\tablefoot{Molecular data adopted from the Leiden Atomic and Molecular Database (LAMDA, \citealt{LAMDA}). $\Delta v$ is the original spectral resolution of our data, before they are smoothed to a common size of 1.0\,km\,s$^{-1}$. The \coet{} and \costft{} have two $\Delta v$ each, which correspond to observations in 2014 and 2016, respectively.
}
\end{flushleft}
\end{center}
\end{table*}

All sources have been selected from the ATLASGAL survey covering 420 deg$^2$ of the inner Galactic plane in the 870\,$\mu$m dust continuum \citep{urquhart2014atlasgal,konig2017atlasgal}. The latest version of the ATLASGAL source catalog contains 5007 clumps spanning a wide range of masses ($M_\mathrm{clump}$) and luminosities ($L_\mathrm{bol}$), and divided into four evolutionary stages -- quiescent, protostellar, young stellar objects, and \mbox{H II} regions \mbox{(H II)}, see \citet{urqu22}.

\begin{figure*}[h!]
  \begin{subfigure}{0.33\textwidth}
      \centering
      \resizebox{0.9\hsize}{!}{\includegraphics{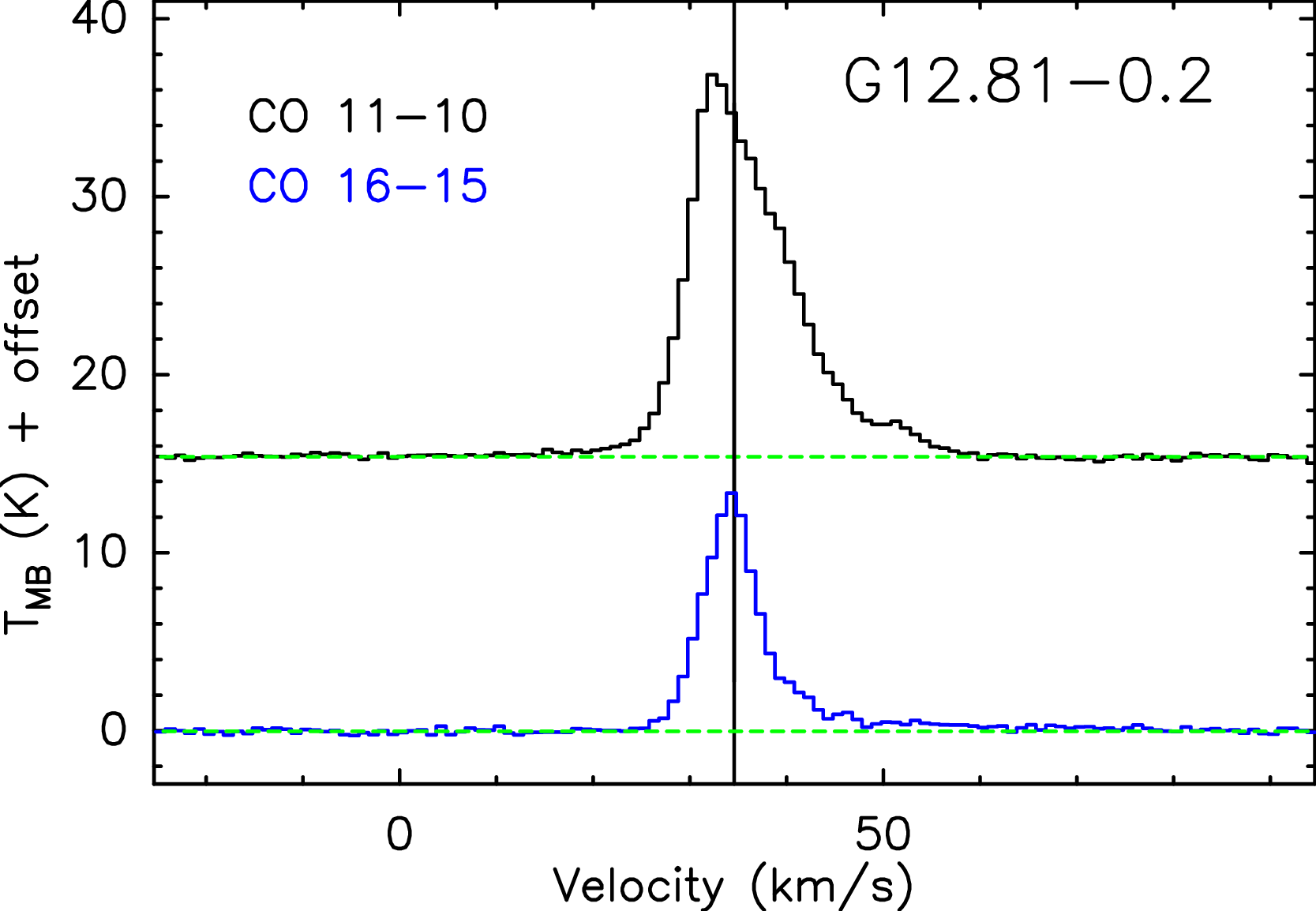}}
   \end{subfigure}%
   \vspace{0.1cm}
   \begin{subfigure}{0.33\textwidth}
      \centering
      \resizebox{0.9\hsize}{!}{\includegraphics{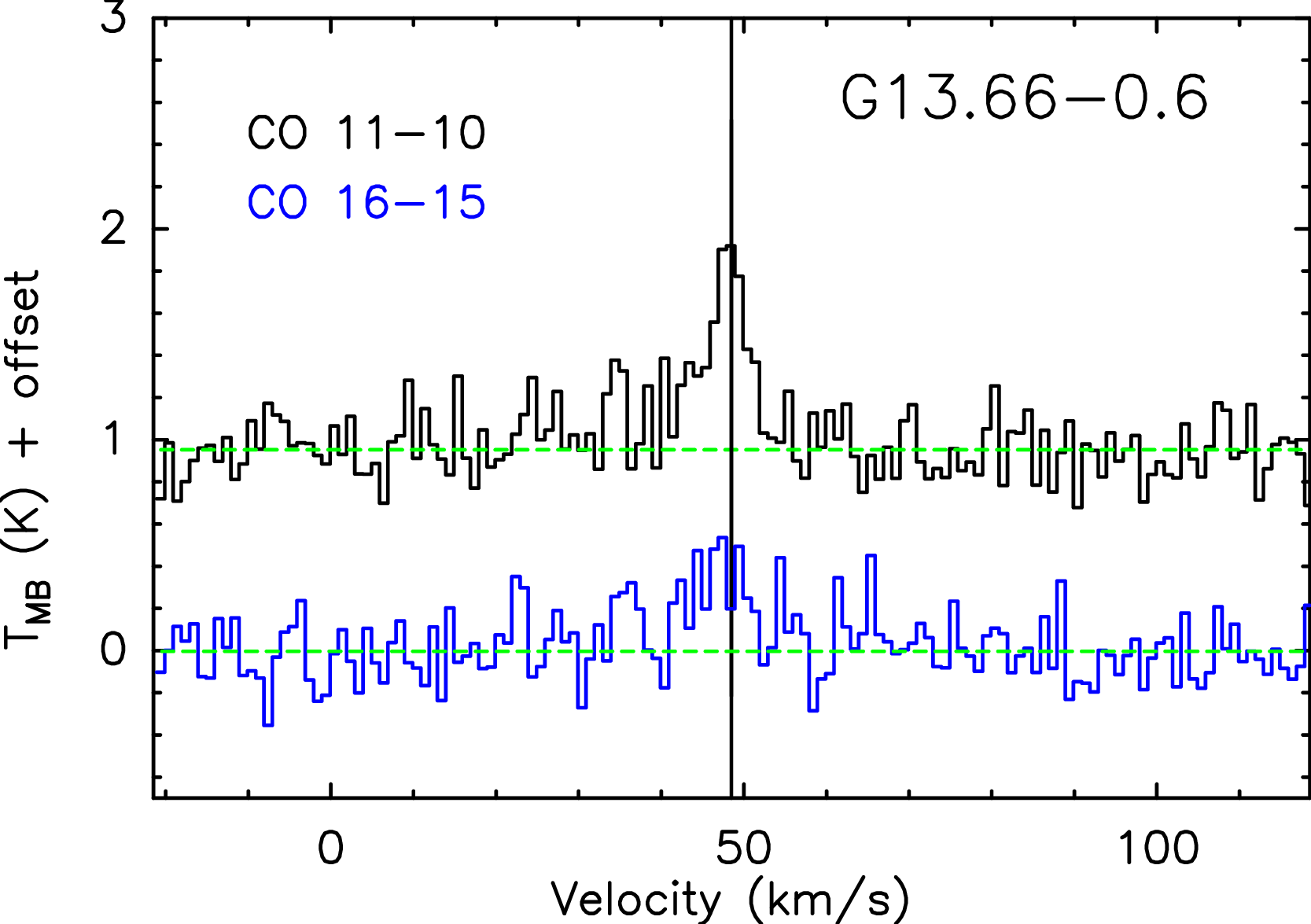}}
   \end{subfigure}%
   \begin{subfigure}{0.33\textwidth}
      \centering
      \resizebox{0.9\hsize}{!}{\includegraphics{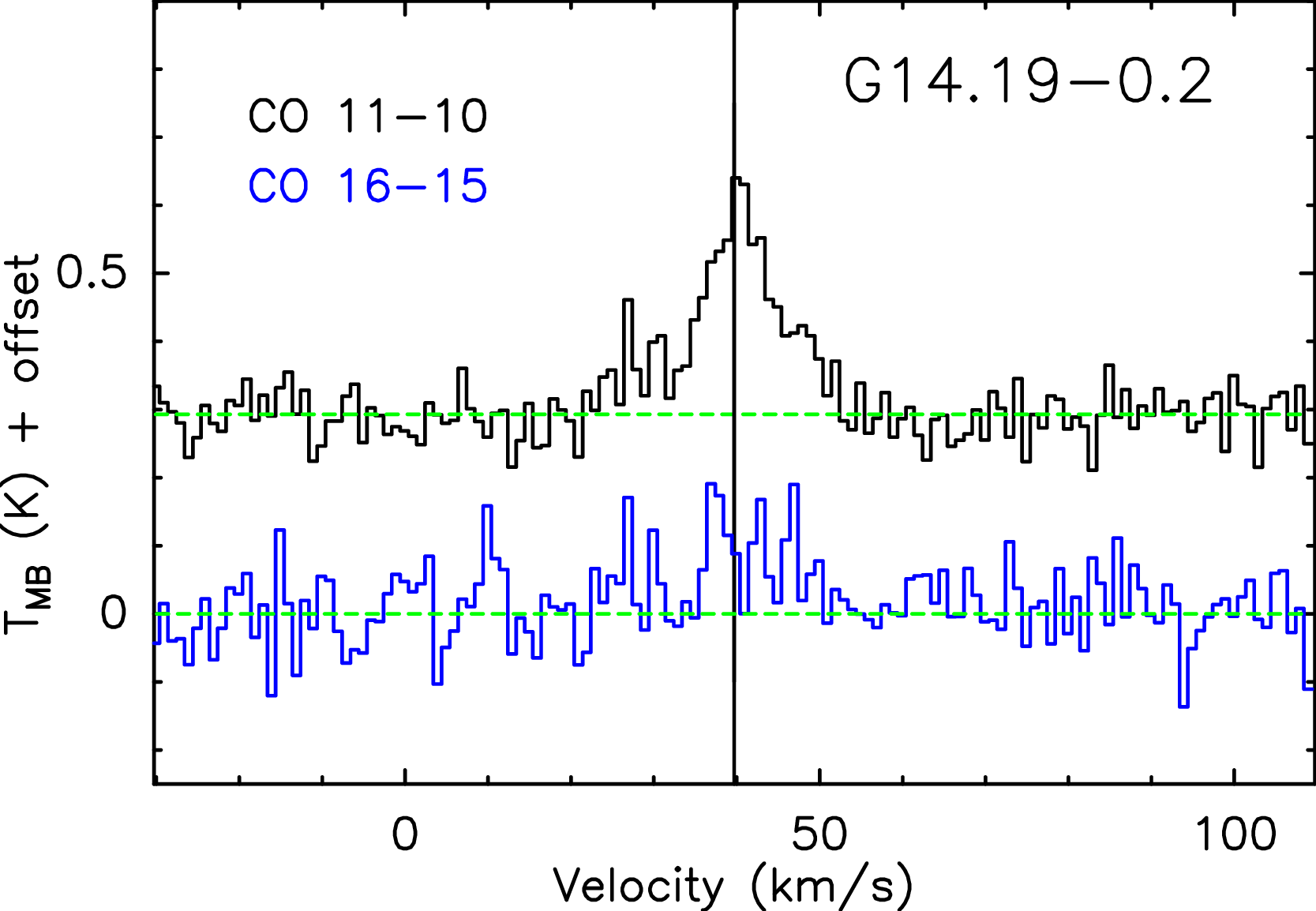}}
   \end{subfigure}
   \begin{subfigure}{0.33\textwidth}
      \centering
      \resizebox{0.9\hsize}{!}{\includegraphics{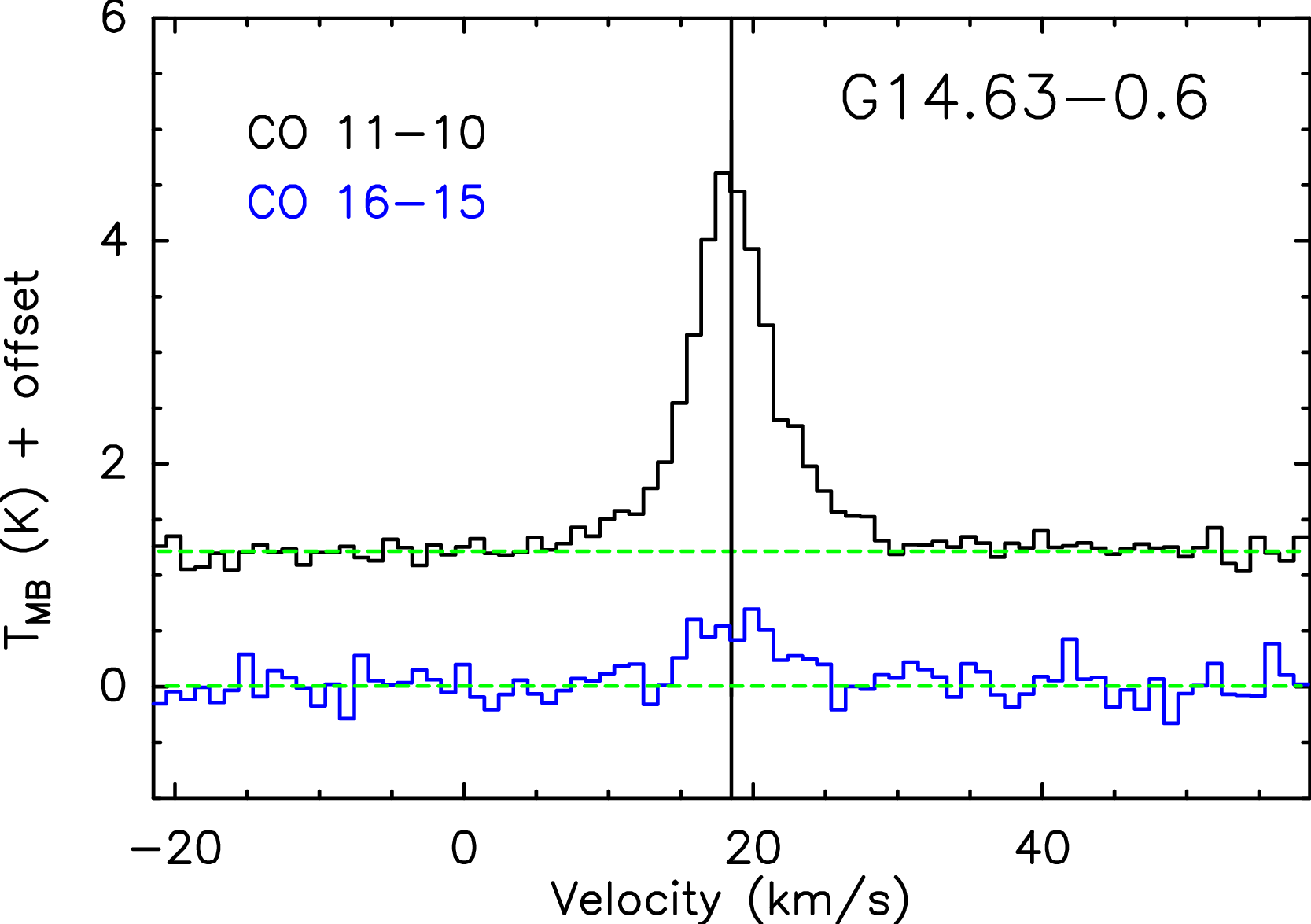}}
   \end{subfigure}%
   \vspace{0.1cm}
   \begin{subfigure}{0.33\textwidth}
      \centering
      \resizebox{0.9\hsize}{!}{\includegraphics{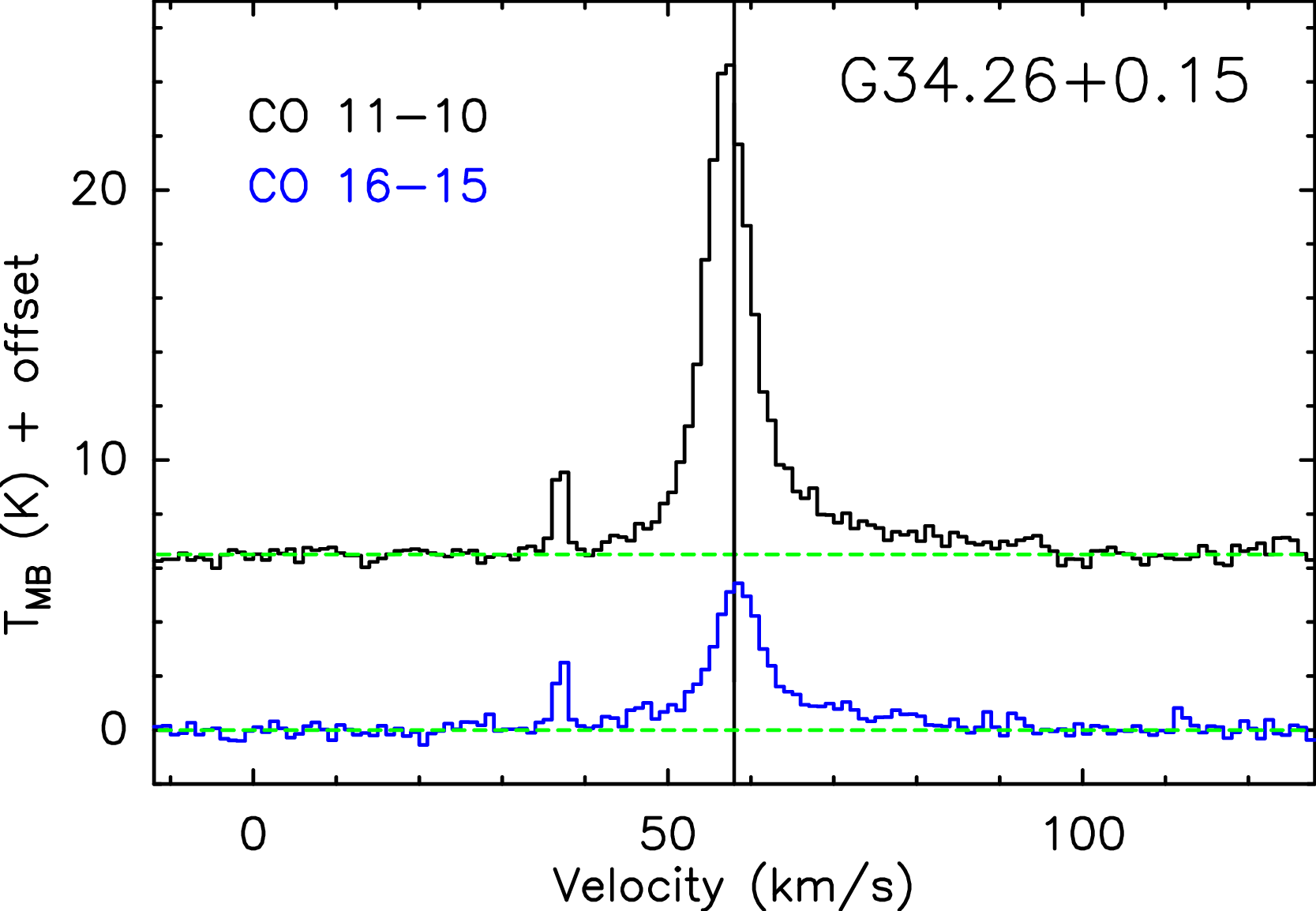}}
   \end{subfigure}%
   \begin{subfigure}{0.33\textwidth}
      \centering
      \resizebox{0.9\hsize}{!}{\includegraphics{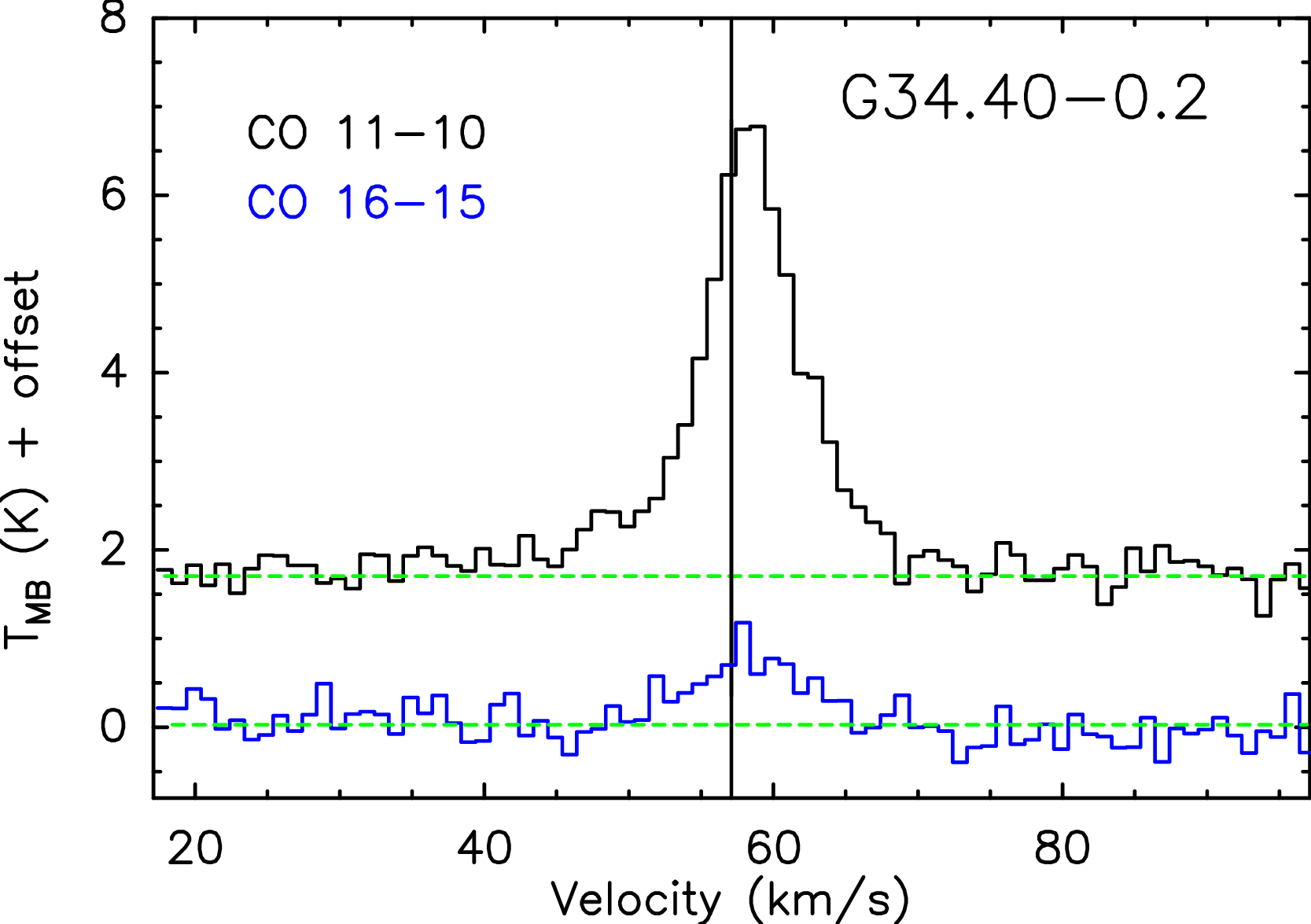}}
   \end{subfigure}
   \begin{subfigure}{0.33\textwidth}
      \centering
      \resizebox{0.9\hsize}{!}{\includegraphics{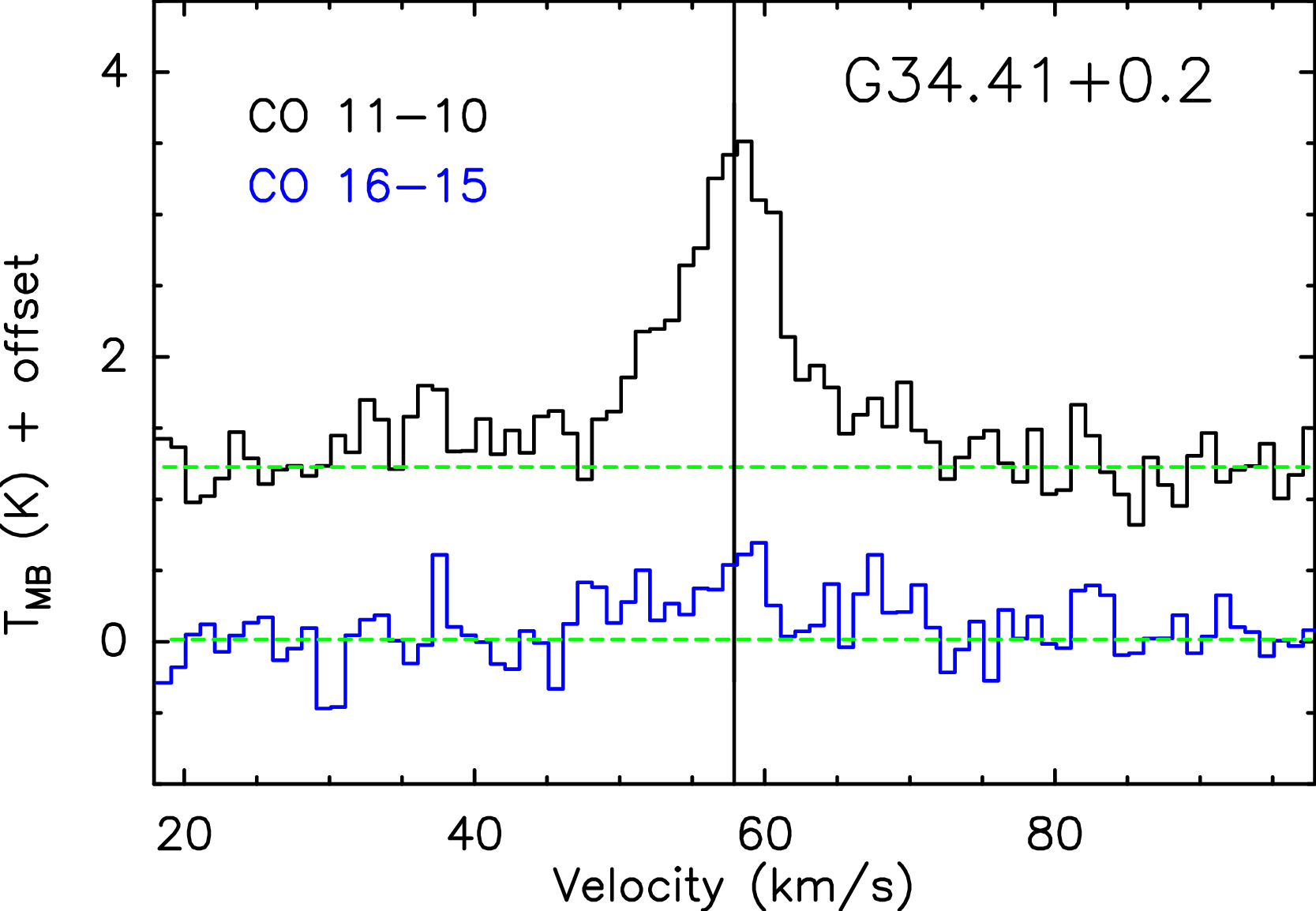}}
   \end{subfigure}%
   \vspace{0.1cm}
   \begin{subfigure}{0.33\textwidth}
      \centering
      \resizebox{0.9\hsize}{!}{\includegraphics{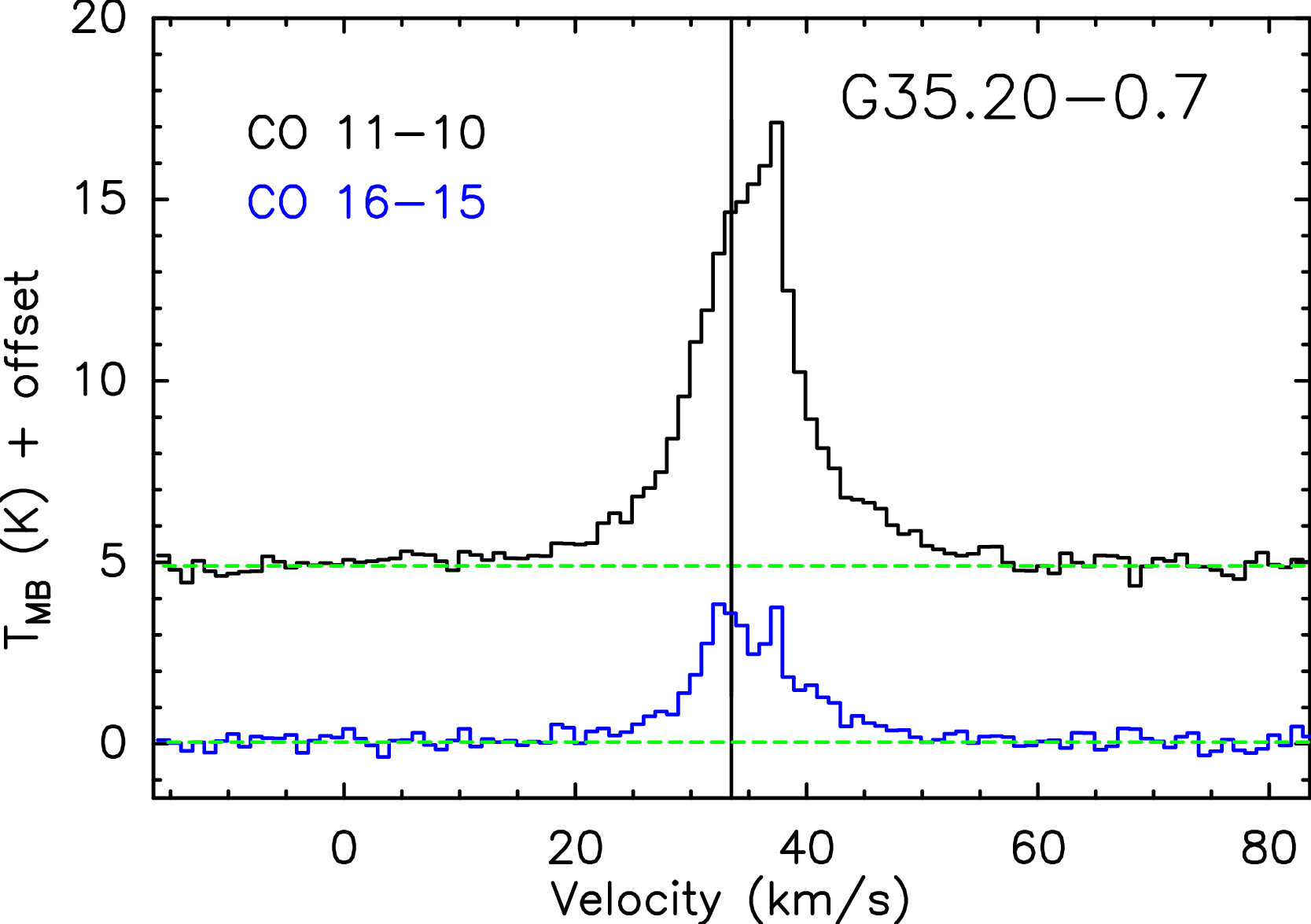}}
   \end{subfigure}%
   \begin{subfigure}{0.33\textwidth}
      \centering
      \resizebox{0.9\hsize}{!}{\includegraphics{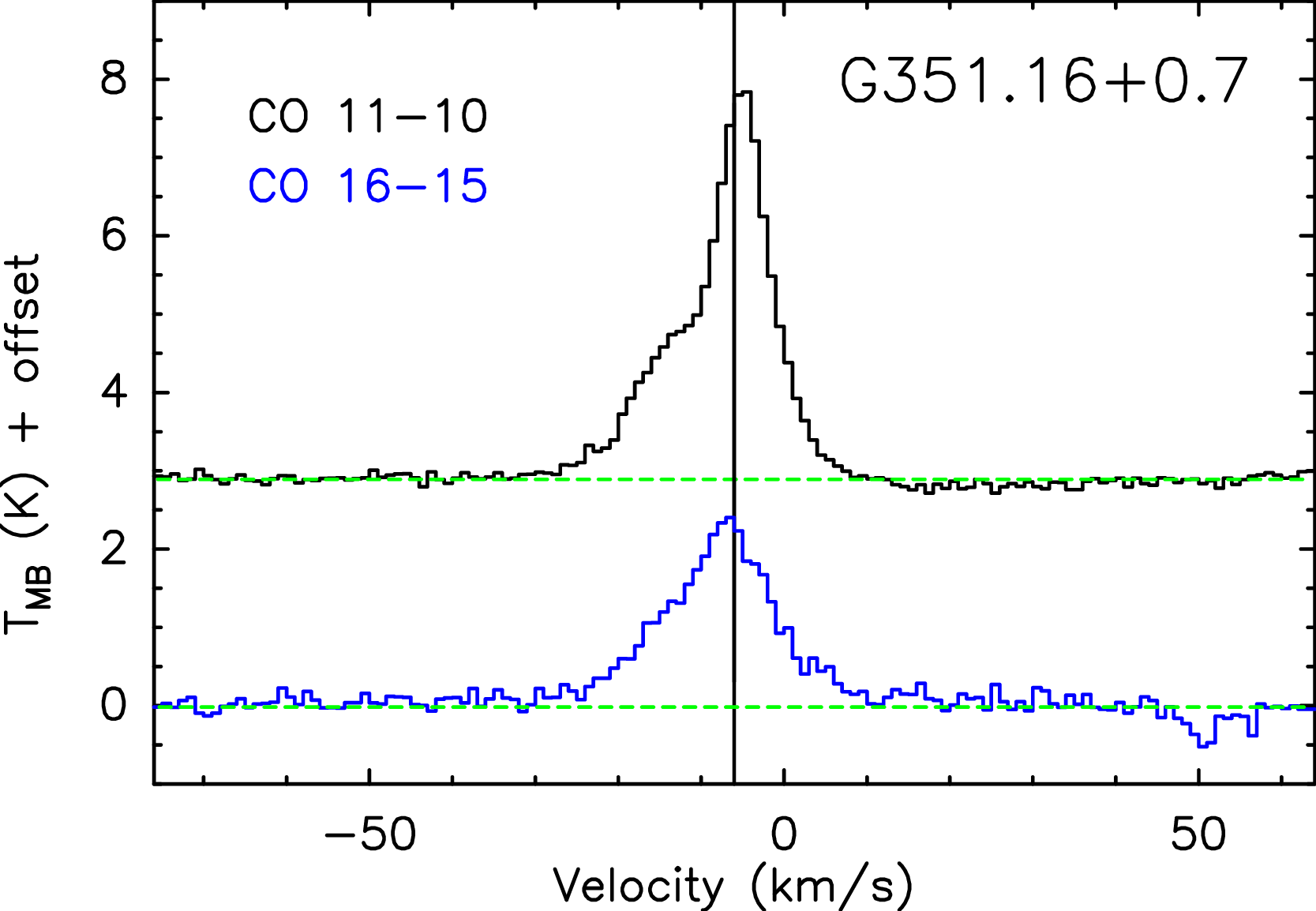}}
   \end{subfigure}
   \begin{subfigure}{0.33\textwidth}
      \centering
       \resizebox{0.9\hsize}{!}{\includegraphics{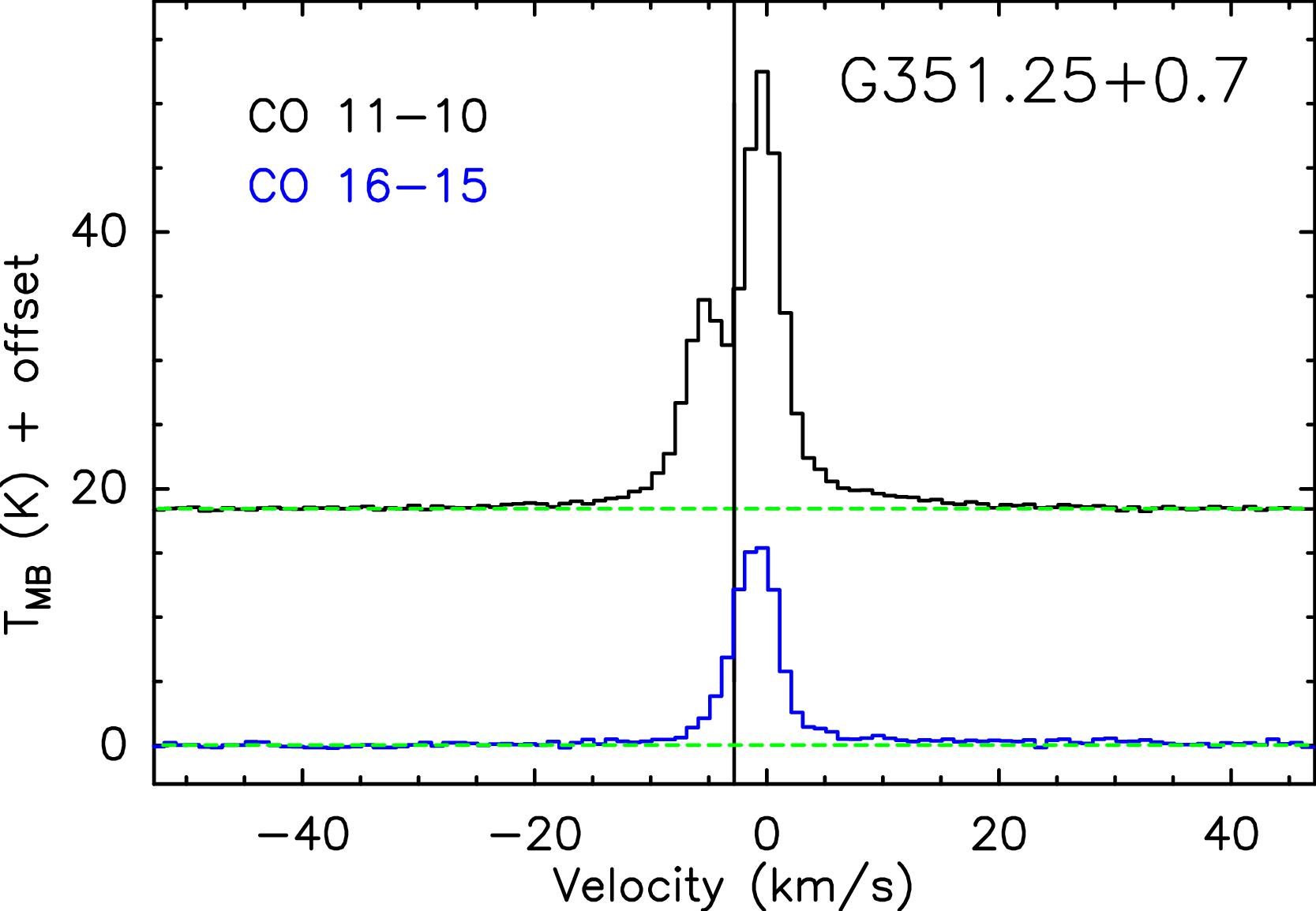}}
   \end{subfigure}%
   \vspace{0.1cm}
   \begin{subfigure}{0.33\textwidth}
      \centering
      \resizebox{0.9\hsize}{!}{\includegraphics{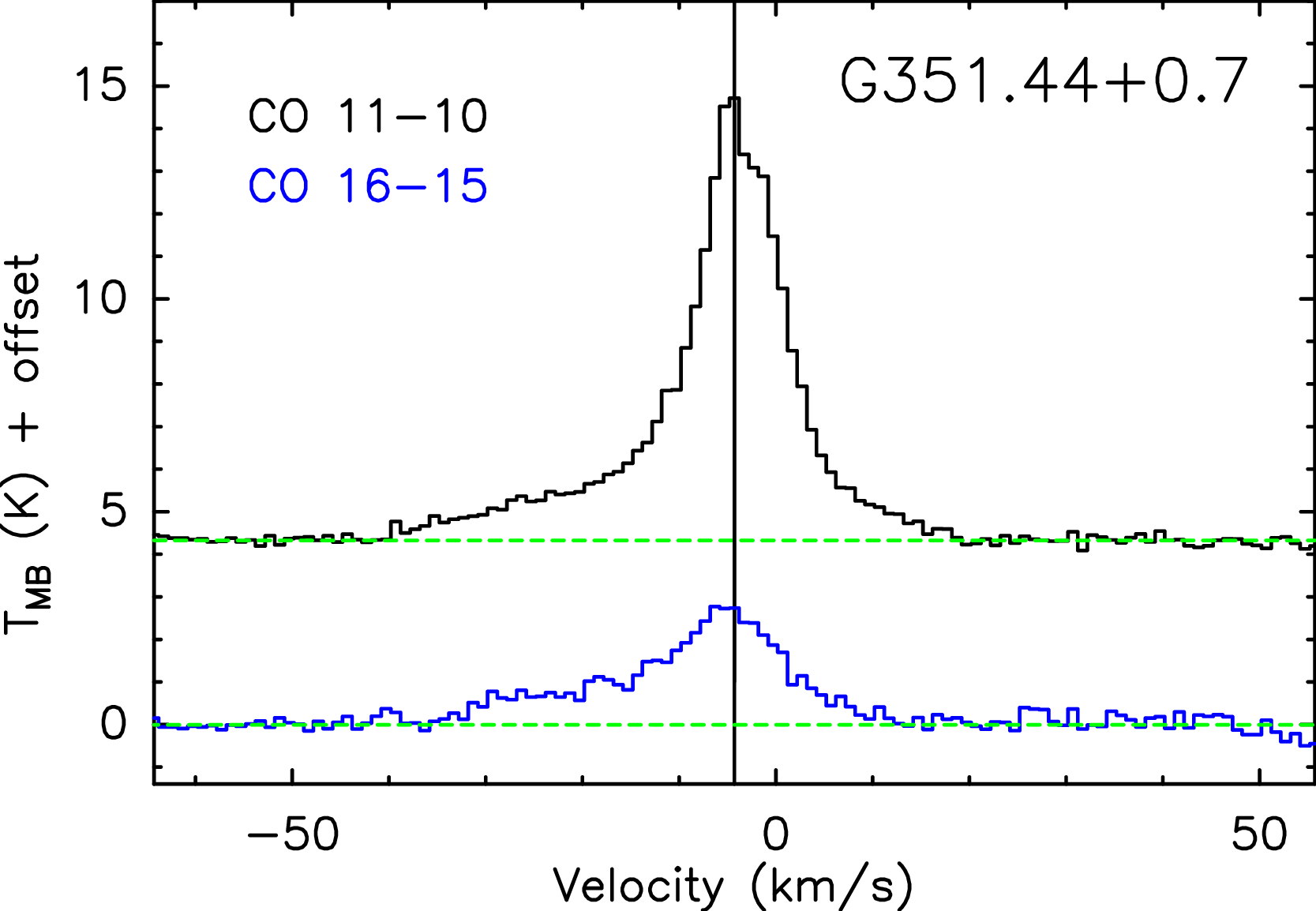}}
   \end{subfigure}%
   \begin{subfigure}{0.33\textwidth}
      \centering
      \resizebox{0.9\hsize}{!}{\includegraphics{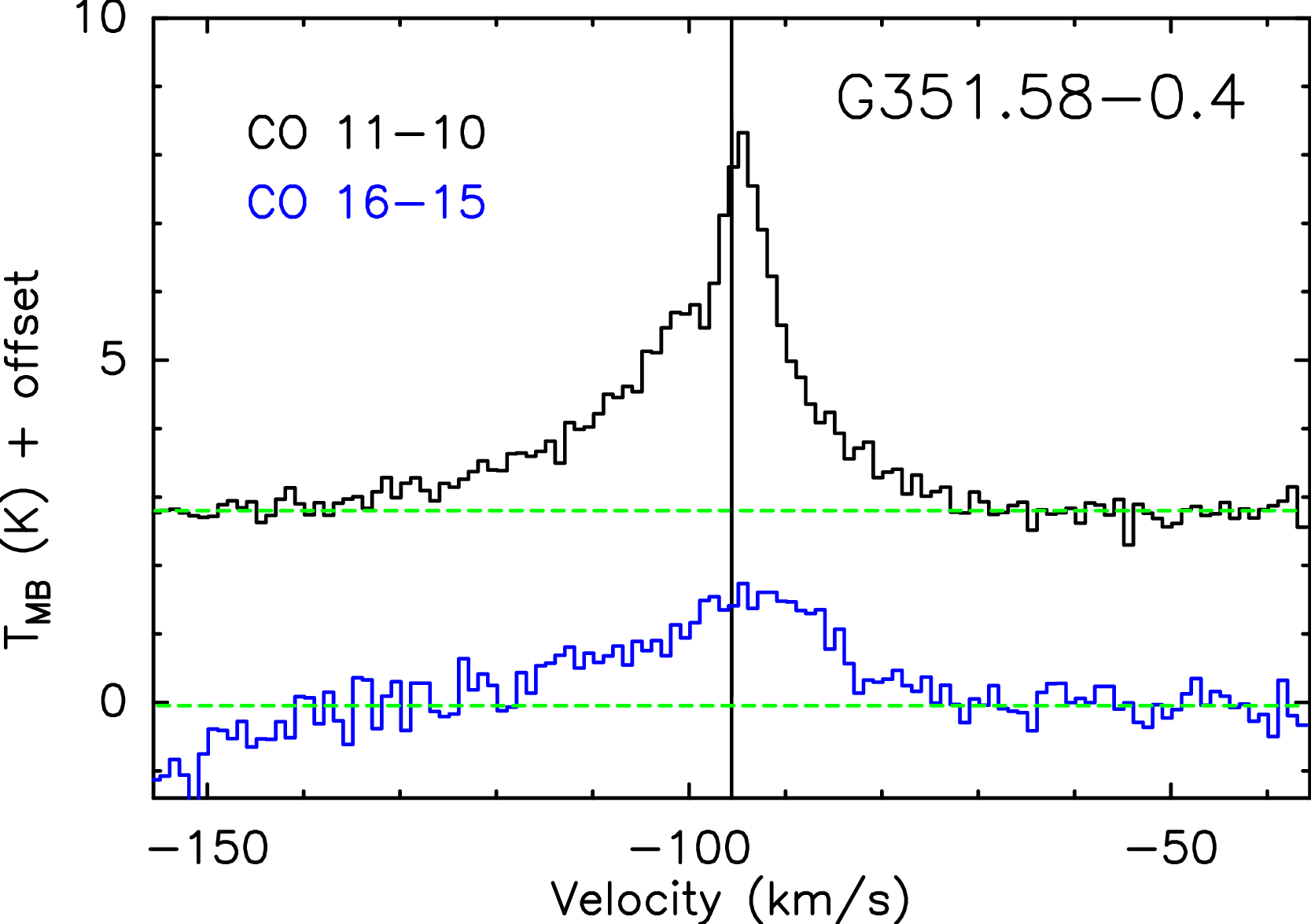}}
   \end{subfigure}
   \begin{subfigure}{0.33\textwidth}
      \centering
      \resizebox{0.9\hsize}{!}{\includegraphics{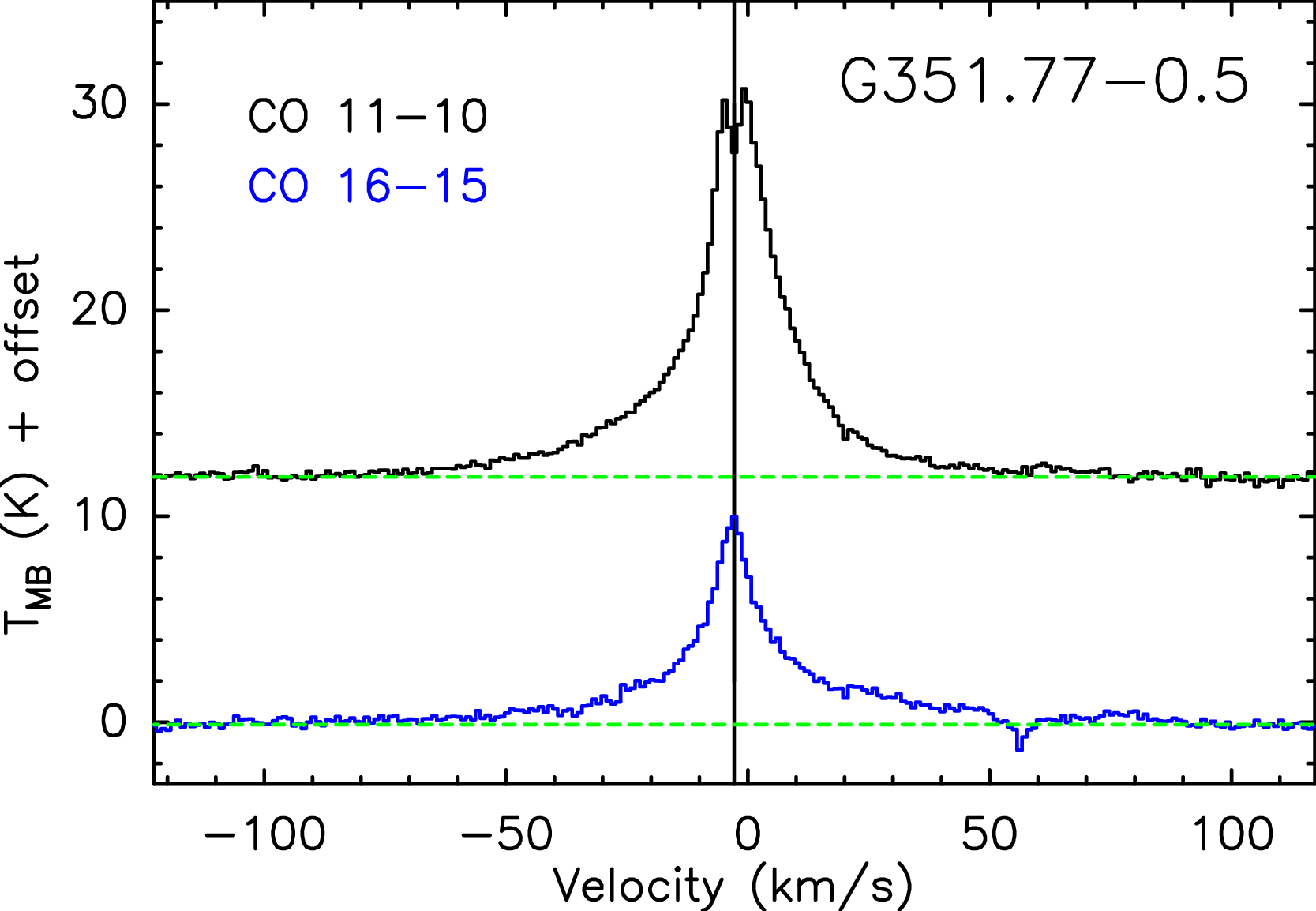}}
   \end{subfigure}%
   \hspace{0.17cm}
   \begin{subfigure}{0.33\textwidth}
      \centering
      \resizebox{0.9\hsize}{!}{\includegraphics{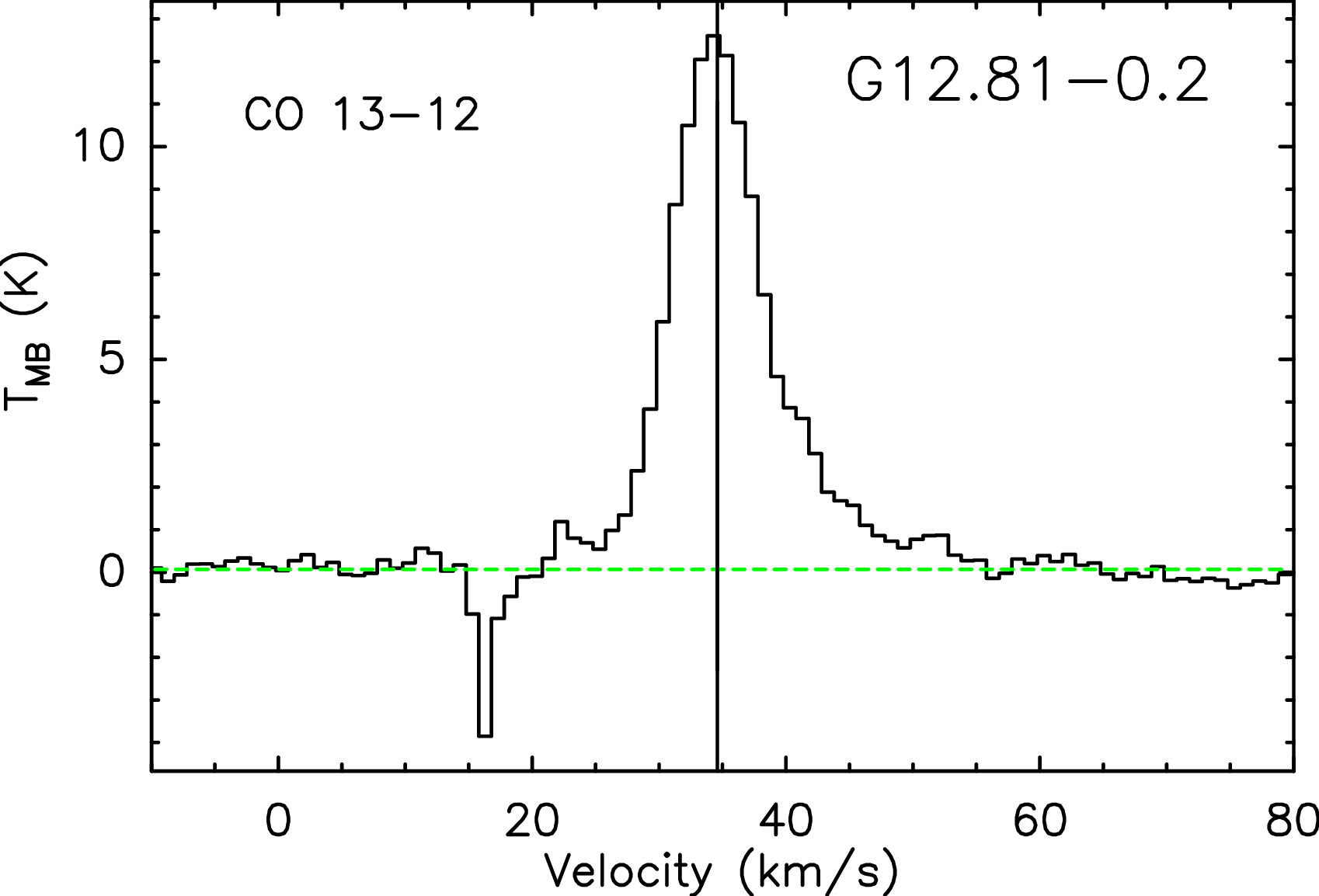}}
   \end{subfigure}%
   \begin{subfigure}{0.33\textwidth}
      \centering
      \resizebox{0.9\hsize}{!}{\includegraphics{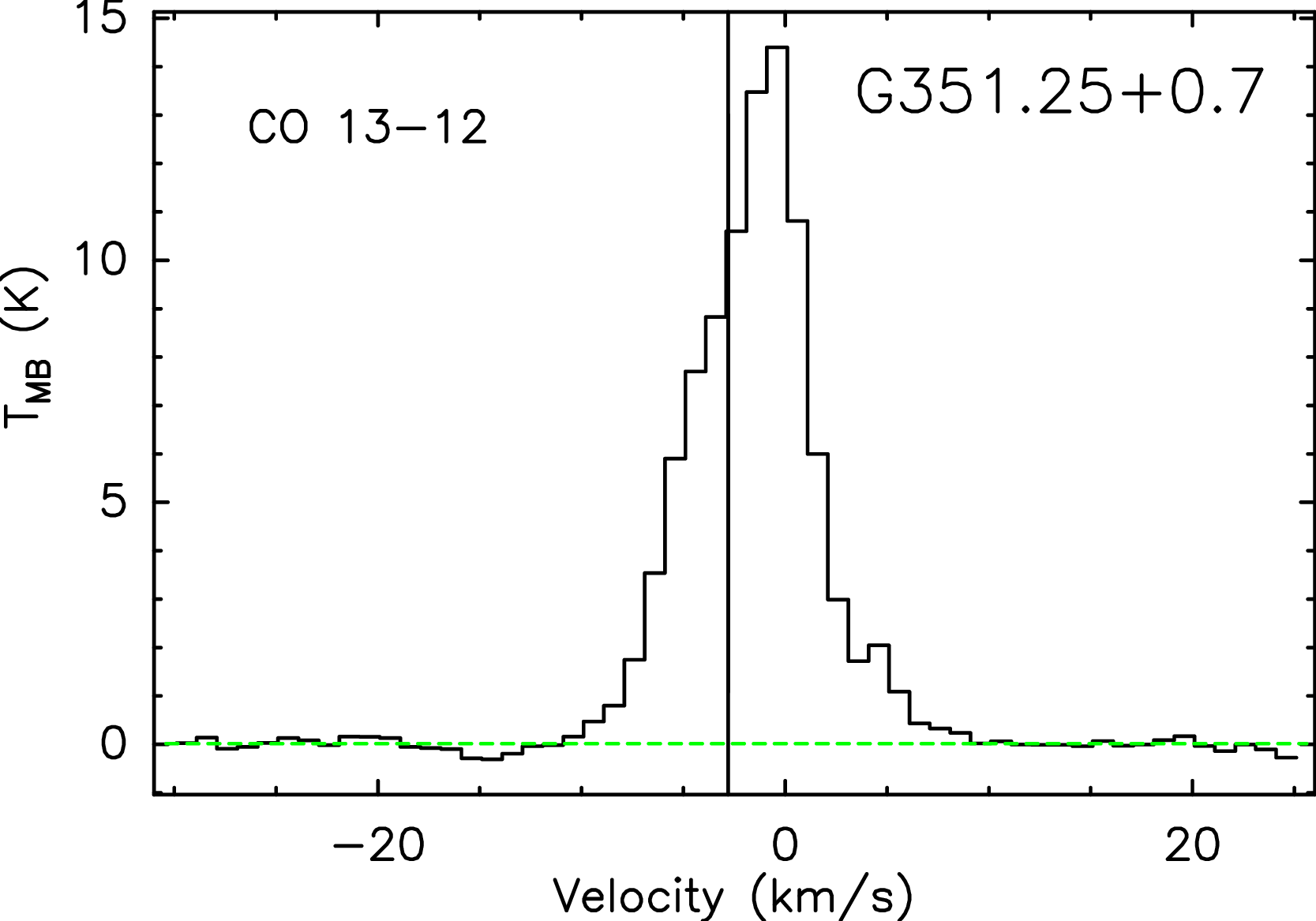}}
   \end{subfigure}
   \caption{SOFIA line profiles of CO $J$=11--10 (black), 16--15 (blue), 13--12 (bottom right) transitions. All spectra are resampled to a common spectral resolution of 1.0\,km\,s$^{-1}$. Black vertical lines show values of $V_{\text{lsr}}$ (see Table\,\ref{tab:catalog}). Green horizontal lines show baselines. }
   \label{fig:present_highJCO}
\end{figure*}

\begin{figure*}[h!]
    \centering
  \begin{subfigure}{0.45\textwidth}
      \resizebox{0.9\hsize}{!}{\includegraphics{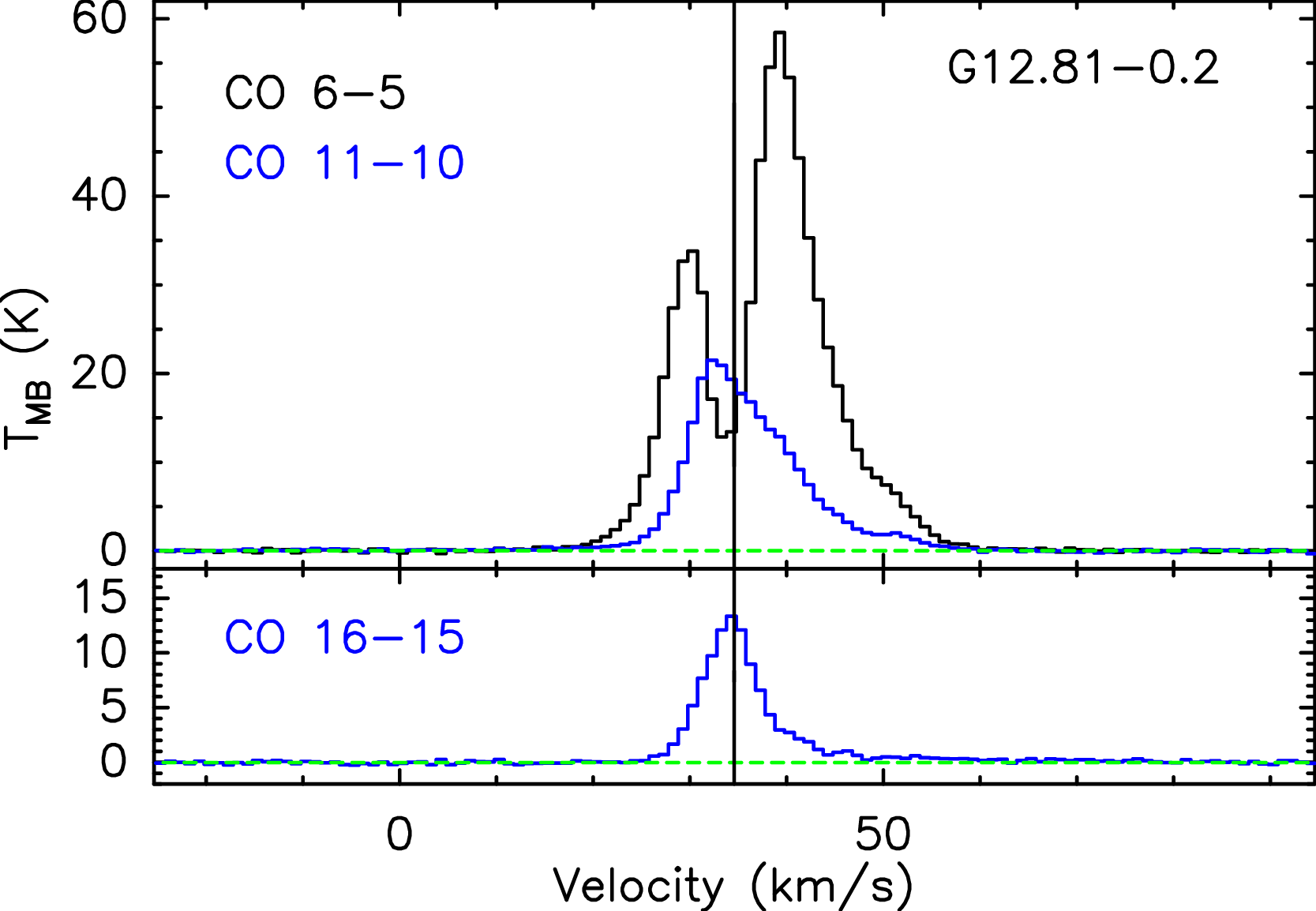}}
   \end{subfigure}%
   \begin{subfigure}{0.45\textwidth}
      \resizebox{0.9\hsize}{!}{\includegraphics{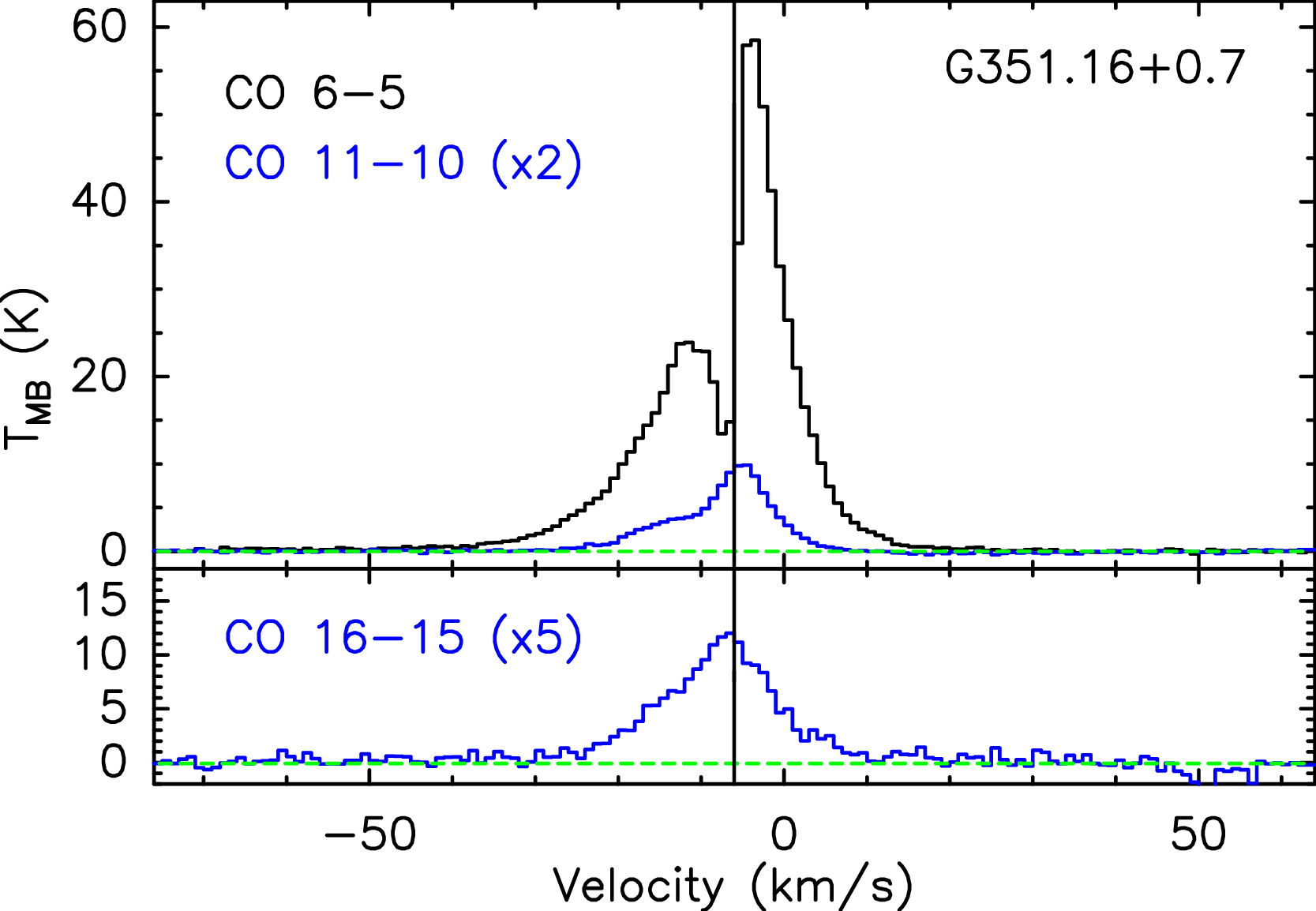}}
   \end{subfigure}
   \begin{subfigure}{0.45\textwidth}
      \resizebox{0.9\hsize}{!}{\includegraphics{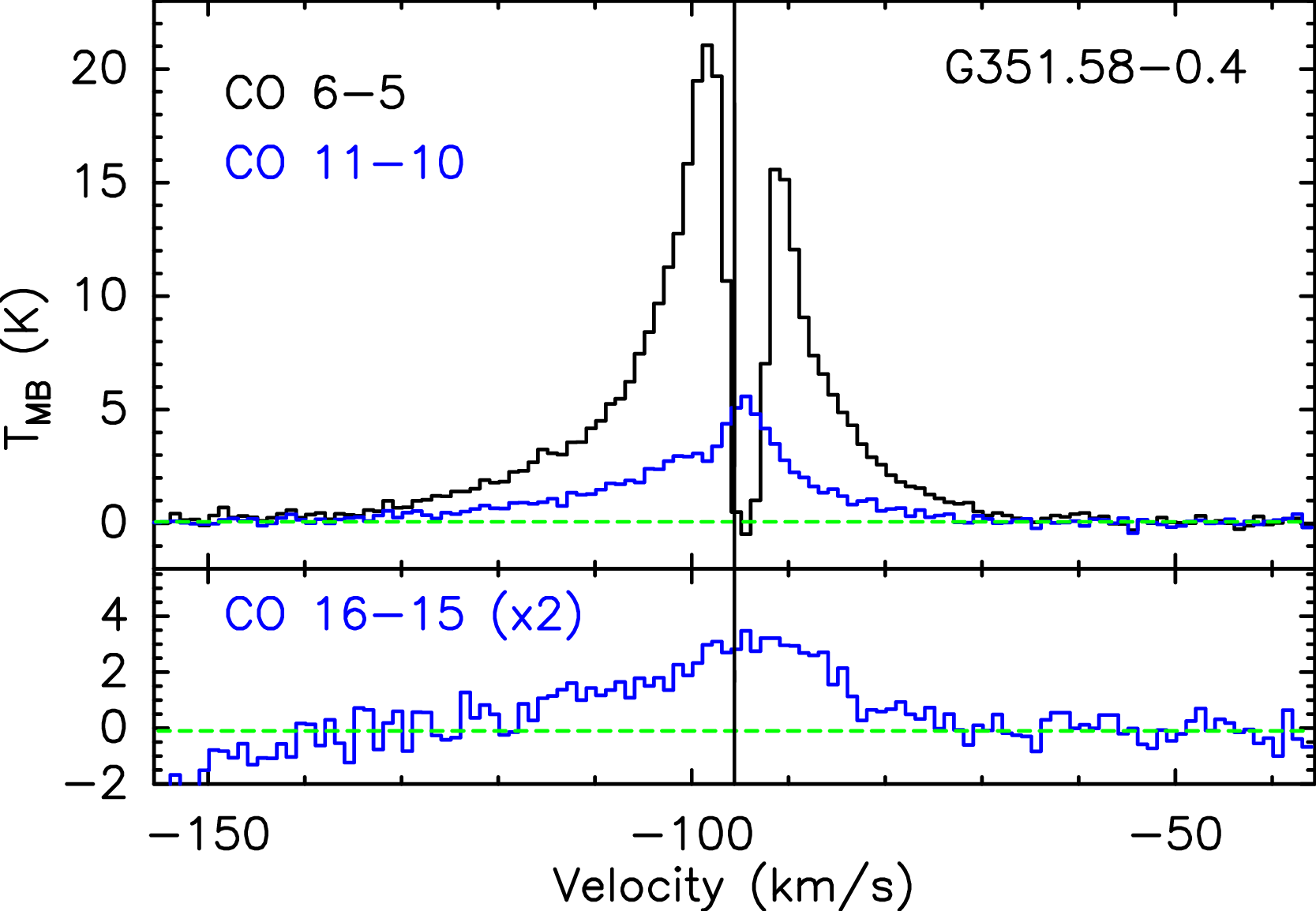}}
   \end{subfigure}
   \begin{subfigure}{0.45\textwidth}%
      \resizebox{0.9\hsize}{!}{\includegraphics{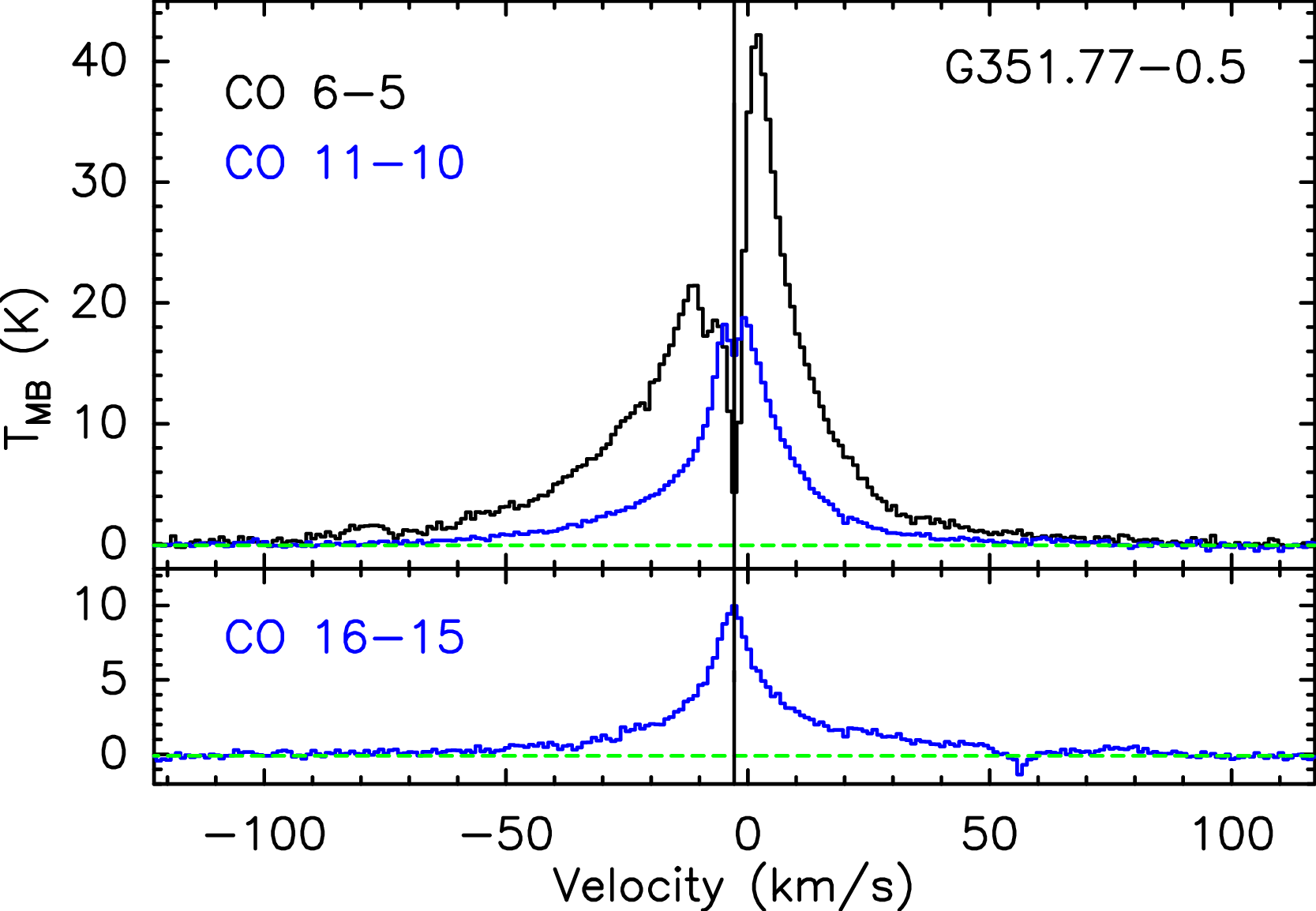}}
   \end{subfigure}
   \caption{SOFIA/GREAT line profiles of the \coet{} and 16--15 lines as well as the \cosf{} lines.
   Source velocities ($V_{\text{lsr}}$) are shown with vertical lines. The lines are smoothed to a common bin of 1.0\,km\,s$^{-1}$.}
   \label{fig:overlay}
\end{figure*}

For this work, we originally selected a representative sample of 20 sources grouped within 4 star-forming regions in the Galactic plane. Among them, 13 sources within 3 regions were successfully observed with SOFIA. Table \ref{tab:catalog} shows the final list of sources with the overview of their properties and evolutionary stages. The sample consists of 3 protostellar (24d), 7 young stellar object (IRb), and 3 H II regions (HII), with $L_\mathrm{bol}$ from $1.6 \times 10^3$ to $4.6 \times 10^5$ L$_{\odot}$ and $M_\mathrm{clump}$ from $1.6 \times 10^2$ to $2.3 \times 10^3$ M$_{\odot}$ \citep{konig2017atlasgal,urquhart19_mopra,urqu22}.

\subsection{SOFIA observations and data reduction}
Observations of the \coet{} and 16--15 lines were collected using the SOFIA/GREAT \citep{heyminck2012great, risacher2015first} and upGREAT receivers \citep{risacher2018}. Our program \lq\lq Probing high-$J$ CO through the evolution of high-mass star forming clumps'' (project IDs 02\_0102 \& 03\_0103; PI: F. Wyrowski) run during Cycle 2 (2014 May) and Cycle 3 (2016 May).

GREAT was a high resolution, dual-color spectrometer ($R\geq10^7$) initially designed for single-beam observations. In 2014, we used its L1 and L2 channels to obtain simultaneous coverage of bands in the 1.25--1.52 THz and 1.80--1.90 THz windows, respectively. In 2016, we combined the GREAT's L1 channel with the upGREAT Low Frequency Array (LFA) which covered the 1.83--2.07 THz window in two polarizations. The 7-pixel hexagonal setup of the LFA provided spatial information about the line emission whereas each pixel had an FWHM beam size of  14.8$\arcsec$ on the sky\footnote{Observer's Handbook for Cycle 3: \url{https://www.sofia.usra.edu/sites/default/files/ObsHandbook-Cy3.pdf}}. The corresponding beam size in the L1 channel was 19.9$\arcsec$ in 2014 and 19.1$\arcsec$ in 2016. The higher frequency L2 channel provided a FoV of 14.1$\arcsec$. The adopted main beam efficiencies ($\eta_\mathrm{MB}$) are 0.7 (in 2014) and 0.66 (in 2016) for the L1 channel, 0.65 for L2 channel, and 0.65 for the central spaxel of LFA. Data are processed and reduced by SOFIA/GREAT staff and released at product level 3 where first order baselines have been subtracted. Most of the data are ready to use, except for \costft{} spectra of G13.66$-$0.6 where an additional third order baseline was subtracted. Spectral resolutions are presented in Table\,\ref{tab:observations}. To perform the analyses without any spectral resolution bias, all spectra were smoothed to a common resolution of 1.0\,km\,s$^{-1}$.

For G12.81$-$0.2 and G351.25+0.7, additional line observations were collected using the SOFIA/4GREAT receiver \citep{duran2020}. The observations were done in 2019 under project \lq\lq high-$J$ CO observations towards high-mass star forming clumps'' (project ID 83\_0711; PI: H. T. Dat). 4GREAT was a single-beam system with four sub-receivers (4G-1 to 4G-4) and could observe four spectral windows simultaneously. The 4G-3 and 4G-4 modules, which cover the 1.24--1.52 THz and 2.49--2.69 THz windows, were tuned to map the \cott{} and 22--21 transitions. The maps were scanned in $5\times5$ grids with centers $6\arcsec$ away from each other. The typical beam sizes for the 4G-3 and 4G-4 modules are $20\arcsec$ and $10.5\arcsec$, respectively \citep{duran2020}. Main beam efficiencies are 0.7 for 4G-3 and 0.57 for 4G-4. Observations of the \cottto{} line are affected by instrumental standing waves that make it difficult to confidently detect line emission. The noise levels of averaged spectra range from 0.40\,K to 0.88\,K at $\Delta v$ of 0.6\,km\,s$^{-1}$.

\begin{figure*}
\centering
    \begin{subfigure}{0.45\hsize}
      \centering
      \resizebox{0.9\hsize}{!}{\includegraphics{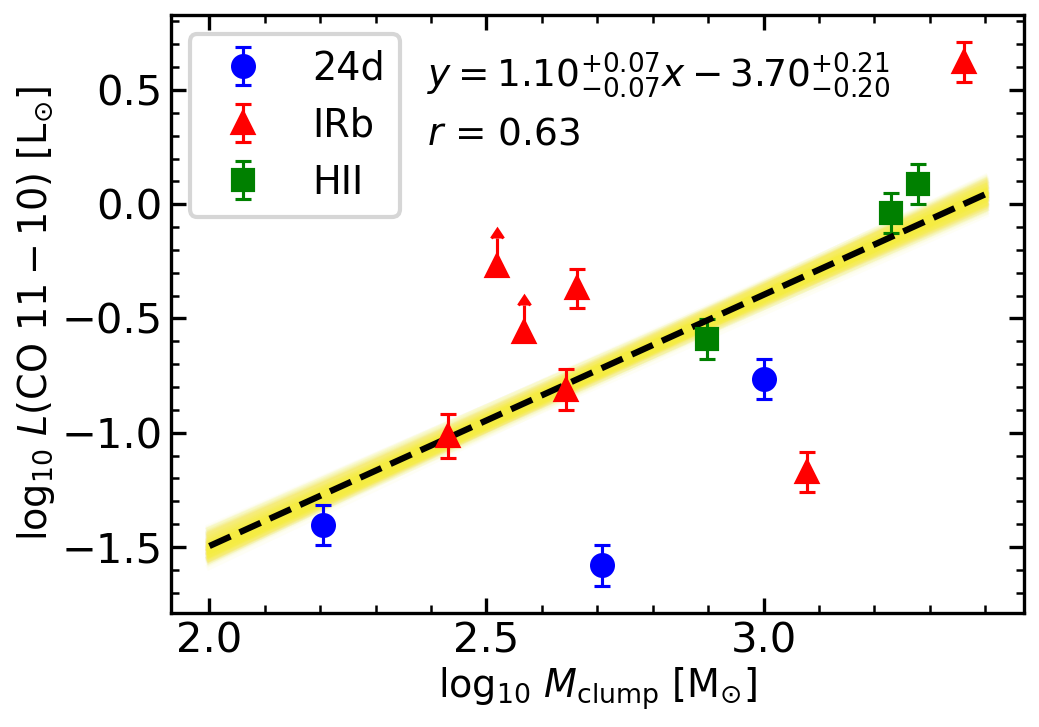}}
   \end{subfigure}%
   \begin{subfigure}{0.45\hsize}
      \centering
      \resizebox{0.9\hsize}{!}{\includegraphics{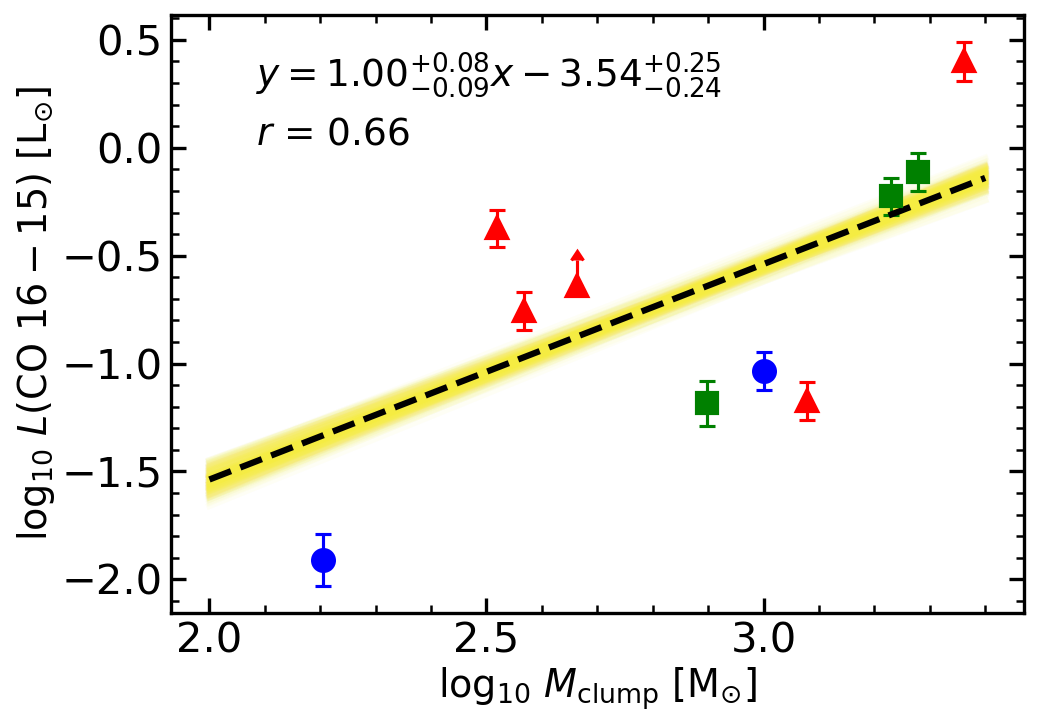}}
   \end{subfigure}
   \begin{subfigure}{0.45\hsize}
      \centering
      \resizebox{0.9\hsize}{!}{\includegraphics{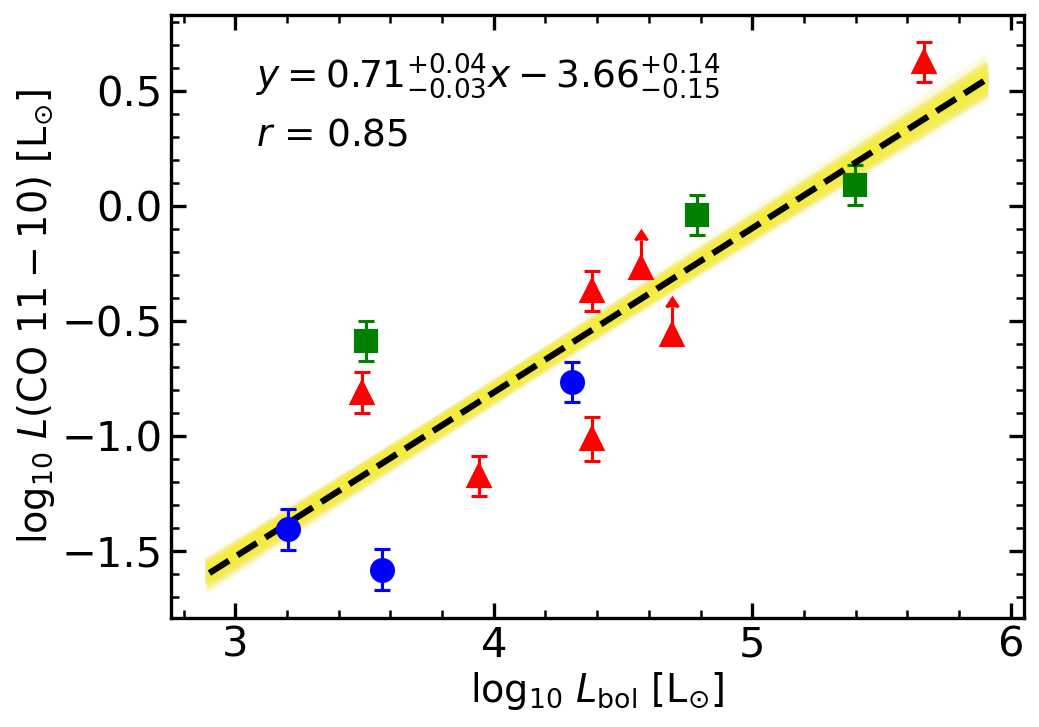}}
   \end{subfigure}%
   \begin{subfigure}{0.45\hsize}
      \centering
      \resizebox{0.9\hsize}{!}{\includegraphics{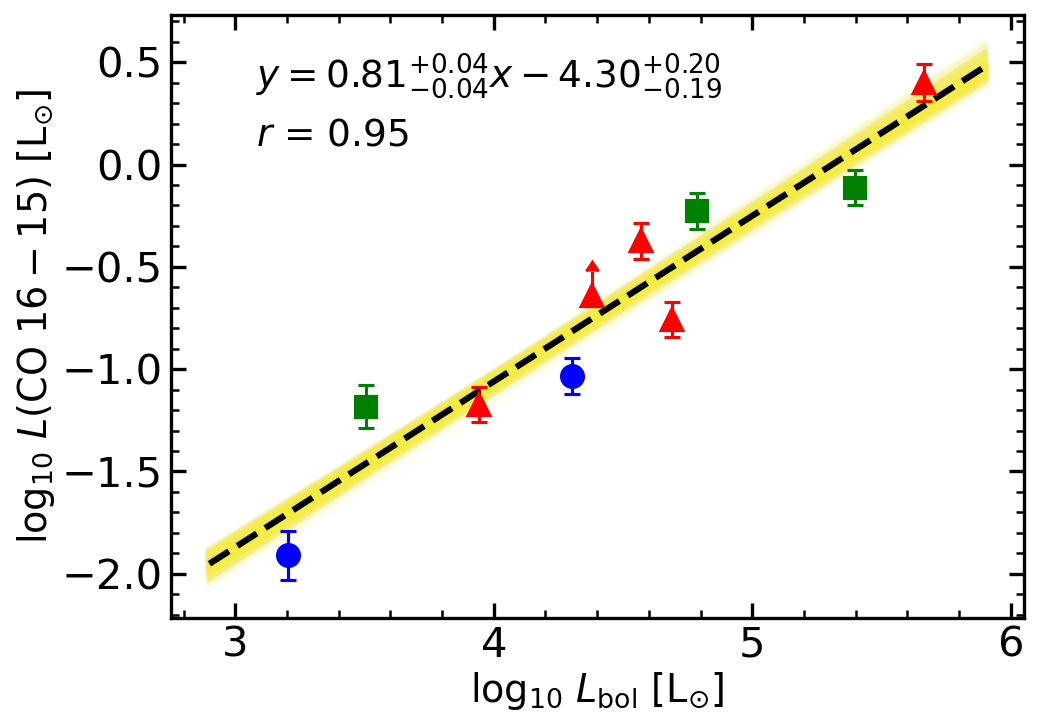}}
   \end{subfigure}
   \caption{Line luminosities of \coet{} and 16--15, as a function of $M_{\mathrm{clump}}$ and $L_\mathrm{bol}$. IR-weak (24d) sources are shown in blue circles, IR-bright (IRb) sources in red triangles, and HII regions (HII) in green squares. Linear regression fit with Markov chain Monte Carlo is shown in dashed black lines and yellow shades. The linear log-log and Pearson correlation coefficients, $r$, are presented on each plot. Objects with self-absorption are shown with an upward arrow, indicating the lower limit for calculated luminosities.}
   \label{fig:lum_vs_srcprop}
\end{figure*}

Data reduction for the \cott{} line was performed with the CLASS program, which is part of the GILDAS\footnote{https://www.iram.fr/IRAMFR/GILDAS/} software developed by the Institut de Radioastronomie Millim{\' e}trique (IRAM). A second order baseline was subtracted, and the spectra were also smoothed to an adequate \mbox{1.0\,km\,s$^{-1}$}. For this study, we extracted averaged spectra within a beam of $20\arcsec$.

\subsection{Additional observations and ancillary data}
Additional single pointing observations of \ttcotn{} and \cetone{} were conducted with the Herschel-Heterodyne Instrument for the Far-Infrared \citep[HIFI,][]{hifi} onboard of the \textit{Herschel} space telescope. Observations for 10 sources (Appendix \ref{app:decomposition-method}) were obtained as part of project \lq A Water survey of massive star forming clumps in the inner Galaxy' (project ID OT2\_fwyrowsk\_3, PI: F. Wyrowski). In addition, archival data for G351.44+0.7 using \textit{Herschel}/HIFI were taken from the \lq Water in star forming region with Herschel' program \citep{irene13,vd21}. Data from the H and V polarizations of the wide-band spectrometer were averaged. Baselines lower than third order were also subtracted. The spectra were converted to a $T_{\mathrm{MB}}$ scale using a forward efficiency of 0.96 and a main beam efficiency of 0.64 for the \ttcotn{} line and 0.74 for the \cetone{} lines, respectively. Finally, the spectra were smoothed to 1.0\,km\,s$^{-1}$. Angular and original spectral resolutions are listed in Table.\,\ref{tab:observations}.

We also use high spectral resolution \ttcosf{} and \cetosf{} data from Dat et al. (in preparation) and \cosf{} from \citet{navarete2019atlasgal}. All three transitions were observed with the CHAMP$^+$ receiver \citep{champ06,champ08} at the Atacama Pathfinder Experiment 12 m submillimeter telescope (APEX) \citep{apex}. The on-the-fly (OTF) scans resulted in datacubes of $80\arcsec \times 80\arcsec$ with angular resolution of $\sim$9$\arcsec$. For comparisons with the higher-$J$ CO observations, averaged spectra with an effective beam size of $20\arcsec$ around the sources were extracted and then smoothed to 1.0\,km s$^{-1}$ for all three lines.

\section{Results and analysis} \label{sec:results_n_analysis}
\subsection{Line detections}  \label{sec:results}
In this section, we examine detection rates of high-$J$ CO lines toward high-mass clumps from ATLASGAL, present their line profiles, and quantify the correlations of integrated intensities with the sources' properties.

Figure\,\ref{fig:present_highJCO} shows the  spectra of high-$J$ CO lines toward the central position of high-mass clumps from our sample (see also Table \ref{tab:catalog}). The pattern of emission is generally compact, based on additional observations offset from the clump centers toward
four sources, see Appendix \ref{app:extent}.

The \coet{} line is detected at 3$\sigma$ or higher levels toward all sources, which span a broad range of evolutionary stages and have diverse properties. The \costft{} line, however, is firmly detected toward 10 out of 13 clumps; G13.66$-$0.6 and G34.41+0.2 show only 2$\sigma$ peaks and G14.19$-$0.2 shows a non-detection. In addition, the \cott{} line was successfully observed and detected toward G12.81$-$0.2 and G351.25+0.7. In Appendix \ref{app:line_parameters}, the peak and integrated intensity of the detected lines are given.

The line profiles of clump central positions exhibit a broad line wing emission, suggesting the presence of outflows (Fig.~\ref{fig:present_highJCO}). The median full width at zero power\footnote{The FWZP is calculated following a procedure described in \cite{irene16}. We first resample the spectra to 3\,km\,s$^{-1}$, and subsequently check the velocity of the channel where the line emission drops below 1$\sigma$.} (FWZP) of 45\,km\,s$^{-1}$ is measured for the \coet{} line and 33\,km\,s$^{-1}$ for the \costft{} line (Appendix C). The broadest profile, with FWZP of 165\,km\,s$^{-1}$, is seen toward G351.77$-$0.5 where high-velocity gas has been detected in CO $2-1$ and $6-5$ lines \citep{leurini2009}. However, multiple pointing observations show a lack of EHV gas component toward the central source; it is only detected at offset outflow positions \citep{leurini2009}, consistent with the analysis of the outflow emission from an intermediate mass protostar Cep E \citep{ruiz12,lefloch15,gusdorf17}. The lack of clear evidence of EHV gas toward our sources may also result from the beam dilution, and could only be addressed using high-angular resolution observations \citep[e.g.,][]{cheng2019}.

The velocity ranges of high-$J$ CO lines resemble those detected in \cosf{} toward the same sources (Fig.\,\ref{fig:overlay}). Self-absorption features are seen in the \coet{} line profiles toward G351.25+0.7 and G351.77$-$0.5. In addition, G12.81$-$0.2 and G35.20$-$0.7 have tilted peaks which could be an indication of self-absorption. The latter source shows also a sign of self-absorption in the \costft{} line. Other profile asymmetries, in particular the triangular blue-wing shape of G351.16+0.7, resemble those of high$-J$ CO emission from a photodissociation region in M17 SW  \citep{jp2015}.

For G34.26+0.15, an additional narrow peak is seen at \mbox{$\sim $38\,km\,$\mathrm{s}^{-1}$} in both the \coet{} and \mbox{\costft{}} spectra. This feature is an artefact due to over-corrected mesospheric CO, which shows the limitations of the adopted atmospheric model \citep[see also, ][]{gusdorf16}.

For G34.40$-$0.2, the line profiles of high-$J$ CO lines seem to be shifted by \mbox{$\sim$1\,km\,$\mathrm{s}^{-1}$} from the source velocity obtained from the \cetone{} line (Fig.\,\ref{fig:present_highJCO}). The uncertainty of the Gaussian fit to the  C$^{18}$O line is smaller than 0.25\,km\,$\mathrm{s}^{-1}$, and thus cannot account for the observed shift, suggesting that it may be caused by  self-absorption. Small velocity-shifts are also present in the line profiles of other objects, e.g. G34.26+0.15 and G351.25+0.7.

We calculate CO line luminosities, $L_{\mathrm{CO}}$, as $4\pi D^2 F^{\mathrm{CO}}_{\lambda}$, where $D$ is the distance to the source (Table \ref{tab:catalog}) and $F^{\mathrm{CO}}_{\lambda}$ is the velocity integrated flux in \mbox{W\,m$^{-1}$}. The flux conversion from \mbox{K\,km\,s$^{-1}$} to \mbox{W\,m$^{-1}$} follows Equation 1 in \cite{indriolo17}. Figure\,\ref{fig:lum_vs_srcprop} shows the correlations between $L_{\mathrm{CO}}$ and source properties (Table \ref{tab:catalog}). The significance of the correlations is quantified by the Pearson correlation coefficient $r$, which depends also on the number of data points $N$ \citep{marseille10}.

\begin{table*}[ht!]
\centering
\small
\caption{Detection of line wing emission in high-$J$ CO lines and their comparison to prior outflow detections
 \label{outflows}}
\label{tab:outflow}
\begin{tabular}{llccccc|ccc}
 \hline
 \hline
No. & Source & \mbox{CO\,4--3}\tablefootmark{a} & \cosf{}\tablefootmark{b} &  \mbox{$^{13}$CO\,3--2}\tablefootmark{c} & \mbox{$^{13}$CO\,2--1}\tablefootmark{d} &  \mbox{SiO\,2--1}\tablefootmark{e} & \coet{}\tablefootmark{f} & \cott{}\tablefootmark{f} & \costft{}\tablefootmark{f}\\
 \hline
1 & G351.16+0.7 & \checkmark & \checkmark &\ldots & -- &   -- & \checkmark & \ldots & \checkmark\\
2 & G351.25+0.7 & \checkmark & \checkmark & \ldots & -- & -- & \checkmark & \checkmark & \checkmark\\
3 & G351.44+0.7 & \checkmark & \checkmark &\ldots & --  & -- & \checkmark &  \ldots & \checkmark\\
4 & G351.58$-$0.4 & \checkmark & \checkmark & \ldots & -- &  -- & \checkmark & \ldots & \checkmark \\
5 & G351.77$-$0.5 & \checkmark &  \checkmark &\ldots & \checkmark &   -- & \checkmark & \ldots &  \checkmark \\
\hline
6 & G12.81$-$0.2 & \checkmark & \checkmark &\ldots & \checkmark &  --  & \checkmark & \checkmark &  \checkmark \\
7 & G14.19$-$0.2 & \checkmark & \checkmark &\ldots  & \checkmark &  \checkmark  & \checkmark & \ldots &  -- \\
8 & G13.66$-$0.6 & \checkmark &  \checkmark & \ldots & -- & \checkmark & -- & \ldots &  -- \\
9 & G14.63$-$0.6 & \checkmark & \checkmark &\ldots & -- &  \checkmark & \checkmark & \ldots &  -- \\
\hline
10 & G34.41+0.2 & \checkmark &  \checkmark &-- & -- &  \checkmark  & \checkmark & \ldots &  -- \\
11 & G34.26+0.15 & \checkmark & \checkmark &\checkmark & -- &  \checkmark & \checkmark & \ldots &  \checkmark \\
12 & G34.40$-$0.2 & \checkmark & \checkmark & -- & -- &  \checkmark  & \checkmark & \ldots &  -- \\
13 & G35.20$-$0.7 & \checkmark & \checkmark & -- & -- &  --  & \checkmark & \ldots &  \checkmark \\
  \hline
\end{tabular}
\begin{flushleft}
\tablefoot{The symbols refer to: '--' non-detection, $\checkmark$ detections, '...' lack of data or unable to detect.
\tablefoottext{a}{Based on single pointing observations from APEX/FLASH$^+$ \citep{navarete2019atlasgal}.}
\tablefoottext{b}{Obtained from \cosf{} maps using APEX CHAMP$^+$ \citep{navarete2019atlasgal}.}
\tablefoottext{c}{Data available only for the sources No. 10--13 in the G34 cloud \citep{yang18}, due to a limited coverage of the CHIMPS survey \citep{rigby16}.}
\tablefoottext{d}{Based on the identification of line wings toward SEDIGISM sources \citep{yang22}.}
\tablefoottext{e}{Sources No. 1--5: Mopra molecular lines survey of ATLASGAL sources \citep{urquhart19_mopra}; Sources No. 6--13: IRAM 30m follow-up of ATLASGAL sources \citep{cse16}.}
\tablefoottext{f}{Wing detection following line decomposition in section \ref{sec:profile-decomposition}.}
}
\end{flushleft}
\end{table*}

\begin{figure*}
    \begin{subfigure}{0.5\hsize}
      \centering
      \resizebox{0.9\hsize}{!}{\includegraphics{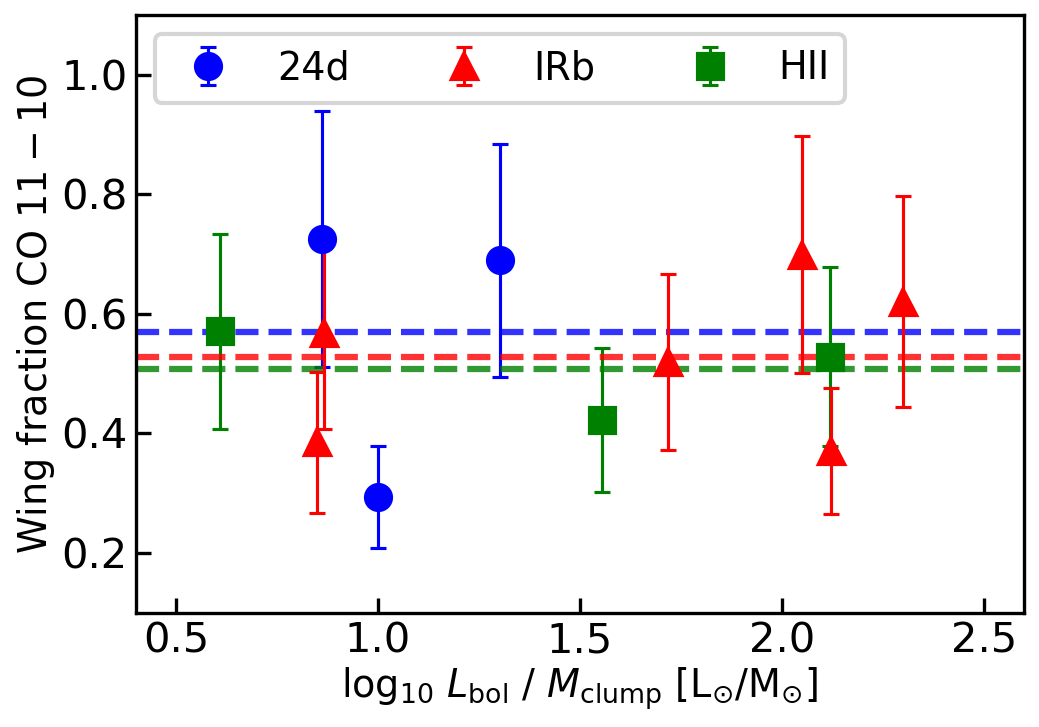}}
   \end{subfigure}%
   \begin{subfigure}{0.5\hsize}
      \centering
      \resizebox{0.9\hsize}{!}{\includegraphics{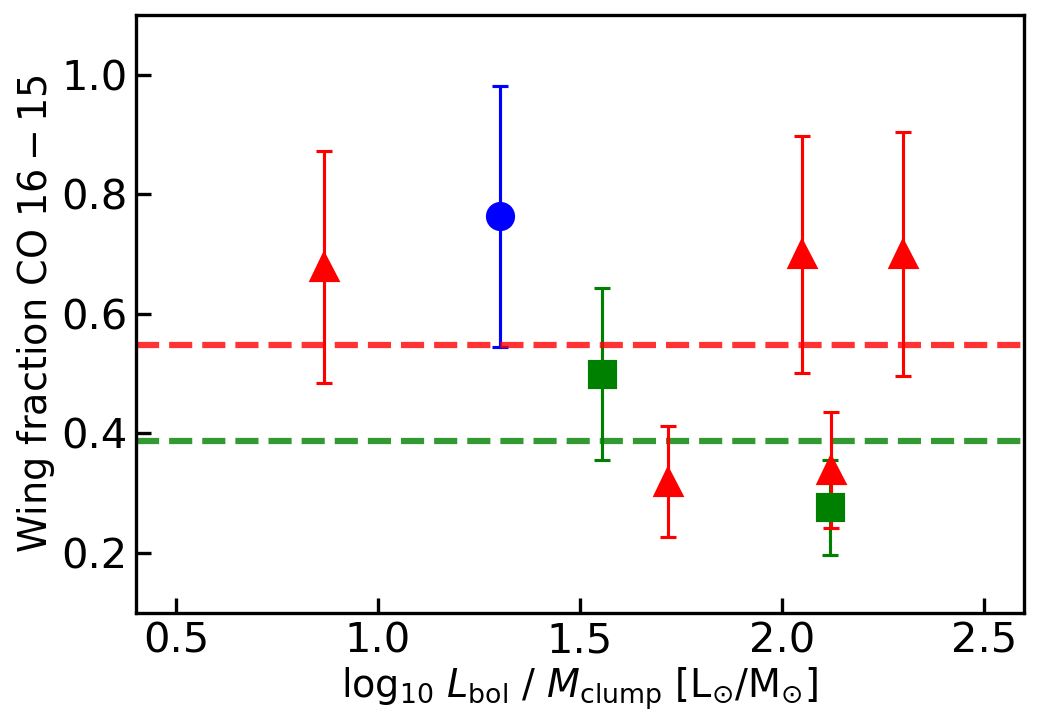}}
   \end{subfigure}
   \caption{The fraction of integrated emission in the line wings of \coet{} (left) and \costft{} (right), as a function of
    $L_\mathrm{bol}/M_\mathrm{clump}$ (Table \ref{tab:catalog}). Dashed horizontal lines show the mean wing fraction at each evolutionary stage. The color-coding is the same as in Fig\,\ref{fig:lum_vs_srcprop}.
    }
   \label{fig:Lmr_vs_wingfraction}
\end{figure*}

\begin{figure*}
\centering
    \begin{subfigure}{0.45\hsize}
      \centering
      \resizebox{0.9\hsize}{!}{\includegraphics{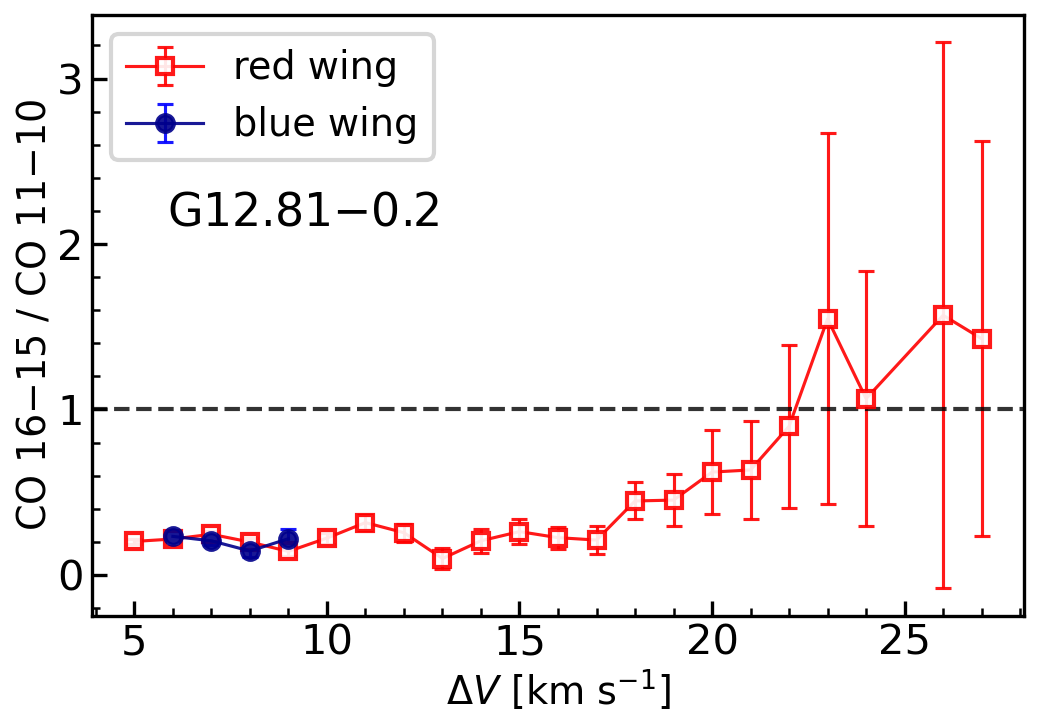}}
   \end{subfigure}%
   \begin{subfigure}{0.45\hsize}
      \centering
      \resizebox{0.9\hsize}{!}{\includegraphics{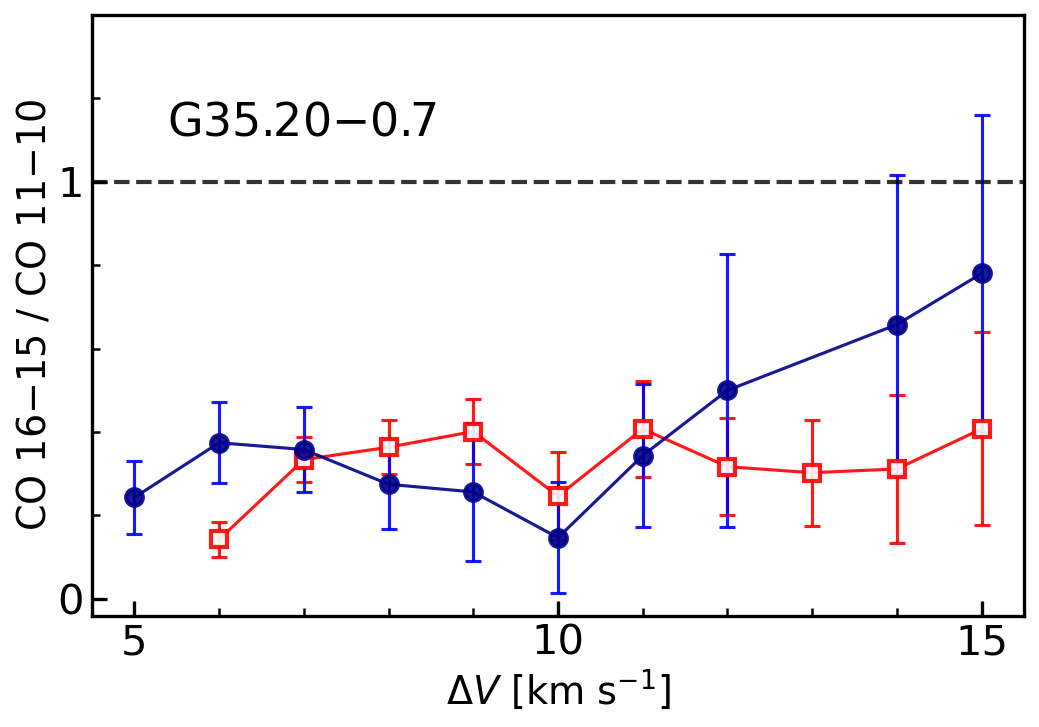}}
   \end{subfigure}
   \begin{subfigure}{0.45\hsize}
      \centering
      \resizebox{0.9\hsize}{!}{\includegraphics{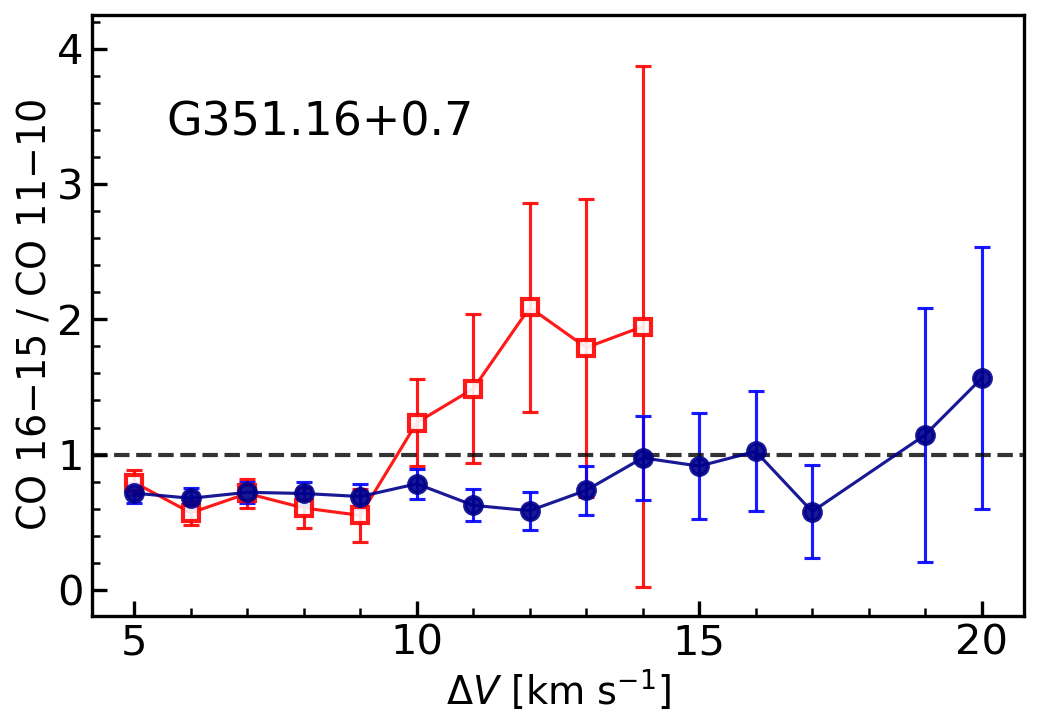}}
   \end{subfigure}%
   \begin{subfigure}{0.45\hsize}
      \centering
      \resizebox{0.9\hsize}{!}{\includegraphics{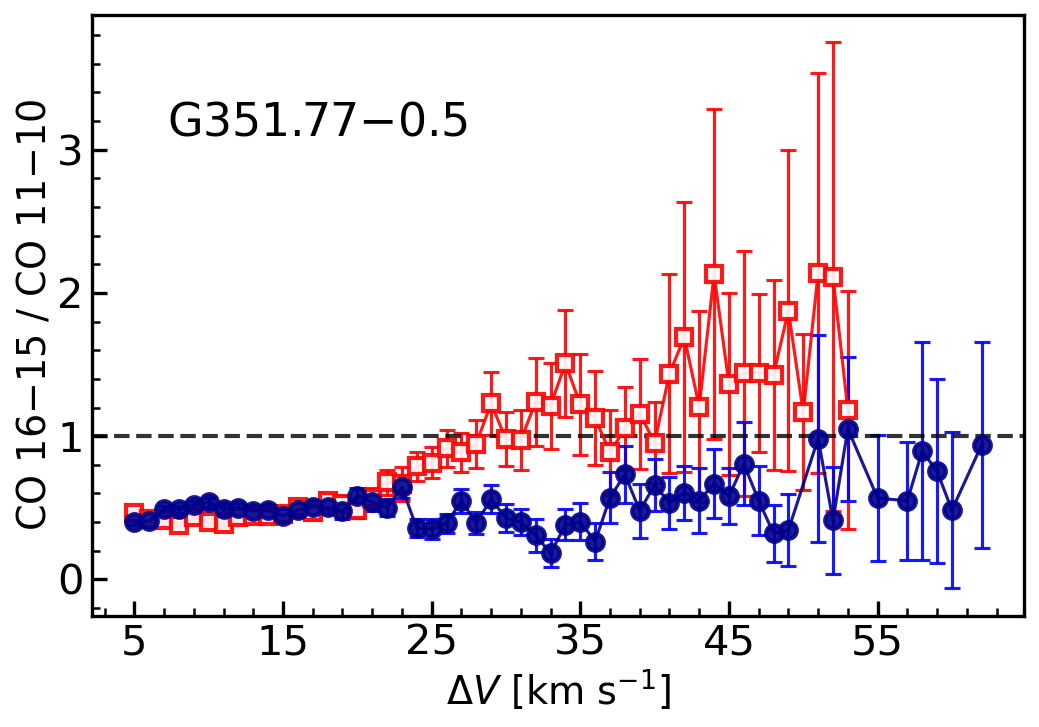}}
   \end{subfigure}
   \caption{The ratio of line wing emission in \costft{} and 11--10 transitions as a function of absolute velocity offset from source velocity. The red-shifted emission is shown in red squares, and the blue-shifted is in blue circles. The dashed horizontal line presents the level above which \costft{} is greater than \coet{}. }
   \label{fig:line_wing_ratio_indiv}
\end{figure*}

Both \coet{} and \costft{} line luminosities show weak correlations ($r$ of 0.63--0.66) with the clump mass, $M_{\text{clump}}$, tracing primarily a cold gas and dust reservoir \citep{konig2017atlasgal}. Stronger correlations ($r$ of 0.85--0.95) are found for the high-$J$ CO line luminosities and clump bolometric luminosities, $L_{\text{bol}}$, in line with previous studies using \mbox{CO\,10--9} (see Section \ref{sec:dis}). Noteworthy, clumps at different evolutionary stages do not show any clear trend in Figure.\,\ref{fig:lum_vs_srcprop}, suggesting similar underlying physical processes are responsible for high-$J$ CO emission from all sources in the sample.

In summary, high-$J$ CO emission is detected in high-mass clumps and correlates most strongly with clump bolometric luminosity. The line shapes show that high-velocity gas most likely associated with the outflows.

\begin{figure}
    \begin{subfigure}{1\hsize}
      \centering
      \resizebox{0.85\hsize}{!}{\includegraphics{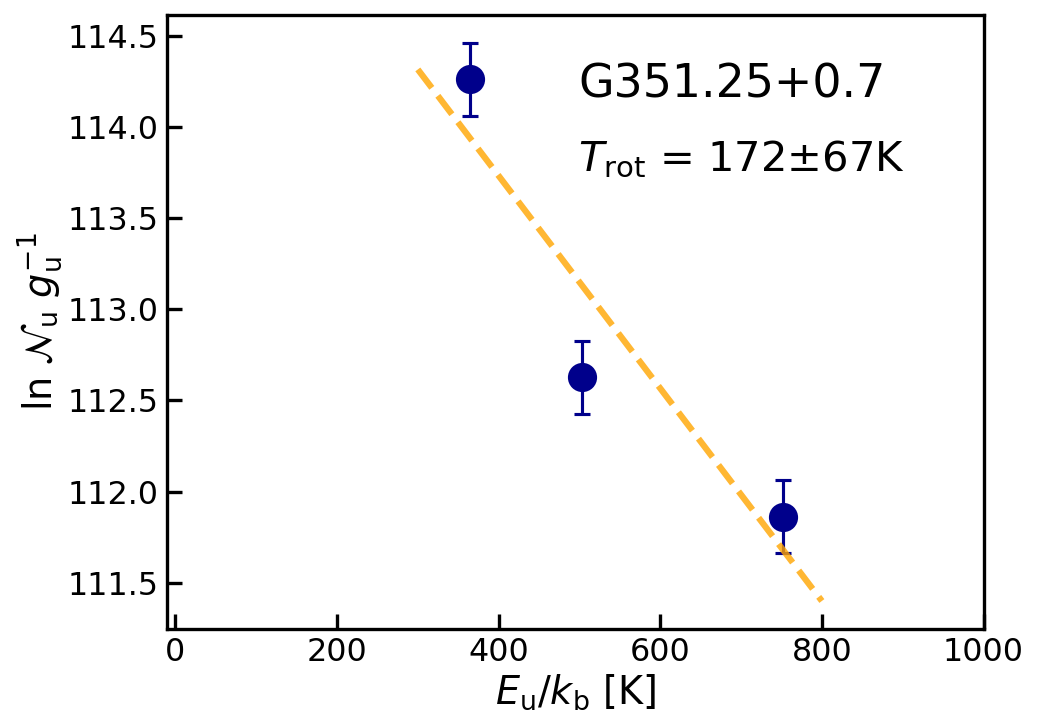}}
   \end{subfigure}
   \begin{subfigure}{1\hsize}
      \centering
      \resizebox{0.85\hsize}{!}{\includegraphics{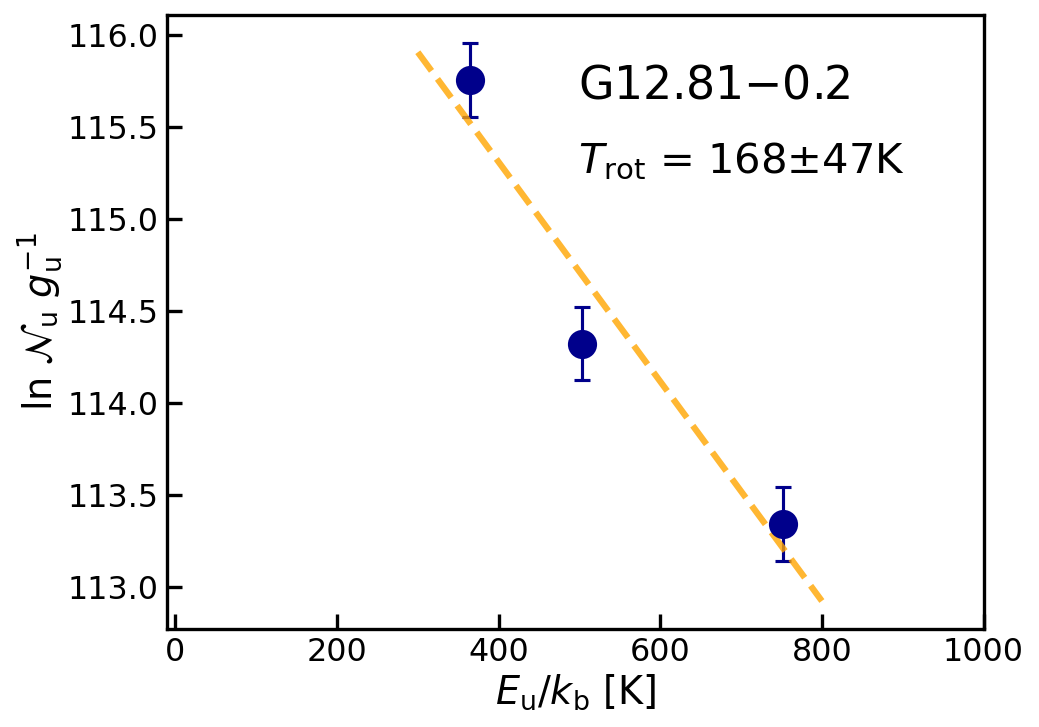}}
   \end{subfigure}
   \caption{Rotational diagrams of CO for G12.81$-$0.2 and G351.25+0.7, which are based on the observations of full profile CO transitions with $J_\text{u}$ of 11, 13, and 16. The natural logarithm of the number of emitting molecules from a level u, $\mathcal{N}_\mathrm{u}$ (dimensionless), divided by the degeneracy of the level, $g_\text{u}$, is shown as a function of the upper level energy, $E_\mathrm{u}$/$k_\mathrm{B}$, in Kelvins. Detections are shown as blue circles. Dashed orange lines show linear regression fits to the data; the resulting rotational temperatures are provided on the plots with the associated errors from the fit.}
   \label{fig:rotation_diagram}
\end{figure}

\begin{figure*}
   \begin{subfigure}{0.5\hsize}
      \centering
      \resizebox{0.9\hsize}{!}{\includegraphics{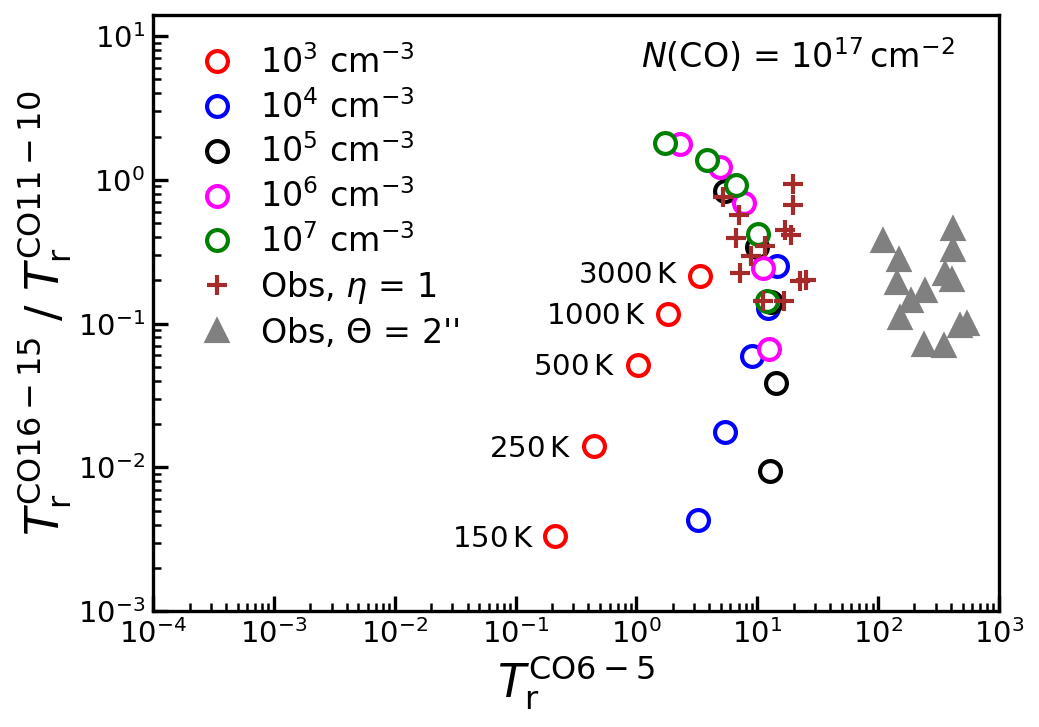}}
   \end{subfigure}%
   \begin{subfigure}{0.5\hsize}
      \centering
      \resizebox{0.9\hsize}{!}{\includegraphics{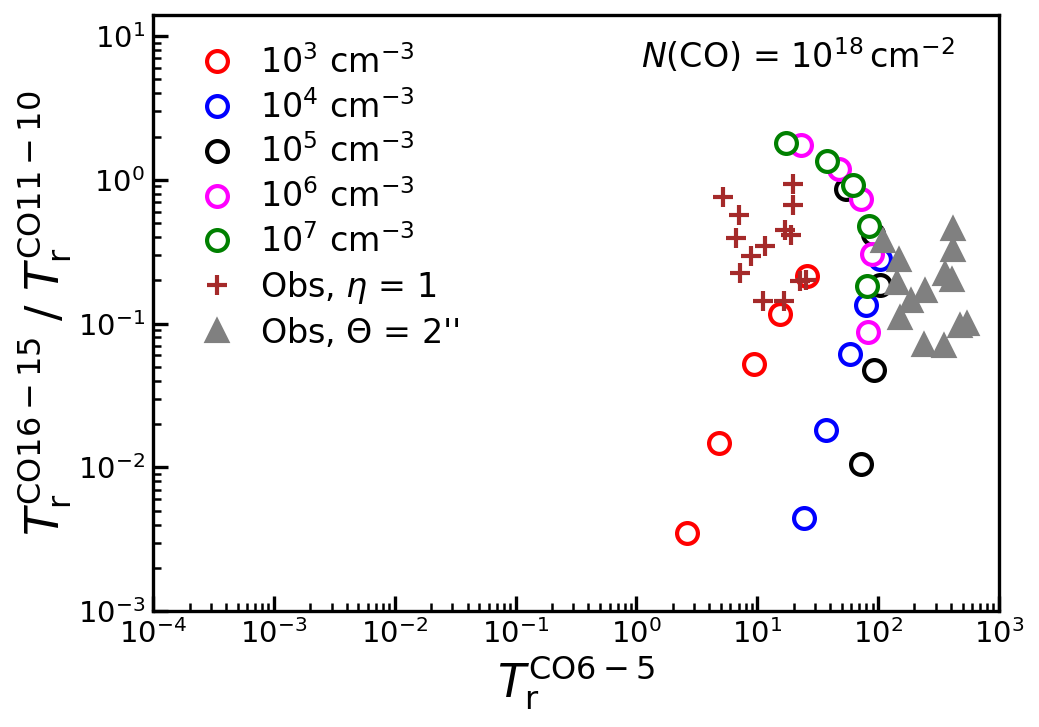}}
   \end{subfigure}%
   \caption{CO excitation conditions from RADEX models versus observations. The plots present models at different $N$(CO) of $10^{17}$ and \mbox{$10^{18}$\,cm$^{-2}$}. The models are presented in empty circles, and their colors correspond to different hydrogen volume density, $n_{\mathrm{H}_2}$, between $10^3$ and $10^7$\,cm$^{-3}$. At each volume density level, four temperatures: 150, 250, 500, 1000, and 3000\,K are sampled. Observations assuming a beam filling factor of 1 are shown in crosses, while observations assuming a tiny source of $2\arcsec$, which correspond to an extreme case of small beam filling factor, are shown in triangles.}
   \label{fig:radexgrid}
\end{figure*}

\subsection{Profile decomposition}\label{sec:profile-decomposition}
We use mid-$J$ ($6 \lesssim J \lesssim 10$) CO rare isotopologue lines to subtract the envelope component from the line profiles of \coet{} and \costft{}. This way, we isolate the high-velocity emission associated with the line wings.

The emission in the line wings is characterised using a decomposition method which is described in detail in Appendix \ref{app:decomposition-method}. Briefly, the decomposition procedure aims to subtract the contribution from the envelope, as traced by rare isotopologue emission, resulting in the residual outflow component \citep{cod04,van2007inferences,deV14,yang18}. This method was initially used for kinematical studies of methanol masers, and subsequently adopted for low-$J$ CO line profiles. Here, we use the version described in \cite{yang18} which do not account for the opacity broadening, because the high-$J$ CO lines are likely optically thin. Rare isotopologue lines are used as a proxy for the envelope emission; here, depending on data availability and detection, we used the emission of the \cetone{} line for eight sources, the \ttcotn{} line for three sources, the \cetosf{} for one source, and the \ttcosf{} for one source (see Appendix \ref{app:decomposition-method}).

We identify line wing emission in the \coet{} line toward all sources except G13.66$-$0.6 (Table \ref{tab:outflow}). The wings in the \costft{} line are seen only toward 8 out of 10 sources with the $3\sigma$ line detection. Properties and profiles of all wing emission are shown in Appendix \ref{app:decomposition-method}.

The ubiquity of line wings is consistent with previous detections of the outflows toward the same sources using lower-$J$ lines of CO and SiO (Table \ref{tab:outflow}). In particular, all sources from our sample show line wings in the \cosf{} line \citep{navarete2019atlasgal}. The non-detection of the \coet{} line wing in G13.66$-$0.6 could be either due to the low S/N (Fig.\,\ref{fig:present_highJCO}) or a lack of recent heating of the outflow gas due to shocks \citep{kar13,kri17}. The \mbox{$^{13}$CO\,2--1} wings have only been seen toward G351.77$-$0.5, G12.81$-$0.2, and G14.19$-$0.2 \citep{yang22} due to limited sensitivity, illustrating the difficulty in detecting line wings in rare CO isotopologues \citep[see also, ][]{ste18,ste19}. Finally, \mbox{SiO\,2--1} have been observed toward our sources \citep{urquhart19_mopra,cse16} and line wings are detected in six of them. All the non-detections, in fact, show line wings in the high-$J$ CO lines (Table \ref{tab:outflow}), indicating that additional factors play a role in the excitation of SiO and CO lines.

Detecting outflows toward distant star-forming clumps is often a hampered by confusion. Background and foreground galactic sources along the line-of-sight might contribute to the wing emission, which may result in false outflow detections. We note, however, that the high detection statistics (> 60\%) of line wings and the wings' smooth shapes in our source sample are very unlikely be explained by source confusion. The high-$J$ CO emission is typically well-confined to the regions with active star formation.

In conclusion, our decomposition method results in the estimate of line wing emission toward 12 and 8 sources in the \coet{} and \costft{} lines, respectively.

\subsection{CO line wing emission}\label{sec:ratio}
Decomposition of the line profiles allows us to quantify the amount of high-$J$ CO emission in the line wings, and its contribution to the entire line profiles. Furthermore, the ratio of the two CO transitions can be studied as a function of gas velocity.

The fraction of emission in the line wings of the \coet{} transition ranges from $\sim$29 to 73\%, whereas the mean fraction of each evolutionary stage is $\sim$50\%, suggesting that there is no dependence with the source evolution (Fig.~\ref{fig:Lmr_vs_wingfraction}). The fraction of emission in \costft{} line wings is similar to the \coet{} transition, and ranges from $\sim$28 to 76\%. These results are consistent with the fraction of line wing emission measured toward two High Mass Protostellar Objects: AFGL 2591 in both \coet{} ($\sim$37\%) and \costft{} \mbox{($\sim$34\%)} from \cite{vdWiel13}, and W3 IRS5 in \mbox{CO\,10--9} ($\sim$50\%) from \cite{irene16}.

The fraction of high-$J$ CO emission has been also estimated for several high-mass YSOs by subtracting the envelope contribution from the total, unresolved line profiles. The \textit{Herschel}/HIFI observations of rare isotopologues of CO were used to constrain models of CO (main isotopologue) emission arising from envelopes \citep[e.g., ][]{herpin12,herpin16,kar14,jacq16}. In case of NGC 7538 IRS1, 70--100\% of velocity-unresolved CO $J=15$--14 emission and 3--22\% of CO $J=22$--21 was attributed to the envelope \citep{kar14}. Thus, the contribution of the outflow component was predicted to increase with the rotational level of CO line.

The increase of the relative contribution of the wing emission from $J_\mathrm{up}=11$ to 16 is indeed measured for 4 out of 8 sources in our sample, for which outflow wings are detected in both CO transitions. The fraction of wing emission increases from $\sim$42\% (\coet{}) to 50\% (\costft{}) for G34.26+0.15, from 57\% to 68\% for G351.16+0.7, from 69\% to 76\% for G351.44+0.7, and from 62\% to 70\% for G351.58$-$0.4. The increase is therefore not as sharp as for the \coftft{} and \cottto{} lines of NGC 7538 IRS1, but consistent with a rising contribution of wing emission in higher-$J$ lines.

The amount of emission in the wings of higher-$J$ CO lines allows us to study the gas excitation conditions in the outflowing gas. Assuming that emission in the line wings is optically thin and thermalized, the higher CO line ratios would correspond to higher gas kinetic temperatures, $T_\mathrm{kin}$ (see Sections \ref{sec:lte} and \ref{sec:nonlte}). Figure\,\ref{fig:line_wing_ratio_indiv} shows the observed ratio of \costft{} and 11--10 lines in the red and blue wings as a function of absolute offset from the source velocity. The ratio is calculated in steps of 1.0\,km\,s$^{-1}$, avoiding the line centers ($\pm$5\,km\,s$^{-1}$), and presented for channels where signal-to-noise is above 2.

The ratio of \costft{} and \coet{} increases as a function of velocity for at least a few sources, e.g., the red wing of G12.81$-$0.2, G351.16+0.7, and G351.77$-$0.5, and the blue wing of G35.20$-$0.7 (Fig.~\ref{fig:line_wing_ratio_indiv}). In most of those cases, the highest-velocity emission is stronger in the \costft{} than in the \coet{}. Such trends are consistent with similar studies using CO\,3--2, 10--9, and 16--15 toward a sample of low- to high-mass protostars (see Section \ref{sec:dis_exc}).

In summary, we find a lack of correlation between the fraction of high-$J$ CO integrated emission in the line wings and the clump evolutionary stage. Yet, the fraction increases with the CO rotational level in half of the sources. The ratio of the wing emission in the \costft{} and \coet{} lines increases with velocity in several sources.

\subsection{Molecular excitation in LTE (full profile + wings)}
\label{sec:lte}
Detection of at least two CO lines allows us to determine the rotational temperature of the outflowing gas detected in the line wings under the assumption of LTE. For comparisons with previous studies with \textit{Herschel}/PACS, the calculations are also performed for the velocity-integrated line profiles.

Emission line fluxes of \coet{} and \costft{} are used to calculate the number of emitting molecules, $\mathcal{N}_\mathrm{u}$, for each molecular transition as:
\begin{equation}
\label{rot1}
\mathcal{N}_\mathrm{u}=\frac{L_{\text{CO}}\lambda}{hcA},
\end{equation}
where $L_\text{CO}$ refers to the line luminosity of CO line at wavelength
$\lambda$, $A$ to the Einstein
coefficient, $c$ to the speed of light, and $h$ to the Planck's constant. Note that for two sources,
G12.81$-$0.2 and G351.25+0.7, additional observations of \cott{} are included. The number of emitting molecules, $\mathcal{N}_\mathrm{u}$, is used instead of column densities, because the size of the emitting region is unresolved.

 \begin{table}[ht!]
\centering
\caption{CO rotational excitation for both the integrated line profiles and line wings only assuming LTE}
\label{tab:rot_diag_results}
\begin{tabular}{lcccc}
 \hline
 \hline
 Source & \multicolumn{2}{c}{Integrated profile} & \multicolumn{2}{c}{Line wings} \\ \cline{2-3} \cline{4-5}
~ & $T_{\text{rot}}$(K) & $\mathrm{log}_\mathrm{10}\mathcal{N}_{\mathrm{tot}}$ & $T_{\text{rot}}$(K) & $\mathrm{log}_\mathrm{10}\mathcal{N}_{\mathrm{tot}}$ \\
\hline
G351.16+0.7 & 199     & 51.7      & 219     & 51.4 \\
G351.25+0.7 & 172(67) & 52.2(0.6) & 165(66) & 51.8(0.6) \\
G351.44+0.7 & 151     & 52.2      & 157     & 52 \\
G351.58$-$0.4 & 158     & 53.6      & 166     & 53.3 \\
G351.77$-$0.5 & 177     & 52.6      & 177     & 52.5 \\
\hline
G12.81$-$0.2  & 168(47) & 52.9(0.4) & 131(31) & 52.8(0.4) \\
G14.63$-$0.6  & 125     & 51.7      & --      & -- \\
G34.26+0.15 & 152     & 53.0      & 162     & 52.6 \\
G34.40$-$0.2  & 111     & 52.7      & --      & -- \\
G35.20$-$0.7  & 141     & 52.7      & 120     & 52.5 \\
\hline
\end{tabular}
\end{table}

The relation between $\mathcal{N}_{\mathrm{u}}$ and the total number of emitting molecules, $\mathcal{N}_{\mathrm{tot}}$, follows the equation:
\begin{equation}
    \mathrm{ln}\left( \frac{\mathcal{N}_{\mathrm{u}}}{g_{\mathrm{u}}}\right) = - \frac{E_{\mathrm{u}}}{T_{\mathrm{rot}}k_{\mathrm{b}}} + \mathrm{ln}\left( \frac{\mathcal{N}_{\mathrm{tot}}}{Q(T_{\mathrm{rot}})} \right),
\end{equation}
where $g_\textrm{u}$ is the statistical weight of the upper level, $E_{\textrm{u}}$ -- the energy of the upper level, $k_{\textrm{b}}$ -- the Boltzmann constant, $T_{\textrm{rot}}$ -- the rotational temperature, and $Q$($T_\mathrm{rot}$) -- the partition function at the temperature $T_{\textrm{rot}}$.

The rotational temperature is calculated from the slope $b$ of the linear fit ($y=ax+b$) to the data in the natural logarithm units, \mbox{$T_\mathrm{rot}=-1/a$}.  The total number of emitting molecules, $\mathcal{N}_\mathrm{tot}$, is determined from the fit intercept $b$ as:
\begin{equation}
\label{rot3}
\mathcal{N}_\mathrm{tot}=Q(T_{\mathrm{rot}})\cdot \mathrm{exp}(b).
\end{equation}

Figure\,\ref{fig:rotation_diagram} shows example Boltzmann diagrams for G351.25+0.7 and G12.81$-$0.2, constructed using the velocity-integrated emission of CO (full profile). Table \ref{tab:rot_diag_results} shows $T_{\text{rot}}$ and $\mathcal{N}_{\text{tot}}$ for all sources with at least two CO line detections, separately for the integrated-profile emission and the line wings (see Section \ref{sec:profile-decomposition}).

The two sources with three CO line detections are characterized by $T_{\text{rot}}$ of $\sim$170\,K using the integrated line emission. The remaining sources show $T_{\text{rot}}$ in the range from $\sim$110\,K to \mbox{200\,K}, with a mean of 152\,K. Similar temperatures are obtained for the wing emission tracing outflow gas, with a mean $T_{\text{rot}}$ of 167\,K. While the wing emission is often responsible for the bulk of the total emission, hot core emission, with typical temperatures of $\sim$ 100--200\,K \citep{fontani2007,taniguchi2023}, might also contribute to the far-IR emission at source velocity.

For G34.26+0.15, $T_{\text{rot}}$ of $\sim$150\,K is significantly lower than $365\pm15$\,K obtained from \textit{Herschel}/PACS \citep{kar14}. We note, however, that the latter temperature was obtained using CO lines with $J_\text{u}$ from 14 to 30, sensitive to both \lq\lq warm'' and \lq\lq hot'' gas components \citep{kar18}. If CO transitions with $J_\text{u}$ from 14 to 16 are used instead, $T_{\text{rot}}$ of \mbox{$244\pm45$\,K} is obtained \citep[adopting values from Table C.1. in][]{kar14}. Even lower $T_{\text{rot}}$ is expected when the \coet{}, tracing colder gas component, is used in the calculation, in line with results obtained for G34.26+0.15. Noteworthy, it is essential to have many CO transitions to determine all the underlying physical conditions.

Finally, we note that the ratio of the total number of emitting molecules ($\mathcal{N}_{\mathrm{tot}}$) in the line wings and the total line profile ranges from $40\%$ to $79\%$ (Table \ref{tab:rot_diag_results}), consistent with the overall fraction of wing emission (Section \ref{sec:profile-decomposition}). In absolute terms, $\mathrm{log}_\mathrm{10}\mathcal{N}_{\mathrm{tot}}$ ranges from 51.7 to 53.6, consistent with the average $52.4(0.1)\pm0.5$ measured for high-mass protostars with \textit{Herschel}/PACS \citep{kar14}.

\subsection{Molecular excitation in non-LTE (wings)}
\label{sec:nonlte}
Due to relatively low densities in the regions of ISM where outflows propagate, the LTE assumption may not hold. Non-LTE modelling is therefore necessary to determine the physical conditions responsible for the observed line emission. Here, we use the well-established code RADEX \citep{radex07} to estimate gas temperatures, densities, and CO column densities, which reproduce the observed line wing emission of three mid- and high-$J$ CO lines: \cosf{}, 11--10, and 16--15.

We calculated model grids for a range of kinetic temperatures, $T_\mathrm{kin}$, from 150 to 3000\,K, H$_2$ number densities, $n_{\mathrm{H}_2}$, from $10^3$ to $10^7$\,cm$^{-3}$, and CO column densities, $N\mathrm{(CO)}$, of 10$^{16}$, 10$^{17}$, and 10$^{18}$\,cm$^{-2}$. We assumed H$_2$ as the only collision partner, and a background temperature of 2.73\,K. The linewidths of all lines were fixed at 19\,km\,s$^{-1}$, based on the observations of \cosf{} (Appendix \ref{app:line_parameters} and \citep{navarete2019atlasgal}).

For comparisons of models with observations, we used peak intensities obtained from RADEX, since the wing emission does not follow a simple Gaussian; we have also converted the observations from $T_{\text{MB}}$ to $T_{\text{r}}$ through $T_{\text{r}} = T_{\text{MB}}/\eta$. Because we do not spatially resolve the line emitting regions, we considered two cases during the computation of the beam filling factor: (i) the source that fills the entire beam ($\eta = 1$); (ii) the source size of $2\arcsec$ or $\eta \sim 8\times 10^{-3}$-- $5\times 10^{-2}$, consistent with size of a source in our sample, G34.26+0.15, which was measured from the Spitzer/IRAC 3.6\,$\mu$m image. The spatial extent of \cosf{} emission is $\sim$4 times larger than the APEX/CHAMP$^+$ beam, according to previous observations \citep{navarete2019atlasgal}.

Figure\,\ref{fig:radexgrid} shows the comparison of non-LTE radiative transfer models with line wing observations of high-$J$ CO lines\footnote{Observations of G351.44+0.7 are not included because we could not obtain its \cosf{} line wing due to the lack of \ttcosf{} opacity (Appendix \ref{app:decomposition-method}).} (Section \ref{sec:profile-decomposition}). The ratio of \costft{} and \coet{} depends on both $T_\mathrm{kin}$ and $n_{\mathrm{H}_2}$, and shows a spread of 3 orders of magnitude. On the other hand, the intensity of \cosf{} is most sensitive to the assumed $N$(CO), and increases by 2--3 orders of magnitude between 10$^{16}$ and 10$^{18}$\,cm$^{-2}$. The impact of the assumed beam filling factor is almost negligible to the \costft{} / \coet{} ratio.

The models match the observations best for the assumed CO column densities of $10^{17}$ and $10^{18}$\,cm$^{-2}$ (Fig.\,\ref{fig:radexgrid}). The solutions for temperature and density are degenerate and can be split into two regimes: (i) lower-density with $n_{\mathrm{H}_2}$ of $10^3$--$10^4$\,$\text{cm}^{-3}$ and $T_\mathrm{kin}$ of at least 1000\,K, and (ii) high-density, moderate-temperature scenario with $n_{\mathrm{H}_2}$ of $10^5$--$10^7$\,$\text{cm}^{-3}$ and $T_\mathrm{kin}$ between 150 and 500\,K. The ratio of high-$J$ CO lines can be well-reproduced for both considered filling factors in the scenario (ii); the best-matching source size is likely larger than $2\arcsec$ but depends on the assumed column density. In scenario (i), the ratio of high-$J$ CO lines can be reproduced for a small fraction of our sample assuming $T_{\mathrm{kin}}$ of 1000\,K. Much higher temperatures would be required to match observations of the majority of targets. In general, the \cosf{} peak intensity increases with gas density: for example, for $n_{\mathrm{H}_2}$ of $10^3$\,$\text{cm}^{-3}$ and $N(\mathrm{CO})$ of $10^{18}$\,$\textrm{cm}^{-2}$, models would match the observations assuming the filling factor of 1, whereas $n_{\mathrm{H}_2}$ of $10^4$ $\text{cm}^{-3}$ and $N(\mathrm{CO})$ of $10^{17}$--$10^{18}$ $\textrm{cm}^{-2}$, point at smaller filling factors.

In conclusion, only scenario (ii) can explain the observations of all targets. The $T_\mathrm{kin}$ range in this scenario is also in better agreement with $T_\mathrm{rot}$ estimated under the LTE condition in Section \ref{sec:lte}, and consistent with detections of molecular species excited exclusively in high-density environments toward other high-mass clumps \citep[e.g.,][]{vdT13,vdT19}. On the other hand, scenario (i) requires temperatures in excess of 3000\,K to explain the observed CO lines ($E_{\mathrm{up}}$\,<\,800\,K) at more than half of our targets; such temperatures are too high even for the outflows from high-mass stars. Therefore, we prefer the high-density, moderate-temperature scenario to describe the physical conditions toward our source sample. We note, however, that our models constrain only the ranges of temperature and density, as we cannot fully break the degeneracy between different models.

To summarize, non-LTE radiative transfer models provide support to the LTE excitation of high-$J$ CO emission in the high-mass clumps. The best match with observations is obtained for gas densities of $10^5$--$10^7$ $\text{cm}^{-3}$, $T_\mathrm{kin}$ between 150 and 500\,K, and CO column densities of $10^{17}$ and $10^{18}$\,cm$^{-2}$. Such conditions are consistent with CO excitation in outflows and will be discussed further in Section \ref{sec:dis_exc}.

\section{Discussion} \label{sec:dis}
High spectral resolution observations from SOFIA/GREAT allow us to disentangle dynamical properties of high-$J$ CO emission toward high-mass star forming clumps. The excitation conditions have been studied in the high-velocity gas component assuming both LTE and non-LTE regimes, supporting the origin in moderate-temperature, high-density gas associated with the outflows (Section \ref{sec:lte}--\ref{sec:nonlte}). Here, we discuss our results in the context of previous observations of high-mass protostars with \textit{Herschel} and SOFIA.

\subsection{High-$J$ CO emission in high-mass clumps} \label{sec:dis_co}
The high-$J$ ($J\gtrsim10$) CO emission in high-mass star-forming regions has been attributed to gas cooling of several physical components, including (i) a warm, dense envelope of central protostars \citep{cec96,doty97}, (ii) UV-irradiated outflow cavity walls \citep{bru09,irene16}, (iii) currently-shocked gas in the outflows \citep{vdWiel13,kar14}, (iv) photodissociation regions \citep{lane90,oss10,oss15,stock15}. A similarity of CO to H$_2$O, both in spatial extent and line shapes, supported the scenario of shock excitation in similar layers composing the outflow cavity walls \citep[see e.g.,][]{irene16,kri17,vd21}.

Broad line profiles of high-$J$ CO lines provide a solid evidence of the outflow origin of a part of CO emission in high-mass clumps from the ATLASGAL survey  \citep[Section \ref{sec:results}, see also ][]{irene13,indriolo17}. Noteworthy, the fraction of CO emission in the line wings with respect to the total line emission is not sensitive to the evolutionary stage of the clumps (Section \ref{sec:ratio}). In fact, a significant fraction of \coet{} is detected in the line wings of clumps at very early evolutionary stages (up to 76\%, Section \ref{sec:ratio}).

The signposts of outflow activity in the IR-weak clumps are in agreement with the ubiquitous detection of broad line wings in the \mbox{SiO\,2--1} line toward ATLASGAL sources spanning all evolutionary stages, including 25\% of infrared-quiet clumps \citep{cse16}. Indeed, molecular outflows are also commonly detected toward \mbox{70\,$\mu$m} dark clumps using other tracers \citep{urqu22,yang22}.

\begin{figure}
    \centering
    \resizebox{0.9\hsize}{!}{\includegraphics{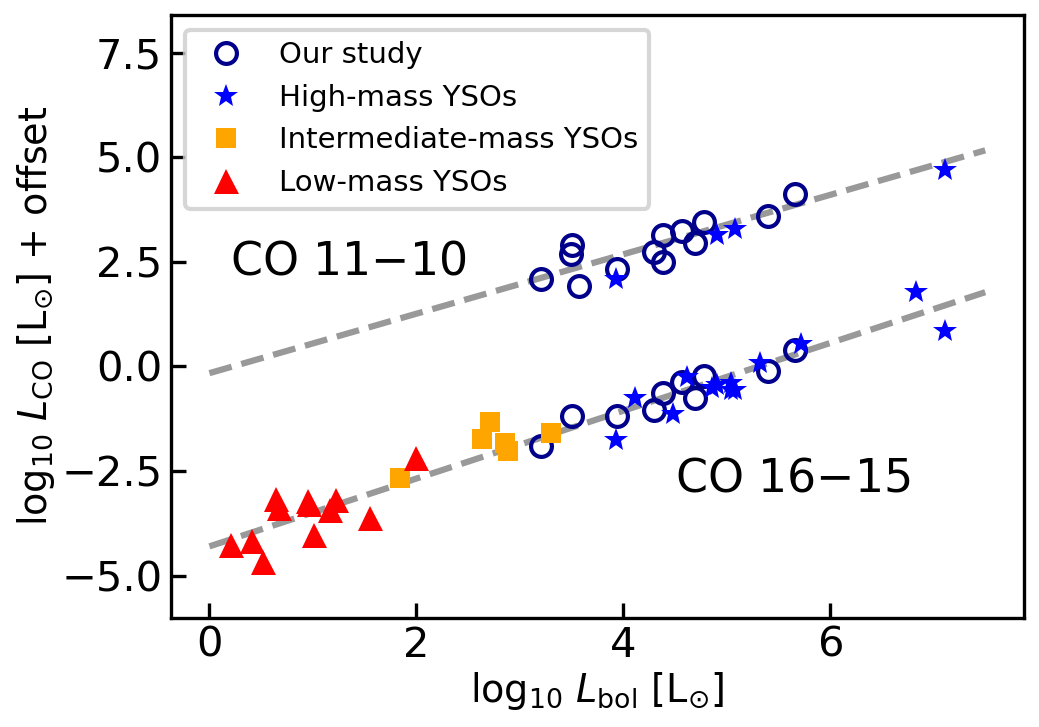}}
   \caption{Velocity-integrated CO line luminosity of 11--10 and \mbox{16--15} transitions versus source bolometric luminosity from low- to high-mass star-forming regions. The dashed lines show a linear fit obtained using only the sources from our study, which are shown in blue empty circles. Blue stars show observations of other high-mass protostars from \citet{kar14,indriolo17,maja}, orange squares present emission from intermediate-mass objects \citep{mat15}, and red triangles show data for Class 0 protostars \citep{kri17}.}
   \label{fig:corr-low2highmass}
\end{figure}

The integrated high-$J$ CO emission shows a strong correlation with the clump bolometric luminosity (Section \ref{sec:results}). The correlation extends even to low- and intermediate-mass protostars (Figure \ref{fig:corr-low2highmass}), suggesting a similar physical mechanism operating over a few orders of magnitude different spatial scales. In deeply-embedded low-mass objects, $L_\mathrm{bol}$ is dominated by accretion luminosity, which in turn is closely related to the amount of mass ejected in the outflows \citep{frank14}. Thus, the tight correlation of high-$J$ CO emission with $L_\mathrm{bol}$ for ATLASGAL clumps suggests an equally high contribution of accretion luminosity in high-mass regions.

The velocity-resolved SOFIA spectra provide strong support for an origin of the bulk high-$J$ CO emission in outflows, during all  evolutionary stages of high-mass clumps. The correlation of CO line fluxes with bolometric luminosity suggest common physical conditions and processes leading to high-$J$ CO emission from low- to high-mass star forming regions.

\subsection{Excitation conditions}
\label{sec:dis_exc}
Observations of multiple CO lines allow to study gas excitation across various source properties and evolutionary stages. In combination with other far-IR lines, they also constrain the properties of shocks responsible for the emission in broad line wings of high-$J$ CO lines.

The rotational temperatures in the high-velocity gas in the ATLASGAL clumps range from $\sim$120\,K to 219\,K, and are similar to the temperatures obtained from the full line profiles (Section \ref{sec:lte}). Five IR-bright clumps show mean $T_\mathrm{rot}$ of $169\pm30$\,K, whereas two \ion{H}{ii} clumps are characterized by $T_\mathrm{rot}$ of $147\pm16$\,K. Thus, a possible decrease of gas temperature in the outflows as the clumps evolve might be present, but for the sources in our sample the difference is not significant.

Rotational temperatures of $\sim$200--210\,K, consistent with our measurements, have been estimated in the line wings of the high-mass protostar DR21(OH) assuming LTE \citep{leu15}. Non-LTE modeling of multiple CO lines indicated $T_\mathrm{kin}$ of 60--200\,K in the outflow gas component of another high-mass source, AFGL 2591 \citep{vdWiel13}. In W3 IRS5, excitation temperatures of $\sim$ 100--210\,K were measured in the \mbox{CO\,10--9} and 3--2 lines' velocities
range covered by an outflow in this region (from 5 to 20\,km\,s$^{-1}$), with the highest temperatures corresponding to the highest velocities \citep{irene13}. A similar trend of increasing gas temperature with velocity is also clearly detected in the outflow wing emission from ATLASGAL clumps observed with SOFIA/GREAT (Section \ref{sec:ratio}), and in the SiO survey of $\sim$430 clumps observed with the IRAM 30\,m telescope \citep{cse16}.

\begin{table}[t!]
\centering
\caption{CO rotational excitation determined from integrated line profiles toward high-mass objects}
\label{tab:summary}
\tiny
\begin{tabular}{lrl}
 \hline
 \hline
 Source & $T_{\text{rot}}$(K) &  Reference \\
 \hline
NGC 7538 IRS1\tablefootmark{a}  & $160(10)$ & \cite{kar14} \\
AFGL 2591 & $130(5)$ & \cite{maja} \\
W49N\tablefootmark{b} & $220(20)$ & \cite{indriolo17} \\
Orion S   & $145(5)$  & \cite{indriolo17} \\
Orion KL  & $180(25)$  & \cite{indriolo17} \\
Sgr B2(M) & $140(20)$ & \cite{indriolo17} \\
\hline
G12.81$-$0.2   & 168(47) & this work\\
G351.25+0.7 & 172(67) & this work \\
ATLASGAL (all)\tablefootmark{c}  & 111--199  & this work \\
\hline
\end{tabular}
\tablefoot{Rotational temperatures obtained from the integrated line profiles
of CO transitions with $J_\mathrm{u}$ of 11, 13, and 16.
\tablefoottext{a}{Calculation using CO lines with $J_\mathrm{u}$ from 14 to 17.}
\tablefoottext{b}{Calculation using CO lines with $J_\mathrm{u}$ from 14 to 16.}
\tablefoottext{c}{Based only on \coet{} and \costft{} detections, see Section \ref{sec:ratio}.}
}
\end{table}

Comparisons of gas excitation using high-$J$ CO lines can be extended to a larger number of sources once the emission in the full line profiles is considered. The velocity-integrated \textit{Herschel}/PACS detections of CO transitions from $J_\text{u}$ of 14 to 30 toward 10 high-mass protostars provided an average $T_\text{rot}$ of $\sim$300$(23)\pm$60\,K \citep{kar14}. Protostars with detections of higher-$J$ lines were generally characterized by higher-$T_\text{rot}$, suggesting the possible presence of an additional $T_\mathrm{rot}\gtrsim700$\,K gas component detected toward low-mass protostars that appears to be \lq\lq hidden'' in their high-mass counterparts, possibly due to the a small beam filling factor of such emission and/or optically thick continuum emission \citep{manoj13,green13,kar13,kar18}. Clearly, any comparisons of $T_\mathrm{rot}$ should consider the similar $J-$levels for their calculation \citep[Section \ref{sec:lte}, and e.g.,][]{neu12,maria17,yang18COPS}.

Table \ref{tab:summary} compares $T_\mathrm{rot}$ measurements for several high-mass protostars with the data of the same or similar CO transitions to our SOFIA/GREAT survey (Section \ref{sec:lte}). All sources with at least 3 observed transitions show $T_\mathrm{rot}$ from 130\,K to 220\,K, consistent with the values determined for ATLASGAL clumps and hot cores. The relatively narrow range of $T_\mathrm{rot}$ is qualitatively similar to that of the universal \lq\lq warm'', $\sim$300\,K gas component based on \mbox{CO\,14--13} to 25--24 transitions toward low-, intermediate-, and high-mass protostars \citep{kar14,mat15,kar18,vd21}. The \coet{} transition in low-mass protostars is typically associated with a \lq\lq cool'' gas component with $T_\mathrm{rot}\sim$100\,K \citep[e.g., ][]{yang18COPS}, and its inclusion in the fit causes the lower values of $T_\mathrm{rot}$ ($<300$\,K).

The CO rotational temperature depends on the gas density and kinetic temperature, and can be characterised with both (i) low-density, high-temperature \citep{neu12,manoj13,yang18COPS}, and (ii) high-density, low-temperature regimes \citep{kar13,kar18,green13,kri17}. The first scenario requires $n_\mathrm{H_2}\sim$10$^3$\,cm$^{-3}$ and $T_\mathrm{kin}\gtrsim2000$\,K, and has the advantage of reproducing the positive curvature of the CO diagrams over a broad range of energy levels \citep{neu12}. In contrast, the second scenario accounts for the similarity of $J\gtrsim14$ CO and H$_2$O emission most evidently seen in low-mass protostars, with high-densities required for H$_2$O excitation \citep{kar13,vd21}.

Non-LTE modeling of massive clumps provides strong support for a high-density scenario of CO excitation (Section \ref{sec:nonlte}). In the regime of moderate gas temperatures ($T_\mathrm{kin}$ from 150 to 500\,K), gas densities of $10^{5}$--$10^{7}$\,cm$^{-3}$ match the data best. Such physical conditions are fully-consistent with the modeling of high-$J$ CO and H$_2$O emission toward high-mass protostars \citep{irene16,vd21}. They are also comparable to the physical conditions determined in the jet, terminal shock and cavities of the intermediate-mass protostar Cep E \citep{lefloch15}.

The underlying mechanism behind the highly-excited CO gas has been investigated for both Cep E and its high-mass counterpart Cep A. Detailed comparisons of CO, in combination with [O I] and OH, suggests the origin in dissociative or UV-irradiated shock models with pre-shock densities above $10^5$ cm$^{-3}$ \citep{gusdorf16,gusdorf17}. Assuming a compression factor of $\sim$100, typical for dissociative shocks \citep{kar13}, such models would be also in agreement with radiative-transfer modeling for high-mass clumps (Section 3.5). However, a fraction of high-$J$ CO emission detected at source velocity could also originate from the central hot core.

\section{Conclusions} \label{sec:conclusions}
We have characterized the SOFIA/GREAT line profiles observed toward 13 high-mass protostars selected from the ATLASGAL survey, which significantly increases the number of high-mass objects that have velocity-resolved high-$J$ CO lines. The velocity information enables to quantify the line components and the properties of their emitting sources. We summarise and draw the following conclusions:

\begin{itemize}
    \item \coet{} emission is detected toward all the sources, as early as the 24d stage. 10 out of 13 clumps also show a clear detection of \costft{}. Additionally, \cott{} is detected toward two sources. The lines exhibit broad line wing emission typical for outflows from YSOs.

    \item We detect wing emission in the \coet{} line from 12 clumps and in the \costft{} line from 8 clumps, implying that the highly excited CO lines originate in outflows. The wing fraction is similar for all clump evolutionary stages. On the other hand, we find no signatures of high-velocity gas (i.e.,
    bullets) in the far-IR spectra.

    \item Under the LTE assumption, we find $T_{\mathrm{rot}}$ of 110--200\,K for the entire line profiles and 120--220\,K for the wing component. Such temperatures are in agreement with gas densities of \mbox{$10^5$--$10^7$\,cm$^{-3}$}, moderate temperatures of 150\,K and 500\,K, and CO column densities of $10^{17}$ and $10^{18}$\,cm$^{-2}$ obtained from the non-LTE models.

    \item Significant correlations between high-$J$ CO emission and bolometric luminosities suggest similar underlying physical processes and conditions across all evolutionary stages of high-mass clumps. The correlation extends also to low-mass protostars, where high-$J$ CO originate in outflow shocks, consistent with our study.
\end{itemize}

High angular-resolution maps of high-mass clumps would be necessary to better characterise the physical structure of the regions with strong high-$J$ CO emission and spatially disentangle outflows and hot cores \citep{goi15}. The MIRI instrument on board the James Webb Space Telescope could pin-point the spatial extent of shocked gas in high-mass star-forming clumps.

\begin{acknowledgements}
The authors thank the anonymous referee for detailed comments that have helped us improve this paper. We thank SOFIA/GREAT staff for collecting and reducing the data. We also thank Dr. Helmut Wiesemeyer for his support in reducing data from the SOFIA/4GREAT observations. AK acknowledges support from the Polish National Agency for Academic Exchange grant No. BPN/BEK/2021/1/00319/DEC/1. AYY acknowledges support from the National Natural Science Foundation of China grants No. 11988101. This work is based on observations made with the NASA/DLR Stratospheric Observatory for Infrared Astronomy (SOFIA). SOFIA is jointly operated by the Universities Space Research Association, Inc. (USRA), under NASA contract NNA17BF53C, and the Deutsches SOFIA Institut (DSI) under DLR contract 50 OK 2002 to the University of Stuttgart. $Herschel$ was an ESA space observatory with science instruments provided by European-led Principal Investigator consortia and with important participation from NASA.  This publication is based on data acquired with the Atacama Pathfinder Experiment (APEX). APEX is a collaboration between the Max-Planck-Institut fur Radioastronomie, the European Southern Observatory, and the Onsala Space Observatory.
\end{acknowledgements}

\typeout{}
\bibliographystyle{aa} 
\bibliography{ref}

\begin{appendix}

\section{Spatial extent of line emission} \label{app:extent}
\begin{figure}
    \begin{subfigure}{1\hsize}
      \centering
      \resizebox{0.75\hsize}{!}{\includegraphics{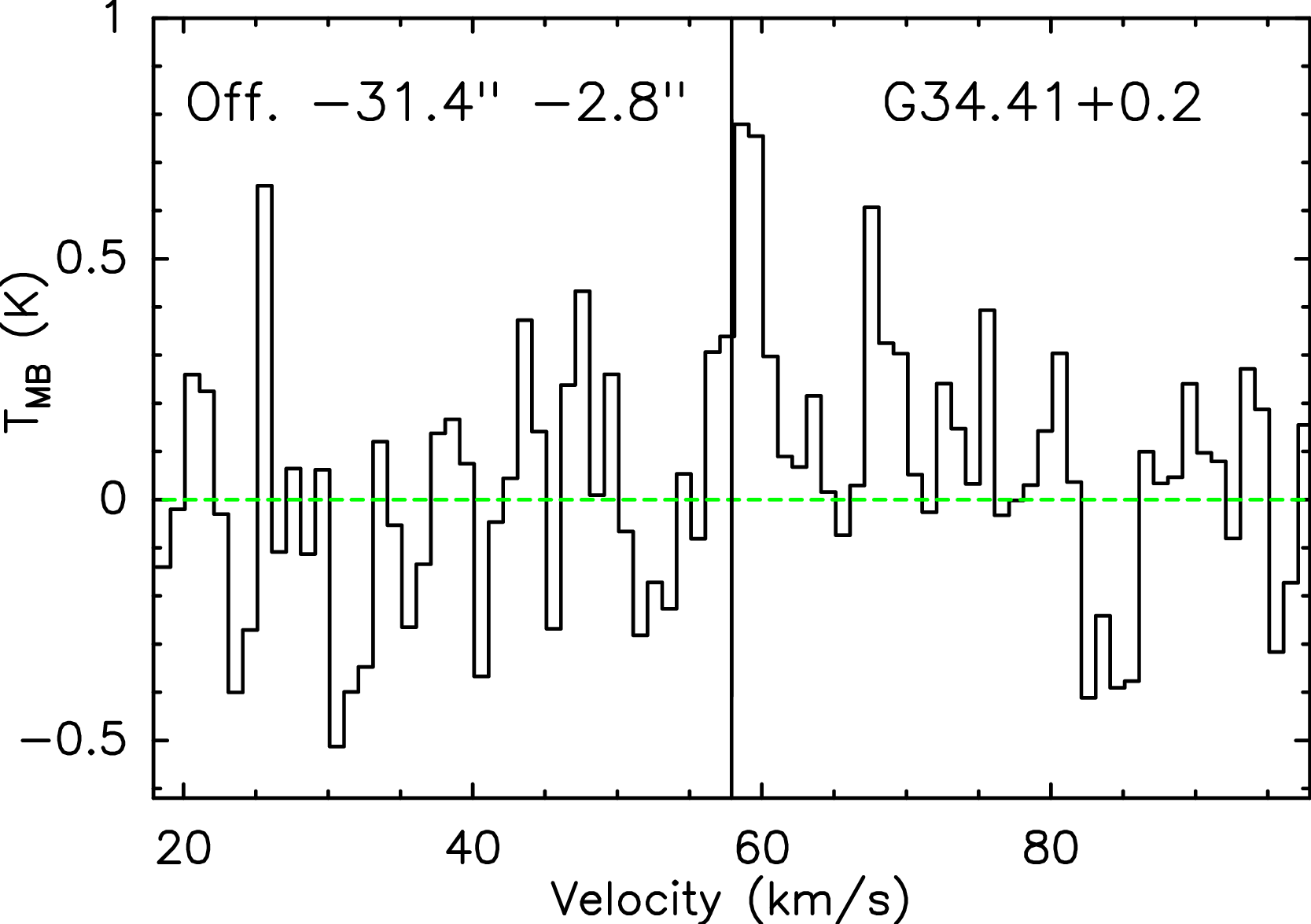}}
   \end{subfigure}\vspace{0.4cm}
   \begin{subfigure}{1\hsize}
      \centering
      \resizebox{0.75\hsize}{!}{\includegraphics{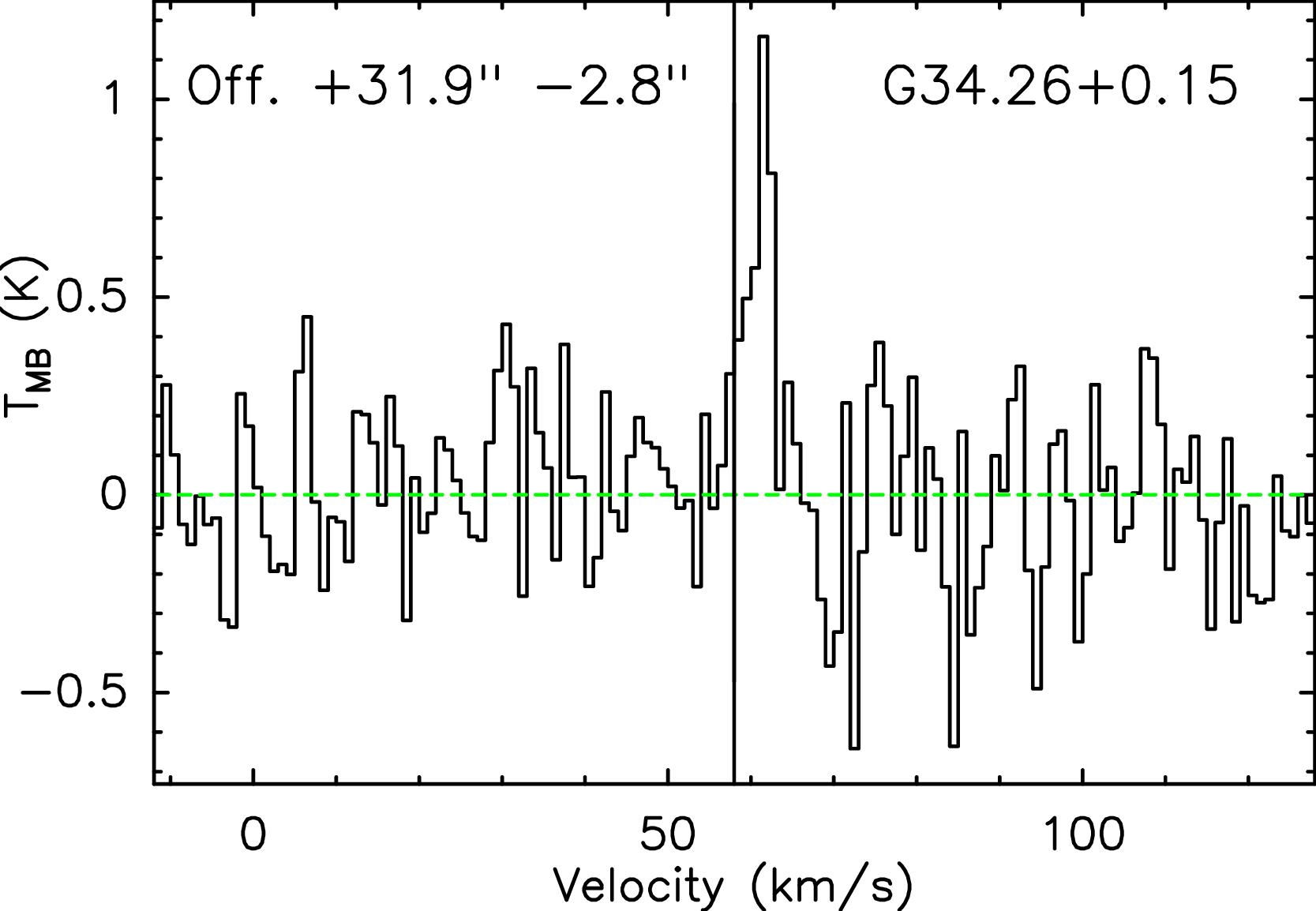}}
   \end{subfigure}
   \caption{Spectra of the \costft{} emission arising from extended positions away from source centers. The offsets in right ascension and declination are shown on each spectrum. Black vertical lines mark position of $V_{\mathrm{lsr}}$. Baselines are shown in green horizontal lines. The spectra are smoothed to 1.0\,km\,s$^{-1}$.}
   \label{fig:ext_co1615}
\end{figure}

For four sources, the GREAT/LFA provided additional \costft{} spectra toward six positions which are at $31\arcsec.8$ \citep{risacher2018} from the source center. We detected no extended emission toward G34.40$-$0.2 and G35.20$-$0.7. On the other hand, G34.41+0.2 and G34.26+0.15 show a weak line emission at one offset position each (Fig.\,\ref{fig:ext_co1615}). For G34.41+0.2, we find similar line intensity peaks at the central and extended positions. However, G34.26+0.15 shows $\sim$5 times weaker emission at the offset positions than in the center. At both objects, we observe no signs of line wing emission at the offset positions.

\section{Line parameters for emission at central position} \label{app:line_parameters}
Table \ref{tab:line_prop} presents peak and integrated intensities, and line luminosities of the high-$J$ CO lines. Table \ref{tab:multigauss_fit} shows results of the multiple Gaussian fit for sources with self-absorption.

\begin{table*}[h!]
\centering
\caption{Line properties of CO lines at different rotational transitions: $J$=11--10, and 16--15.}
\label{tab:line_prop}
\begin{tabular}{ll |ccc|ccc|ccc}
 \hline
 \hline
 & & \multicolumn{3}{c|}{\coet{}} & \multicolumn{3}{c|}{\cott{}} & \multicolumn{3}{c}{\costft{}} \\ \cline{3-11}
  No. & Source & $P$ & $S_{\text{int}}$\tablefootmark{a} & $L$ & $P$ & $S_{\text{int}}$\tablefootmark{b} & $L$ & $P$ & $S_{\text{int}}$\tablefootmark{c} & $L$ \\
          & & (K) & (K\,km\,s$^{-1}$) & (L$_{\odot}$) & (K) & (K\,km\,s$^{-1}$)  & (L$_{\odot}$) & (K) & (K\,km\,s$^{-1}$) & (L$_{\odot}$)\\
 \hline
 1 & G351.16+0.7 & 5.0 & 58.2(0.3) & $6.7 \times 10^{-2}$  & -- & -- & -- & 2.4 & 37.7(0.6) & $6.7 \times 10^{-2}$ \\
 2 & G351.25+0.7 & 34.5 & 238.5(1.2) & $2.8 \times 10^{-1}$  & 14.4 & 92.4(0.8) & $1.3 \times 10^{-1}$ & 15.6 & 98.1(1.3) & $1.7 \times 10^{-1}$ \\
 3 & G351.44+0.7 & 11.0 & 148.1(1.1) & $1.7 \times 10^{-1}$  & -- & -- & -- & 3.0 & 51.9(1.2) & $9.2 \times 10^{-2}$ \\
 4 & G351.58$-$0.4 & 5.7 & 95.7(1.4) & $4.2 \times 10^{0}$  & -- & -- & -- & 1.8 & 37.3(1.6) & $2.5 \times 10^{0}$ \\
 5 & G351.77$-$0.5 &  18.9 & 465.8(2.5) & $5.4 \times 10^{-1}$ & -- & -- & -- & 10.0 & 237.3(1.8) & $4.2 \times 10^{-1}$ \\
 \hline
 6 & G12.81-02 & 21.6 & 265.8(1.1) & $1.2 \times 10^{0}$ & 12.6 & 126.2(1.6) & $6.8 \times 10^{-1}$ &  13.8 & 108.2(1.4) & $7.7 \times 10^{-1}$ \\
 7 & G14.19$-$0.2 & 0.4 & 4.0(0.2) & $2.6 \times 10^{-2}$ & -- & -- & -- & --  & -- & -- \\
 8 & G13.66$-$0.6 & 1.2 & 7.0(0.7) & $9.7 \times 10^{-2}$ & -- & -- & -- & -- & -- & -- \\
 9 & G14.63$-$0.6 & 3.6 & 25.5(0.6) & $3.9 \times 10^{-2}$ & -- & -- & -- & 0.6  & 5.2(1.0) & $1.2 \times 10^{-2}$ \\
 \hline
 10 & G34.41+0.2  & 2.5 & 29.1(1.6) & $1.5 \times 10^{-1}$ & -- & -- & -- & --   & -- & -- \\
  11 & G34.26+0.15 & 18.6 & 172.1(1.9) & $9.1 \times 10^{-1}$ & -- & -- & -- & 5.5 & 60.9(1.6) & $5.9 \times 10^{-1}$ \\
 12 & G34.40$-$0.2  & 5.4 & 48.7(1.3) & $2.6 \times 10^{-1}$  & -- & -- & -- & 1.3 & 6.7(0.9) & $6.5 \times 10^{-2}$ \\
 13 & G35.20$-$0.7  & 13.1 & 140.3(1.8) & $4.3 \times 10^{-1}$  & -- & -- & -- & 3.9 & 40.7(1.1) & $2.3 \times 10^{-1}$ \\
 \hline
\end{tabular}
\begin{flushleft}
    \tablefoot{
    \tablefoottext{a,b,c}{The numbers within brackets represent statistical uncertainty of integrated intensity.}}
\end{flushleft}
\end{table*}

\begin{table*}[h!]
\centering
\caption{Results of multiple Gaussian decomposition for the \cosf{}, 11--10, and 16--15 lines with self-absorption.}
\label{tab:multigauss_fit}
\begin{tabular}{ll |ccc|ccc|ccc}
 \hline
 \hline
 & & \multicolumn{3}{c|}{First component} & \multicolumn{3}{c|}{Second component} & \multicolumn{3}{c}{Third component}\\
 \cline{3-11}
 No. & Source & $V_{\mathrm{central}}$ & $FWHM$ & $T_{\mathrm{peak}}$ & $V_{\mathrm{central}}$ & $FWHM$ & $T_{\mathrm{peak}}$ & $V_{\mathrm{central}}$ & $FWHM$ & $T_{\mathrm{peak}}$  \\
          & & (km s$^{-1}$) & (km s$^{-1}$) & (K)  & (km s$^{-1}$) & (km s$^{-1}$) & (K)  & (km s$^{-1}$) & (km s$^{-1}$) & (K) \\
 \hline
 \hline
& & \multicolumn{9}{c}{\cosf{}} \\ 
 \hline
 1 & G351.16+0.7 & -5.4 & 9.0 & 46.3 & -8.5 & 22.7 & 18.7 & -- & -- & -- \\
 2 & G351.58$-$0.4 & -95.4 & 10.2 & 19.2 & -98.9 & 32.0 & 6.1 & -- & -- & -- \\
 3 & G351.77$-$0.5 & -1.2 & 12.1& 27.6 & -2.8 & 33.6 & 16.1 & -14.1 & 77.0 & 5.5 \\
 4 & G12.81$-$0.2 & 50.0 & 8.6 & 4.9 & 34.6 & 10.8 & 36.7 & 37.0 & 12.8 & 42.6 \\
 5 & G34.26+0.15 & 58.2 & 8.2 & 73.3 & 57.9 & 20.6 & 7.0 & 73.0 & 39.9 & 2.2 \\
 6 & G34.40$-$0.2  & 58.0 & 6.3 & 19.0 & 55.2 & 12.3 & 7.3 & 60.0 & 12.4 & 10.9 \\
 7 & G35.20$-$0.7  & 34.7 & 10.3 & 38.3 & 47.5 & 11.1 & 1.7 & 29.8 & 19.0 & 12.2\\
 \hline
  & & \multicolumn{9}{c}{\coet{}} \\ 
 \hline
 1 & G351.25+0.7 & -6.3 & 3.7 & 5.9 & -1.9 & 5.0 & 50.8 & -0.5 & 21.5 & 2.4\\
 2 & G351.77$-$0.5 & -2.6 & 7.9 & 8.6 & -0.3 & 19.4 & 8.3 & -6.8 & 56.4 & 3.9\\
 \hline
  & & \multicolumn{9}{c}{\costft{}{}} \\ 
 \hline
 1 & G35.20$-$0.7 & 34.7 & 4.8 & 5.0 & 35.4 & 17.9 & 1.5 & --  & -- & -- \\
 \hline
\end{tabular}
\end{table*}

\section{Full velocity spectra at central position} \label{app:full_spectra}
\begin{figure*}
    \centering
    \resizebox{1\hsize}{!}{\includegraphics{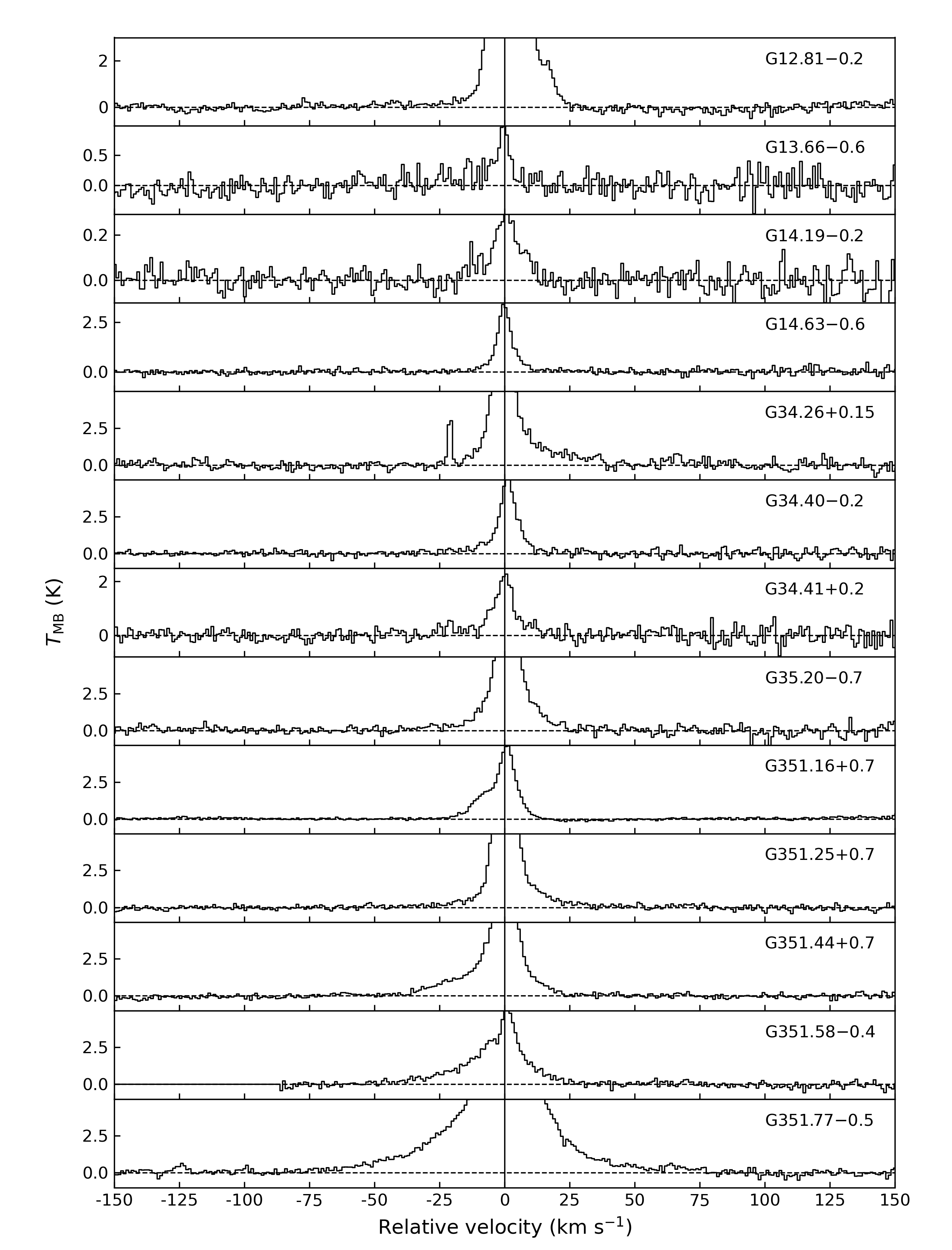}}
   \caption{\coet{} spectra on a $300$\,km\,s$^{-1}$ window, in which source velocities are shifted to 0\,km\,s$^{-1}$ and marked by the solid vertical line. Horizontal dashed lines present zero intensity level.}
   \label{fig:fullspectra11}
\end{figure*}

\begin{figure*}
    \centering
    \resizebox{1\hsize}{!}{\includegraphics{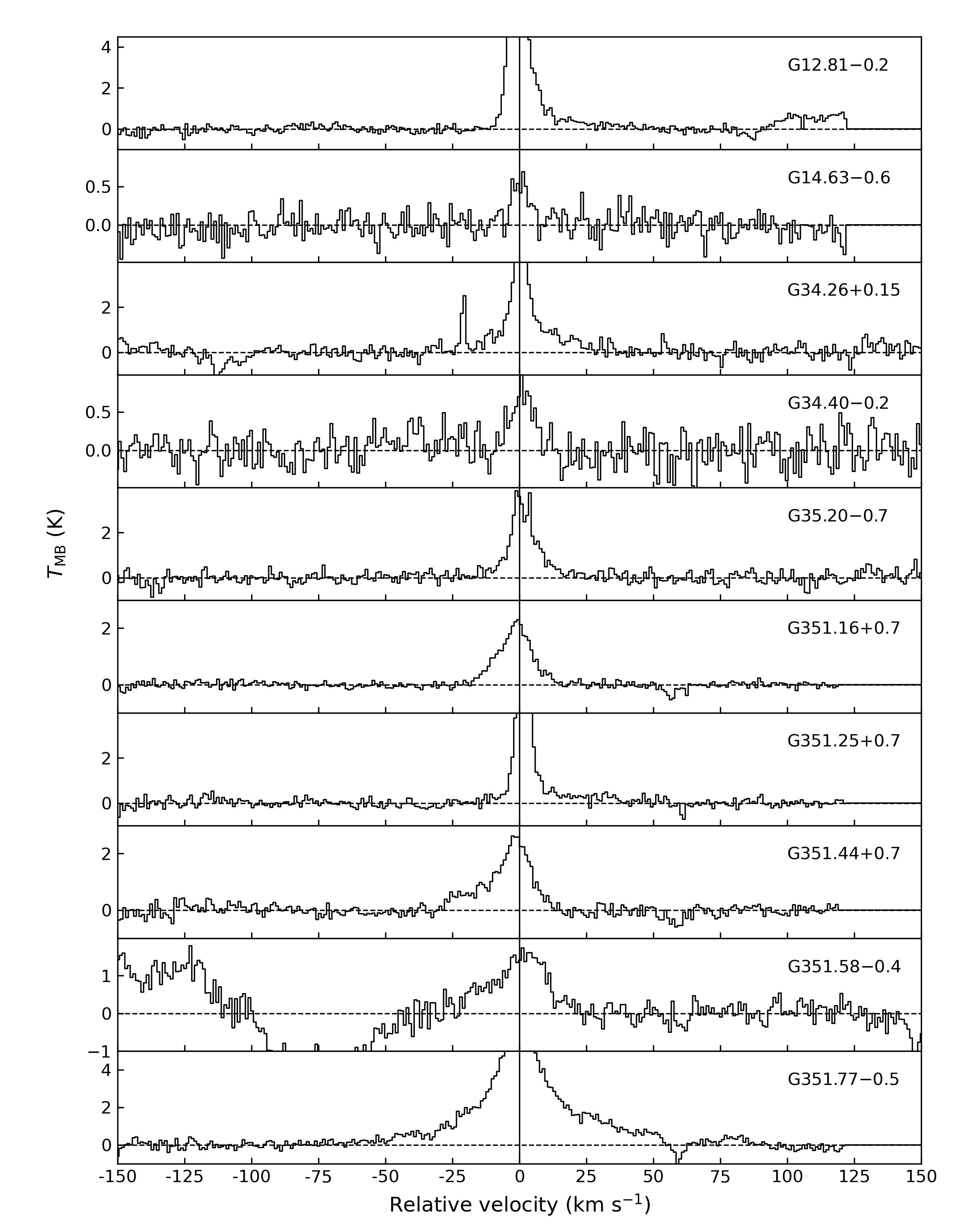}}
   \caption{\costft{} spectra on a window of $300$\,km\,s$^{-1}$. The source velocities are shifted to 0\,km\,s$^{-1}$. Vertical and horizontal lines are similar to those in Fig.\,\ref{fig:fullspectra11}.}
   \label{fig:fullspectra16}
\end{figure*}

Figure \ref{fig:fullspectra11} and \ref{fig:fullspectra16} present the \coet{} and \costft{} spectra from Section \ref{sec:results}, but covering a velocity range of $\pm150$\,km\,s$^{-1}$. The spectra illustrate a lack of EHV emission or any other significant high-velocity gas components toward high-mass clump. A 2-$\sigma$ feature at -125 km s$^{-1}$ appears to be tentatively detected toward G351.77$-$0.5 which is known to drive multiple outflows associated with EHV gas detected in CO\,$2-1$ and $6-5$ \citep{leurini2009}. Noteworthy, the EHV gas component is exclusively detected toward the outflow positions of G351.77$-$0.5, and absent from the on-source spectra \citep{leurini2009}, consistent with observations toward an intermediate-mass protostar Cep E \citep{ruiz12,lefloch15,gusdorf17}. Thus, we do not consider this feature as real.

\section{Details of the profile decomposition method} \label{app:decomposition-method}
\subsection{Line wing emission in high-$J$ CO lines} \label{sec:wing-extract-highJ}
\begin{figure}[h!]
    \centering
   \begin{subfigure}{0.4\textwidth}
      \centering
      \resizebox{0.9\hsize}{!}{\includegraphics{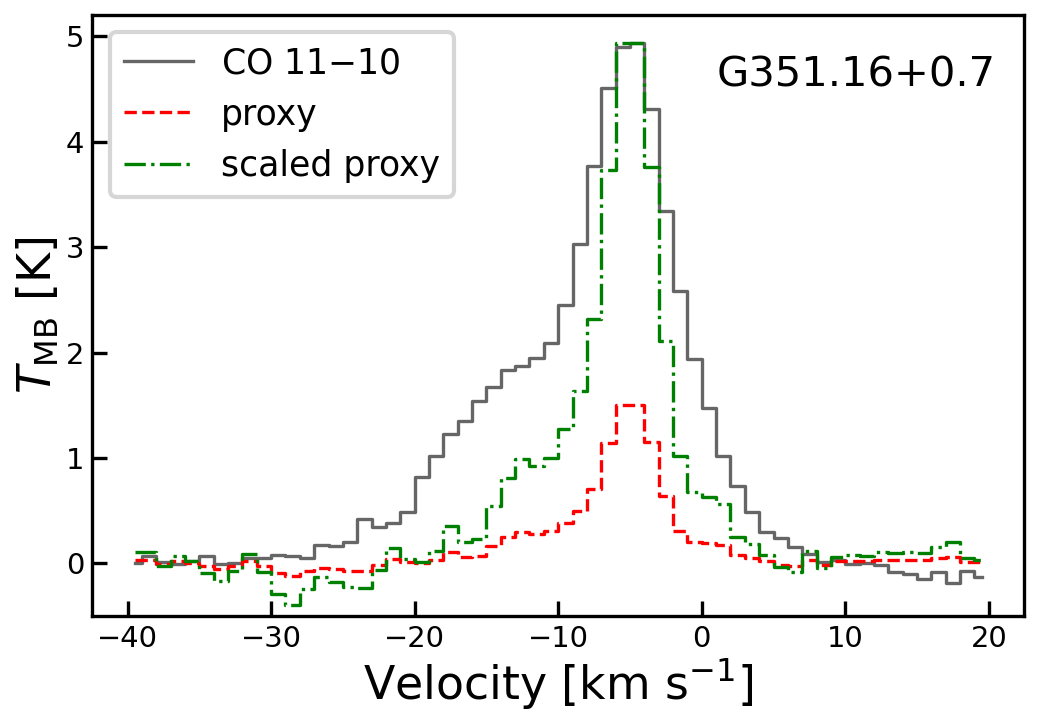}}
   \end{subfigure}
   \begin{subfigure}{0.4\textwidth}
      \centering
      \resizebox{0.9\hsize}{!}{\includegraphics{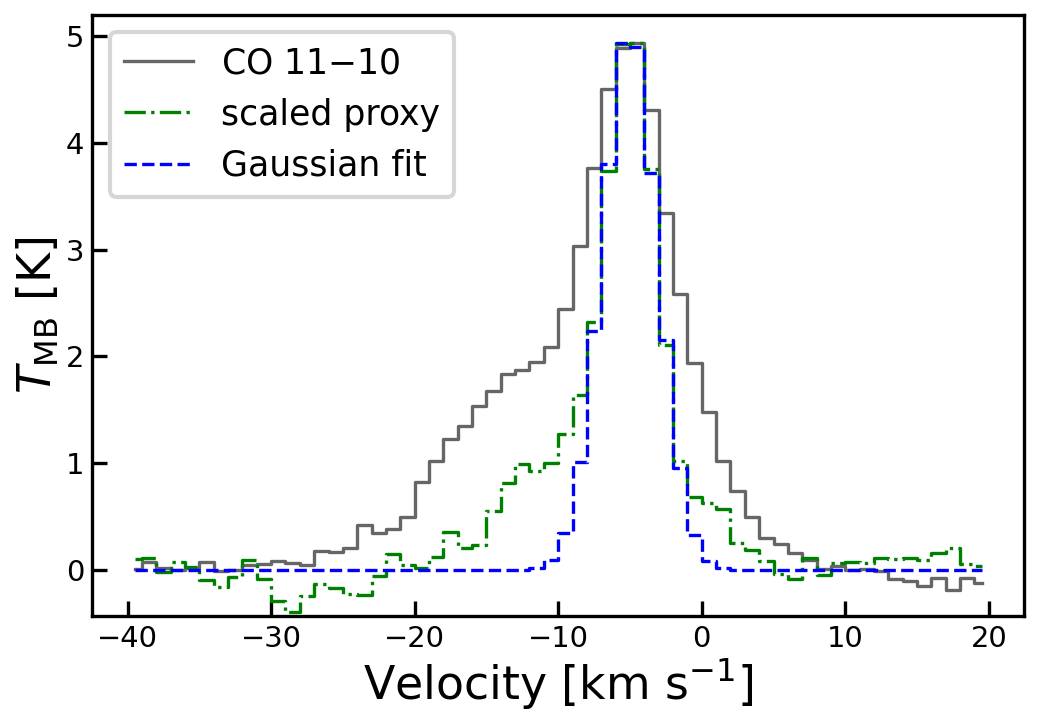}}
   \end{subfigure}
   \begin{subfigure}{0.4\textwidth}
      \centering
      \resizebox{0.9\hsize}{!}{\includegraphics{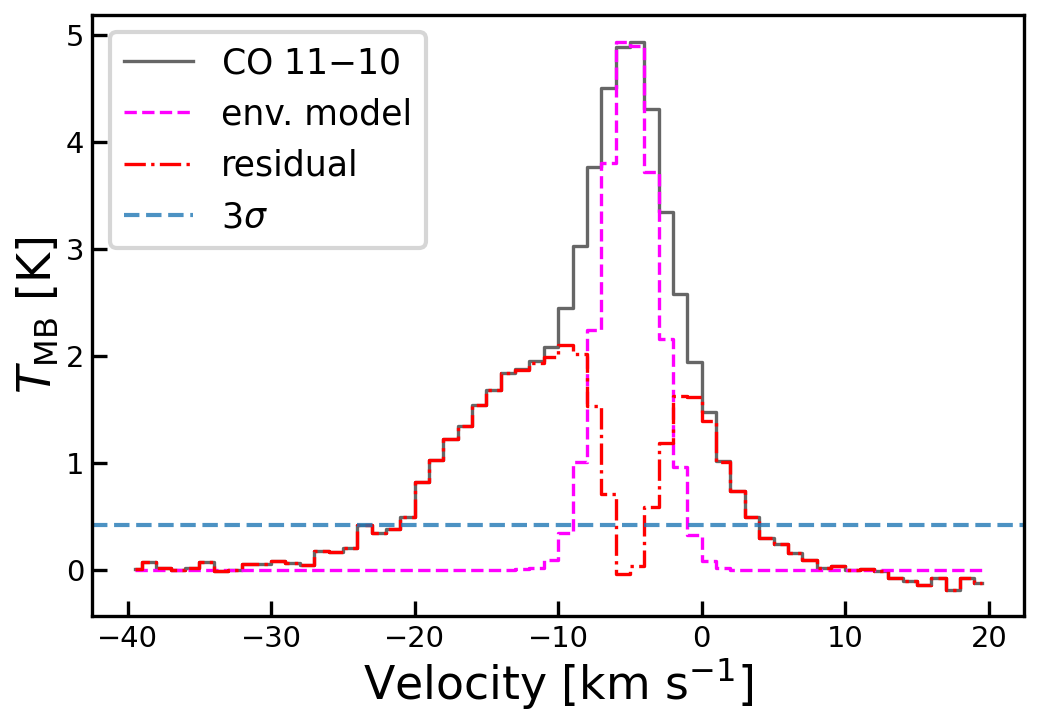}}
   \end{subfigure}
   \caption{Wing extraction steps for an example high-$J$ CO line from top to bottom.}
   \label{fig:wing_detect_highj}
\end{figure}

Assuming that the high-$J$ CO lines are optically thin, the detailed steps to identify wing emission are as follows:

\begin{enumerate}[label={(\arabic*)}]
    \item Scale the isotopologue line which is used as a proxy for envelope emission (called proxy line hereafter) to the height of the high-$J$ CO profiles (see Fig.\,\ref{fig:wing_detect_highj}, top). Table\,\ref{tab:proxyline_list} presents the envelope tracer used for each source.

    \item Fit a Gaussian to the peak of the scaled proxy line (see Fig.\,\ref{fig:wing_detect_highj}, middle) and use it as a model for the envelope emission. In some cases, proxy lines have non-Gaussian shapes with additional wing feature. The Gaussian fit helps remove contamination from the wing emission. The fit is iterated following \citet{van2007inferences}, with high velocity channels being gradually removed until a reasonable fit is obtained with a significant improvement of reduced-$\chi^2$.

    \item Shift the envelope model to velocity of the high-$J$ CO peaks. This step is skipped for the \coet{} line of G12.81$-$0.2 and G035.20-0.7 as their line peaks are tilted, suggesting that a part of the line profile is attenuated.

    \item Subtract the envelope model from the high-$J$ CO profiles (see Fig.\,\ref{fig:wing_detect_highj}, bottom), leaving wing emission in the residual. If the residual show significant emission above three times the spectral noise, a wing detection is confirmed.
\end{enumerate}

Properties of line wing emission are presented in Table \ref{tab:wing_properties}. From step (2), the Gaussian fit to the scaled proxy lines provide source radial velocities, $V_{\text{lsr}}$. Although $V_{\text{lsr}}$ was previously determined with other tracers (e.g., \cstott{} in \citet{giannetti2014atlasgal}), low opacity proxy lines provide more reliable estimations as they probe thoroughly the central gas envelope.

\begin{figure}[h!]
    \centering
    \begin{subfigure}{0.4\textwidth}
      \centering
      \resizebox{0.9\hsize}{!}{\includegraphics{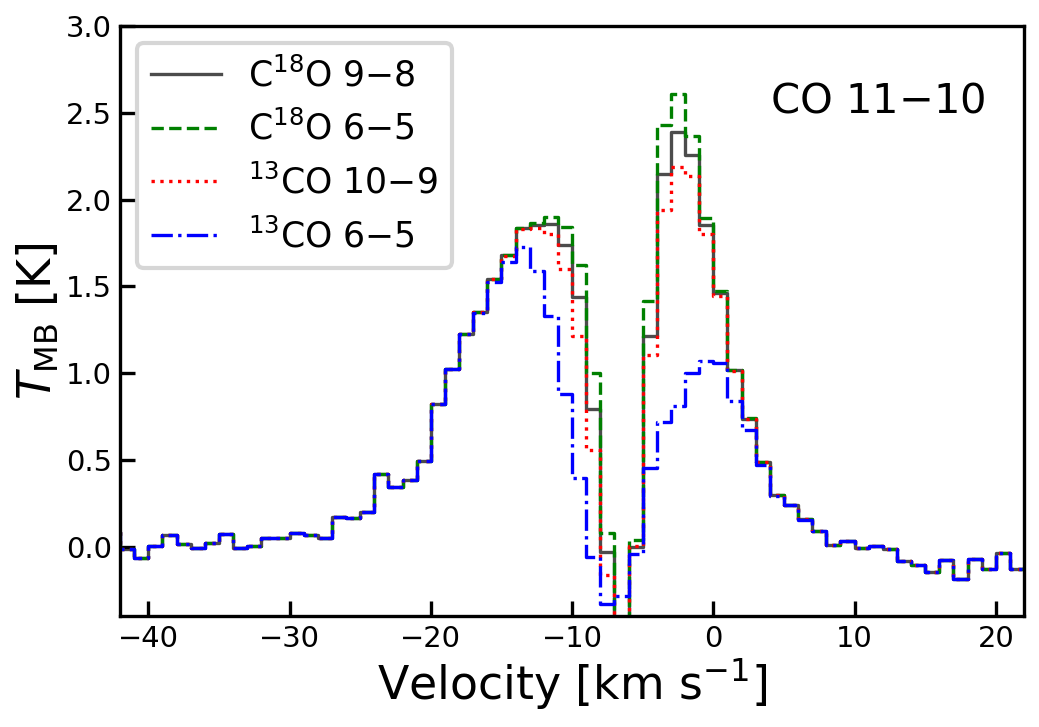}}
   \end{subfigure}
   \caption{\coet{} wing emission of G351.16+0.7 obtained by using different envelope tracers.}
   \label{fig:G35116_many_wings}
\end{figure}

The use of different envelope tracers in Table\,\ref{tab:proxyline_list} could pose a bias on the extracted outflow emission. To assess this issue, we decomposed the \coet{} line of G351.16+0.7 using all four envelope tracers (see Fig.\,\ref{fig:G35116_many_wings}). Table\,\ref{tab:G35116_different_wing} shows that wings extracted by using the C$^{18}$O lines and the \ttcotn{} agree well with each other with differences less $10\%$. Using the \ttcosf{}, however, returns a much less wing emission. The \mbox{mid-$J$} $^{13}$CO line might have higher opacity than the others; thus, having wider width and smaller wing component at the \coet{} line. In this work, we use \ttcosf{} for only one source. Therefore, the bias is minimised. In fact, wing emission derived for that sources is a lower limit for the actual outflow emission, thereby the wing detection statistics is not affected.

\begin{table}[]
    \centering
    \caption{Line wing emission of \mbox{\coet{}} toward G351.16+0.7 using various envelope tracers.}
    \begin{tabular}{cccc}
        \hline
        \hline
        Transition & $S_{\textrm{int}}$\tablefootmark{a} & $T_{\mathrm{peak}}^{\mathrm{blue}}$ & $T_{\mathrm{peak}}^{\mathrm{red}}$ \\
        & (K\,km\,s$^{-1}$) & (K) & (K) \\
        \hline
         \cetone{}  &  33.9 & 1.9 & 2.4 \\
         \cetosf{}  &  35.4 ($4\%$) & 1.9 ($0\%$) & 2.6 ($8\%$) \\
         \ttcotn{}  &  32.5 ($4\%$) & 1.8 ($5\%$) & 2.2 ($8\%$) \\
         \ttcosf{}  &  23.5 ($31\%$)& 1.7 ($11\%$) & 1.1 ($54\%$) \\
        \hline
    \end{tabular}
    \begin{flushleft}
    \tablefoot{Numbers inside brackets show the difference with respect to wing parameters obtained using the \cetone{} transition (top row). \tablefoottext{a}{Total integrated intensity of both blue and red wings.}
    }
    \end{flushleft}
    \label{tab:G35116_different_wing}
\end{table}

\begin{figure}[h!]
    \centering
    \begin{subfigure}{0.4\textwidth}
      \centering
      \resizebox{0.9\hsize}{!}{\includegraphics{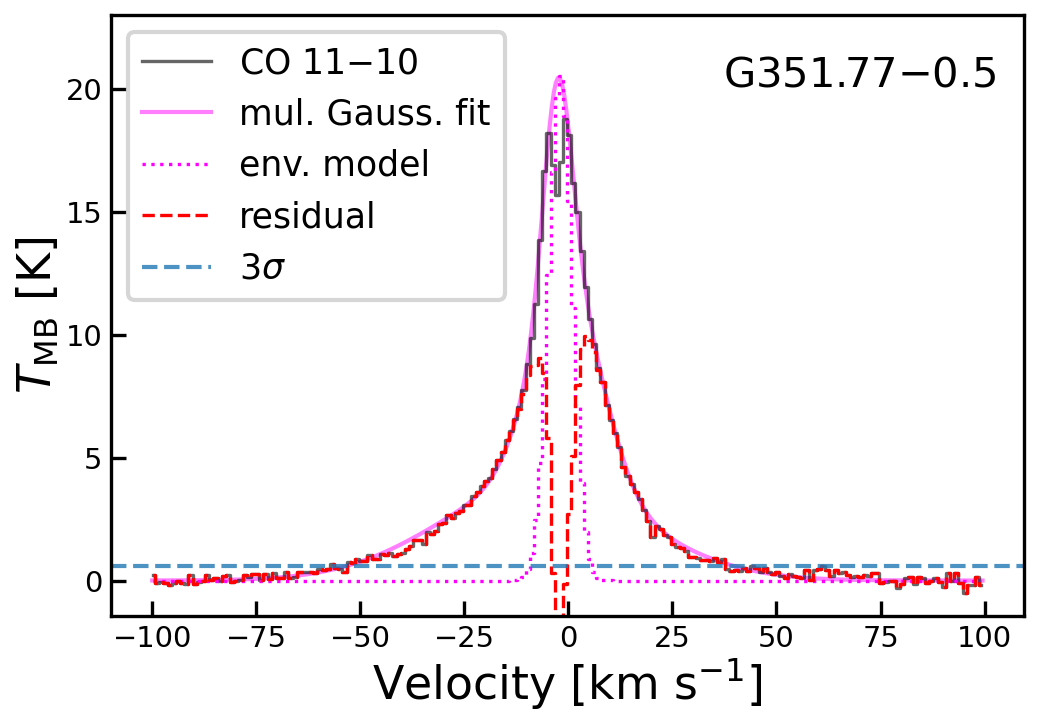}}
   \end{subfigure}
   \caption{Line wing extraction for a self-absorption line in which line peak is determined from a multiple Gaussian fit.}
   \label{fig:AG35177_wing_extract}
\end{figure}

To extract wing emission from self-absorbed lines, we need an adjusted approach (Fig\,\ref{fig:AG35177_wing_extract}). The line peak is underestimated because the emission is absorbed by cooler gas in front of the bulk emission. To recover the line peak, we fit a multiple-Gaussian profile to the line with the absorbed parts being masked out. The fitting process is adopted from \citet{navarete2019atlasgal}. At first, a single Gaussian is fit to the full profiles. A second component is added if there are significant emission above $3 \sigma$ in the residual. Subsequently, a third Gaussian would be added if the residual still show emission. A limit is set at maximum three Gaussian components can be used. Results of the multiple-Gaussian fit are shown in Table \ref{tab:multigauss_fit}. Adopting the fit peak, we can do the line decomposition following the four steps above. In step (3), the envelope model is shifted to velocity of the fit peak.

\begin{table*}
\begin{center}
\caption{Tracers of dense gas envelopes at each regions.}
\label{tab:proxyline_list}
\renewcommand{\footnoterule}{}  
\begin{tabular}{llcccc}
\hline\hline
No. & Source & \cetone{} & \ttcotn{} & \cetosf{} & \ttcosf{} \\
\hline
1 & G351.16+0.7 & \checkmark &  &  &  \\
2 & G351.25+0.7 & \checkmark &  &  &  \\
3 & G351.44+0.7 & \checkmark &  &  &  \\
4 & G351.58$-$0.4 & \checkmark &  &  &  \\
5 & G351.77$-$0.5 &  & \checkmark &  &  \\
\hline
6 & G12.81$-$0.2 & \checkmark &  &  &  \\
7 & G14.19$-$0.2 &  &  &  & \checkmark \\
8 & G13.66$-$0.6 & & \checkmark &  &  \\
9 & G14.63$-$0.6 & & \checkmark &  &  \\
\hline
10 & G34.41+0.2 & \checkmark &  &  &  \\
11 & G34.26+0.15 &  &  & \checkmark &  \\
12 & G34.40$-$0.2 & \checkmark &  &  &  \\
13 & G35.20$-$0.7 & \checkmark &  &  &  \\
\hline
\end{tabular}
\end{center}
\end{table*}

\begin{figure*}[h!]
   \begin{subfigure}{0.33\textwidth}
      \centering
      \resizebox{0.9\hsize}{!}{\includegraphics{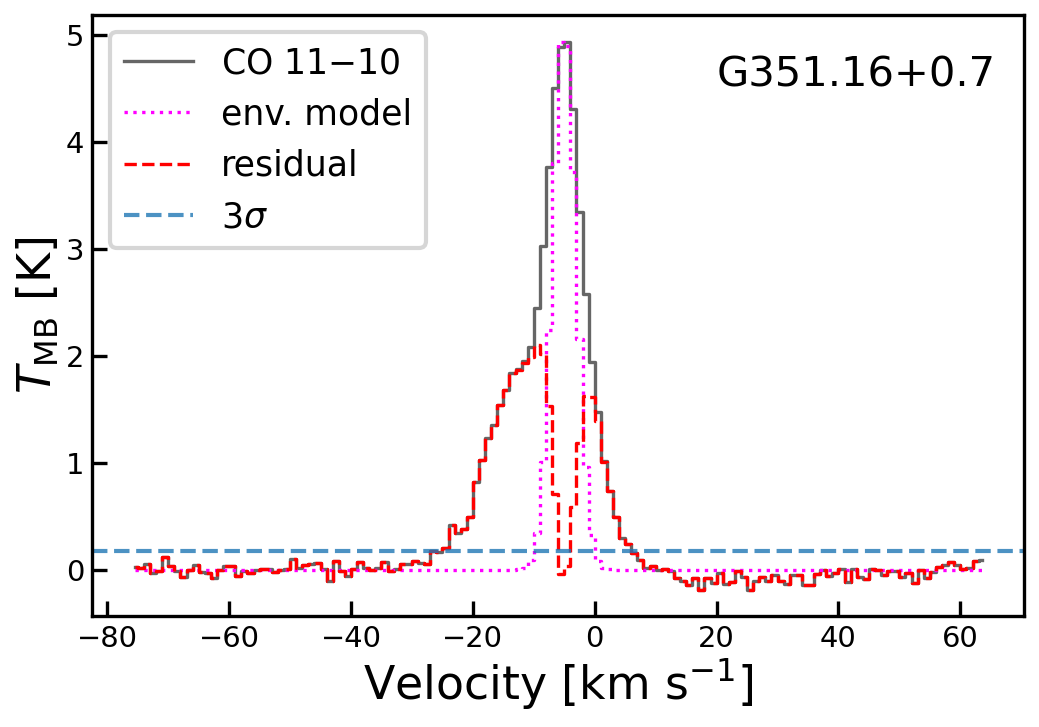}}
   \end{subfigure}%
   \begin{subfigure}{0.33\textwidth}
      \centering
      \resizebox{0.9\hsize}{!}{\includegraphics{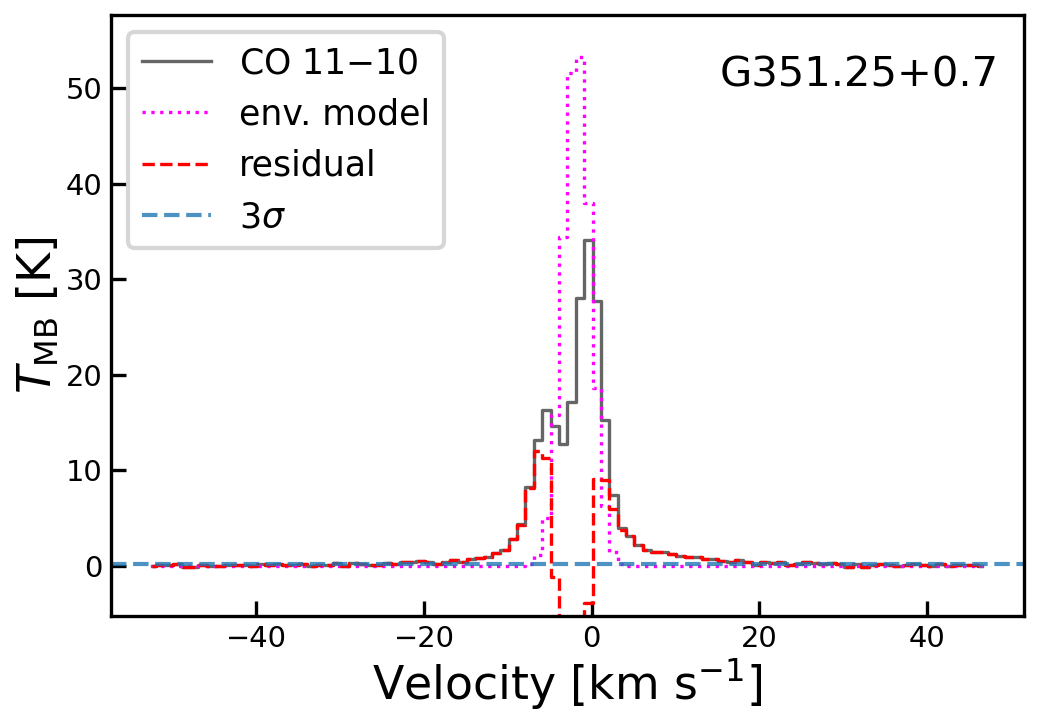}}
   \end{subfigure}%
   \begin{subfigure}{0.33\textwidth}
      \centering
      \resizebox{0.9\hsize}{!}{\includegraphics{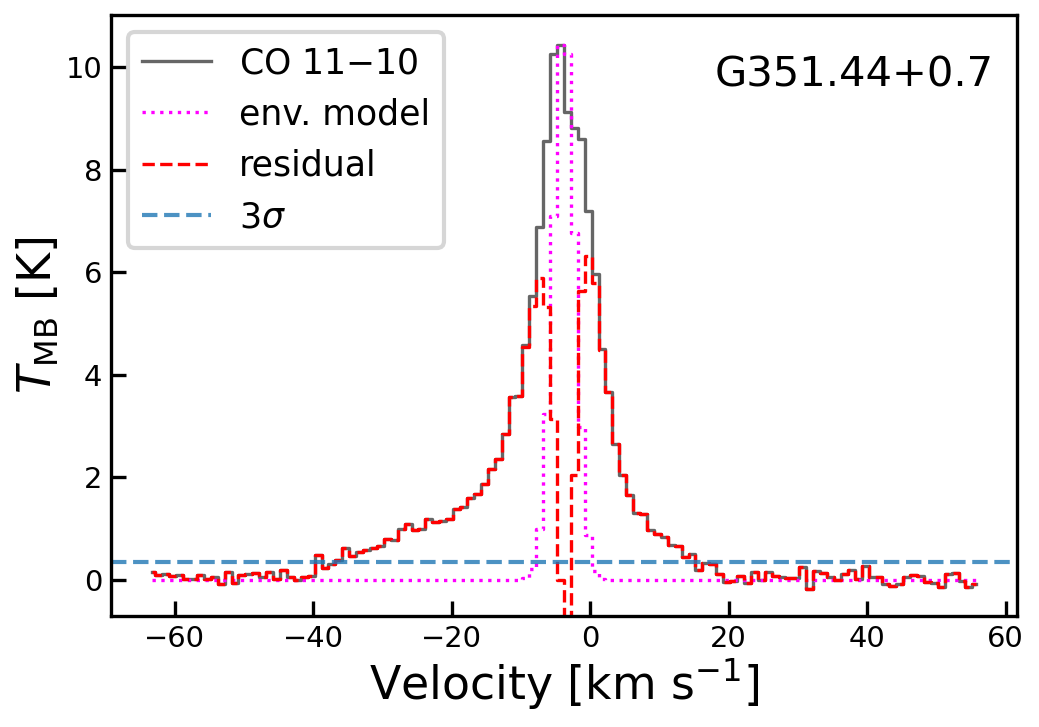}}
   \end{subfigure}
   \begin{subfigure}{0.33\textwidth}
      \centering
      \resizebox{0.9\hsize}{!}{\includegraphics{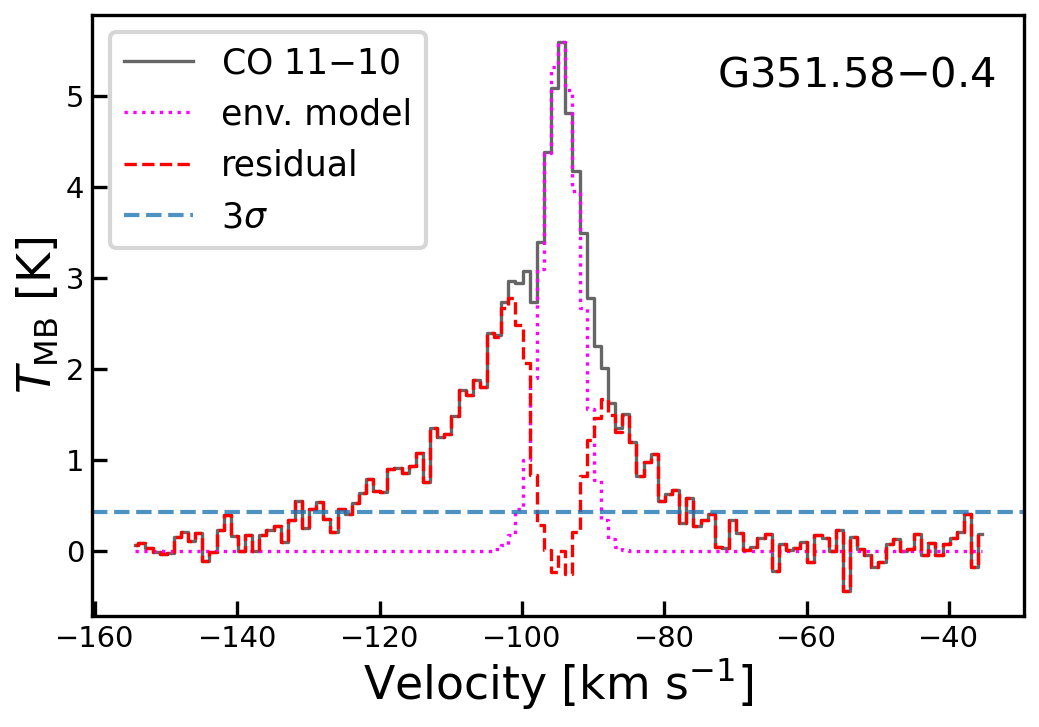}}
   \end{subfigure}%
    \begin{subfigure}{0.33\textwidth}
      \centering
      \resizebox{0.9\hsize}{!}{\includegraphics{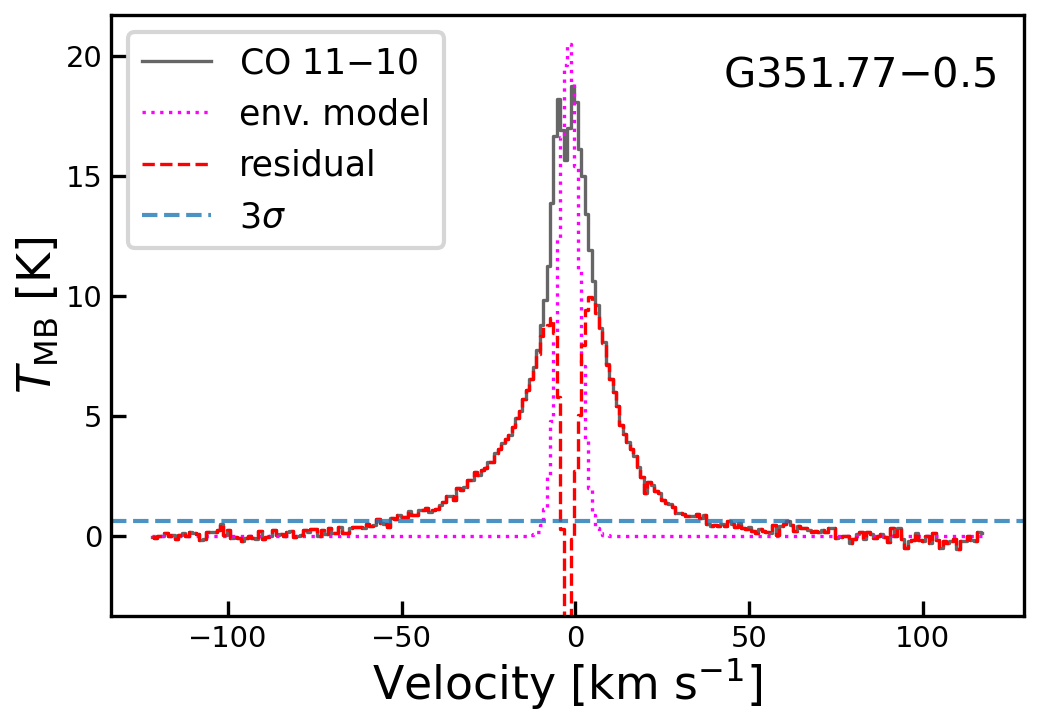}}
   \end{subfigure}%
   \begin{subfigure}{0.33\textwidth}
      \centering
      \resizebox{0.9\hsize}{!}{\includegraphics{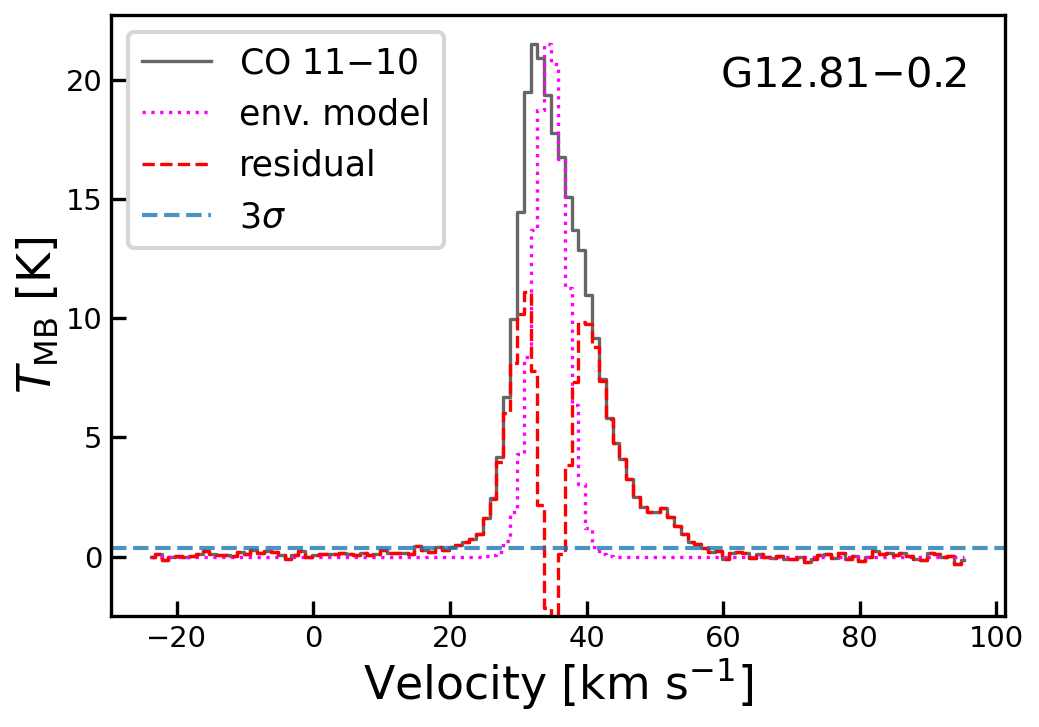}}
   \end{subfigure}
   \begin{subfigure}{0.33\textwidth}
      \centering
      \resizebox{0.9\hsize}{!}{\includegraphics{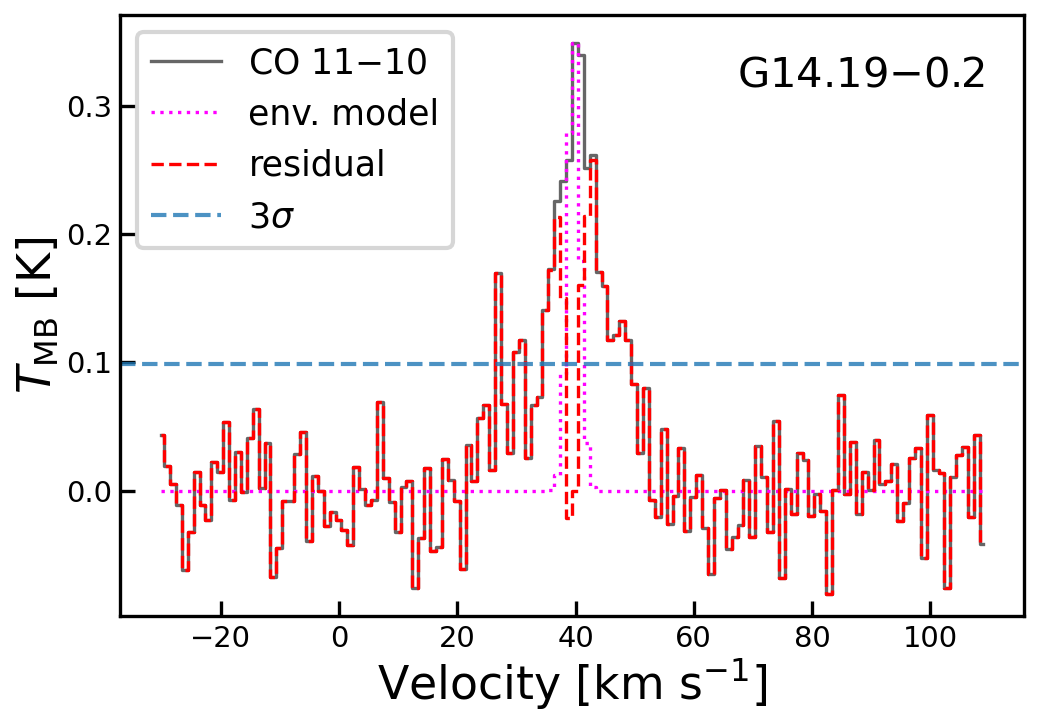}}
   \end{subfigure}%
      \begin{subfigure}{0.33\textwidth}
      \centering
      \resizebox{0.9\hsize}{!}{\includegraphics{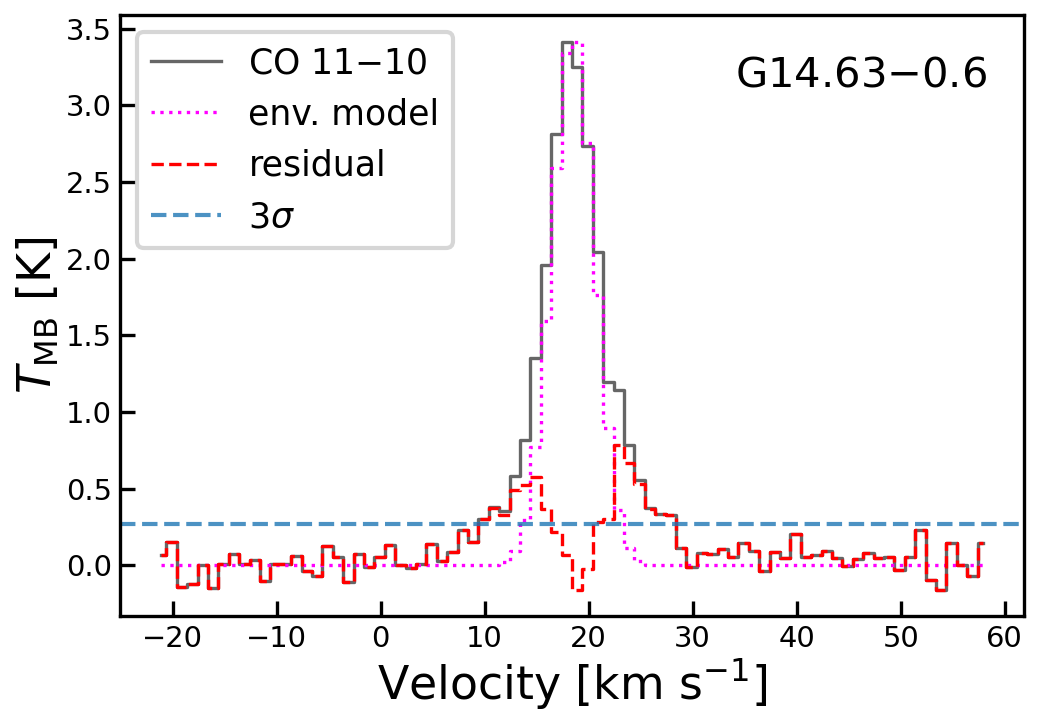}}
   \end{subfigure}%
   \begin{subfigure}{0.33\textwidth}
      \centering
      \resizebox{0.9\hsize}{!}{\includegraphics{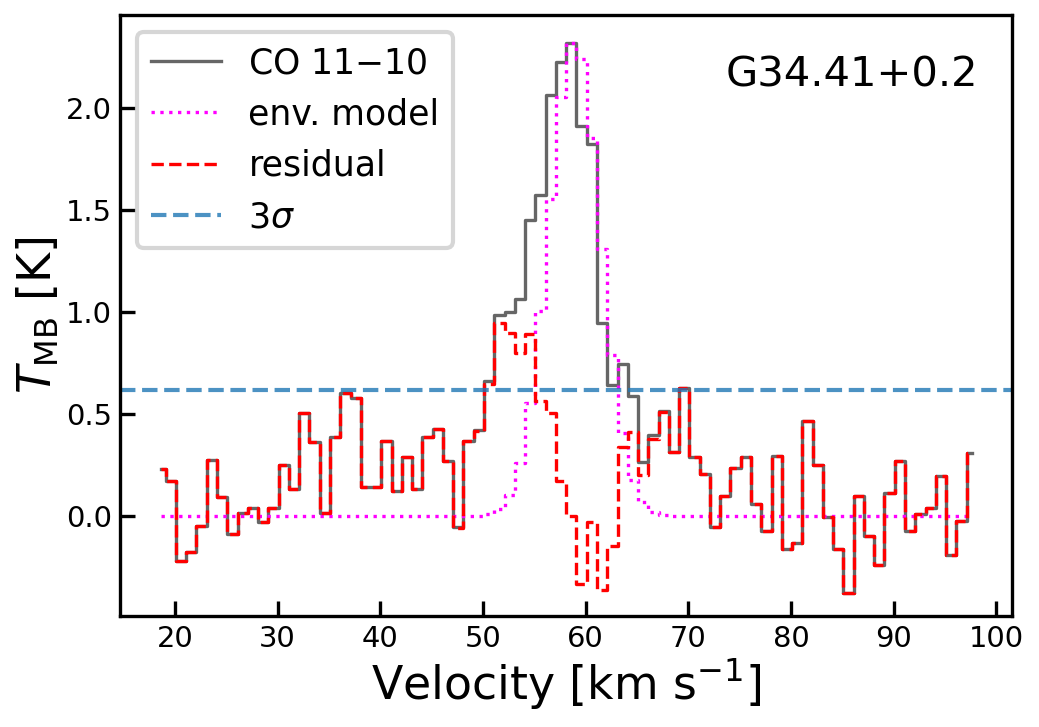}}
   \end{subfigure}
   \begin{subfigure}{0.33\textwidth}
      \centering
      \resizebox{0.9\hsize}{!}{\includegraphics{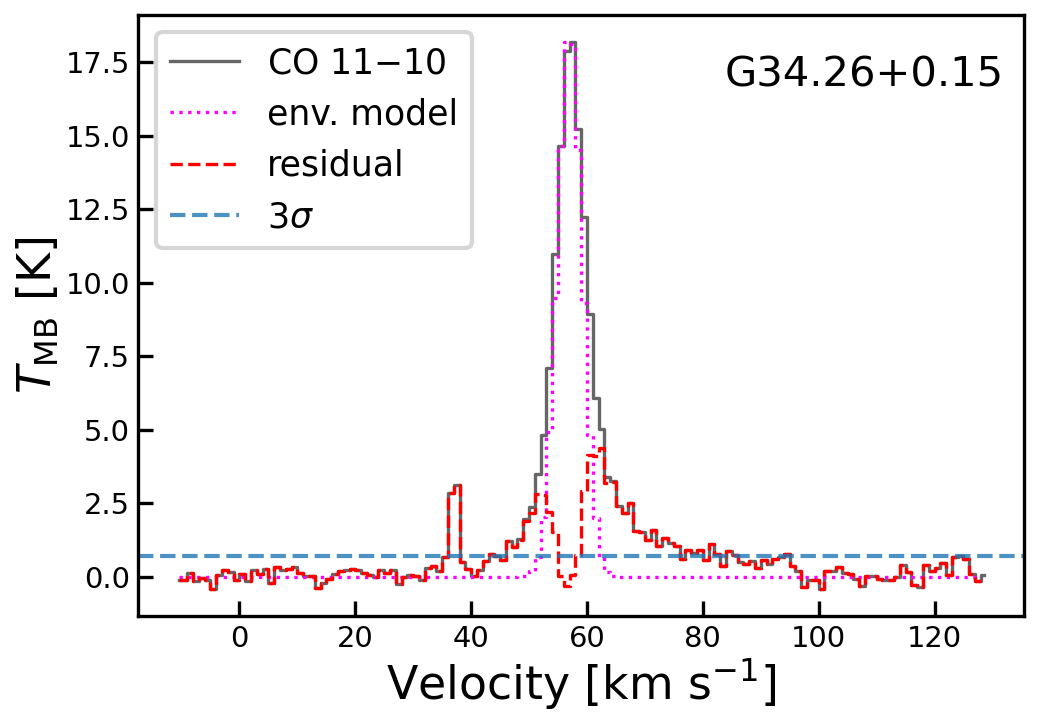}}
   \end{subfigure}%
    \begin{subfigure}{0.33\textwidth}
      \centering
      \resizebox{0.9\hsize}{!}{\includegraphics{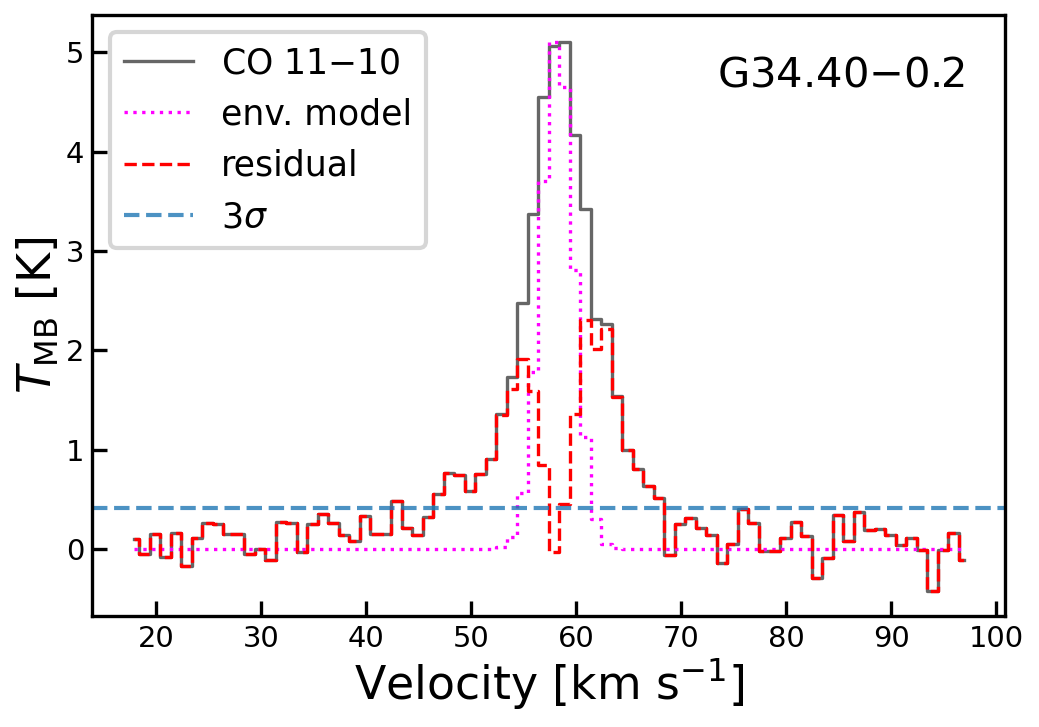}}
   \end{subfigure}%
   \begin{subfigure}{0.33\textwidth}
      \centering
      \resizebox{0.9\hsize}{!}{\includegraphics{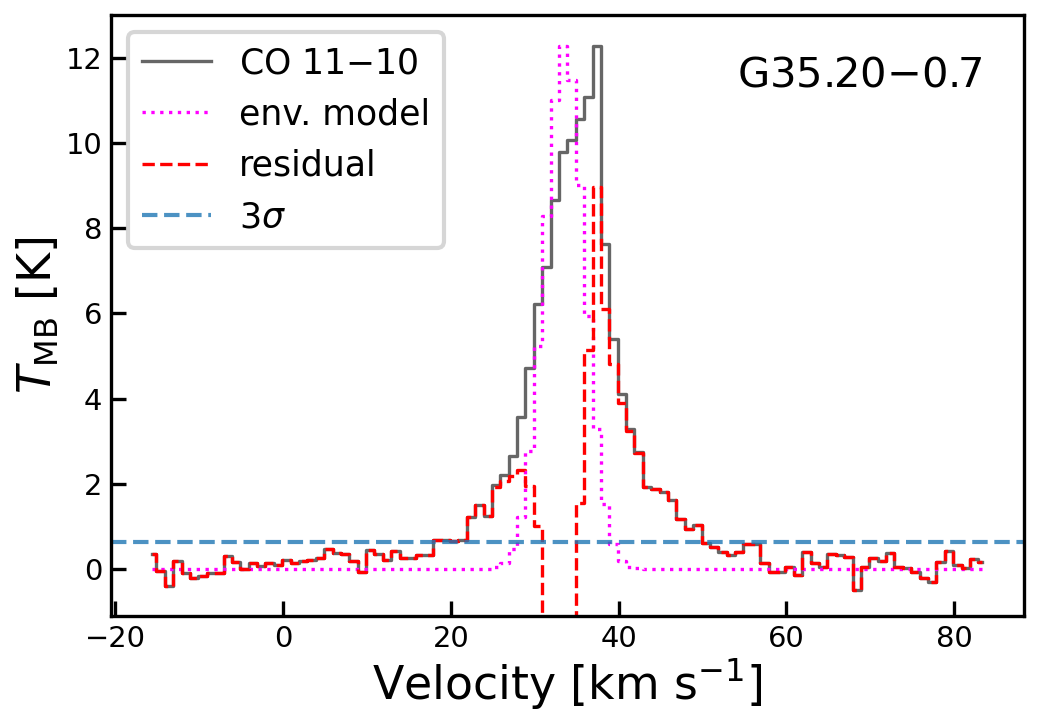}}
   \end{subfigure}%
   \caption{\coet{} wing emission (dashed red profile) extracted from the full line (solid black profile). Dotted magenta line is a model of envelope emission. The detection 3$\sigma$ threshold is shown with a dashed blue line.}
   \label{fig:all_11_wing}
\end{figure*}

\begin{figure*}[h!]
   \begin{subfigure}{0.33\textwidth}
      \centering
      \resizebox{0.9\hsize}{!}{\includegraphics{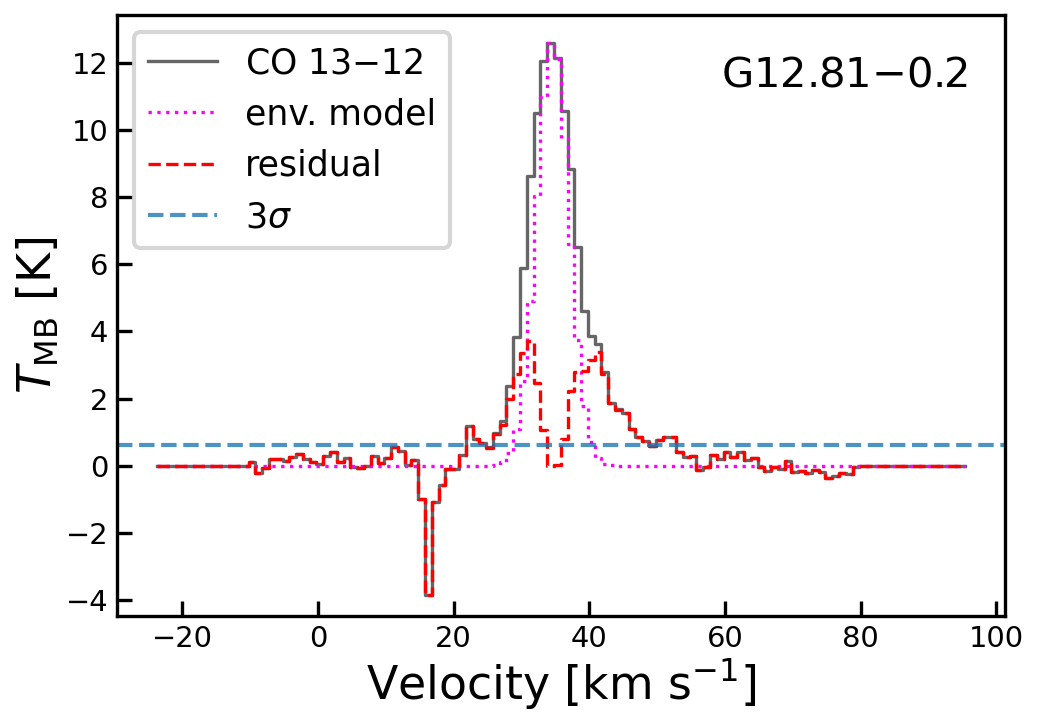}}
   \end{subfigure}%
   \begin{subfigure}{0.33\textwidth}
      \centering
      \resizebox{0.9\hsize}{!}{\includegraphics{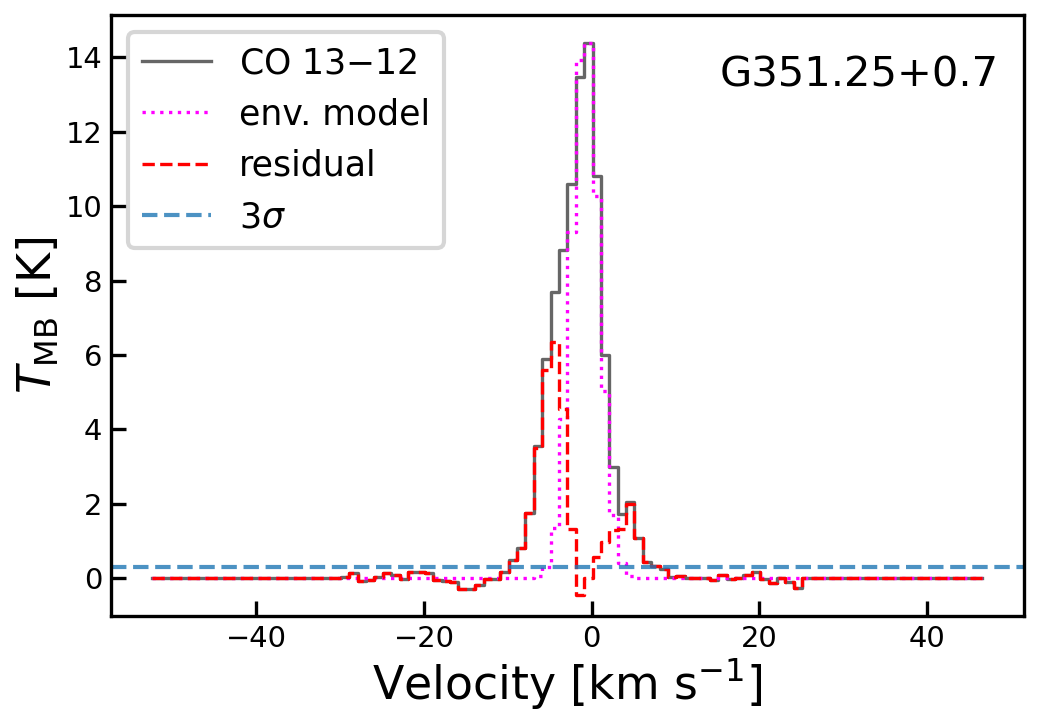}}
   \end{subfigure}%
   \caption{\cott{} wing emission (dashed red profile). The color-coding and line styles are the same as in Fig.\,\ref{fig:all_11_wing}.}
   \label{fig:all_13_wing}
\end{figure*}

\begin{figure*}[h!]
   \begin{subfigure}{0.33\textwidth}
      \centering
      \resizebox{0.9\hsize}{!}{\includegraphics{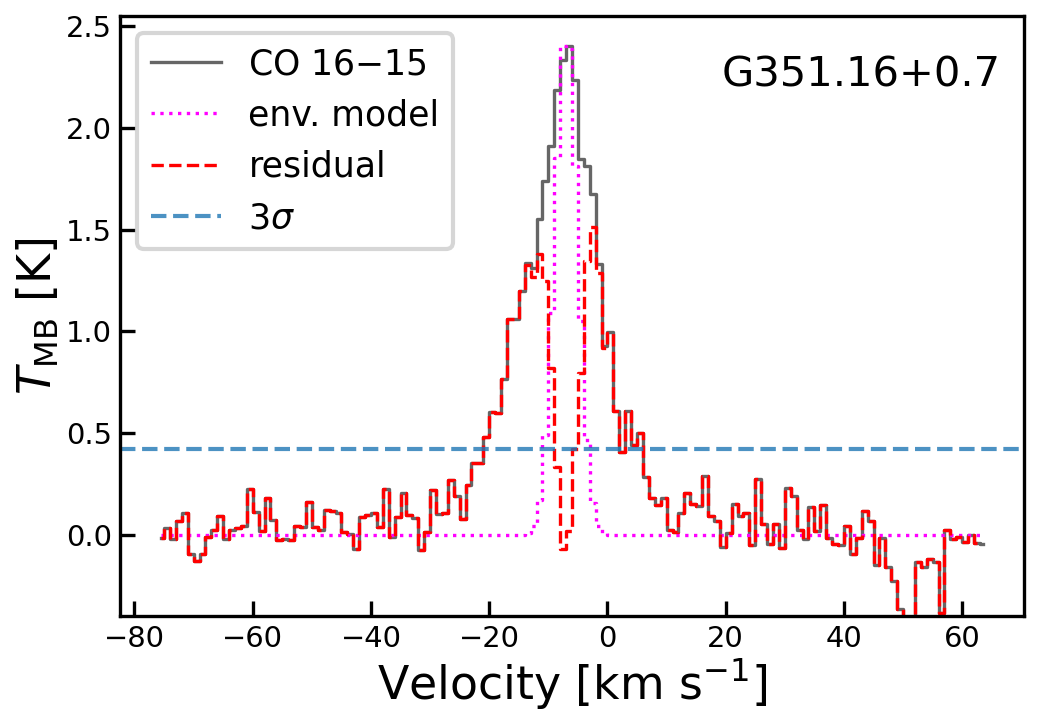}}
   \end{subfigure}%
   \begin{subfigure}{0.33\textwidth}
      \centering
      \resizebox{0.9\hsize}{!}{\includegraphics{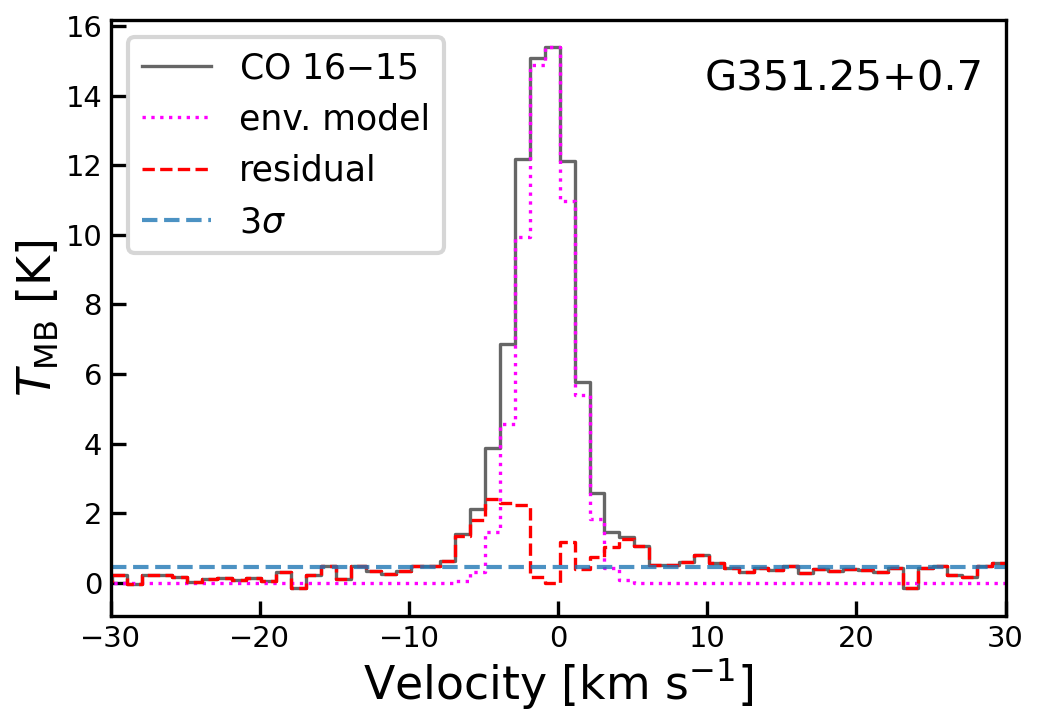}}
   \end{subfigure}%
   \begin{subfigure}{0.33\textwidth}
      \centering
      \resizebox{0.9\hsize}{!}{\includegraphics{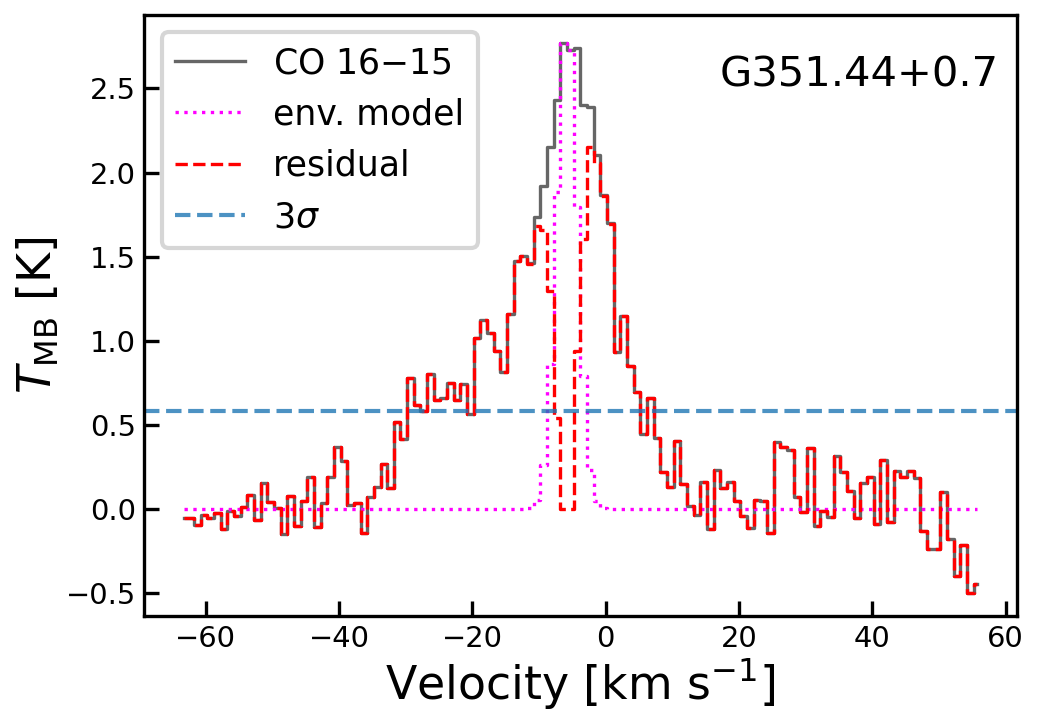}}
   \end{subfigure}
   \begin{subfigure}{0.33\textwidth}
      \centering
      \resizebox{0.9\hsize}{!}{\includegraphics{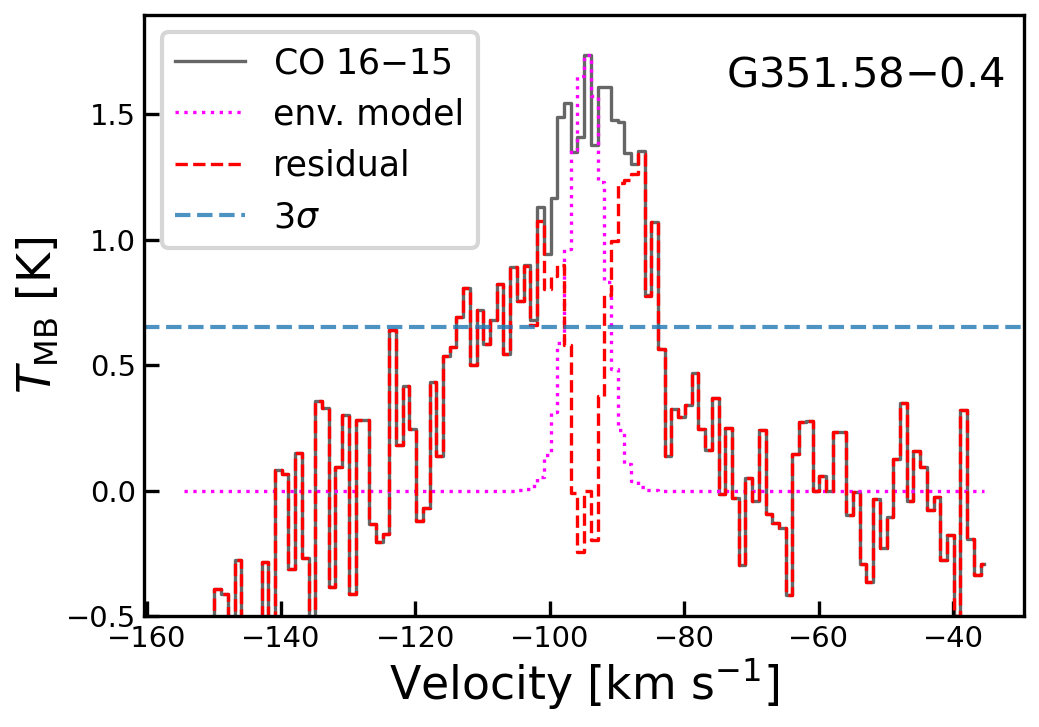}}
   \end{subfigure}%
   \begin{subfigure}{0.33\textwidth}
      \centering
      \resizebox{0.9\hsize}{!}{\includegraphics{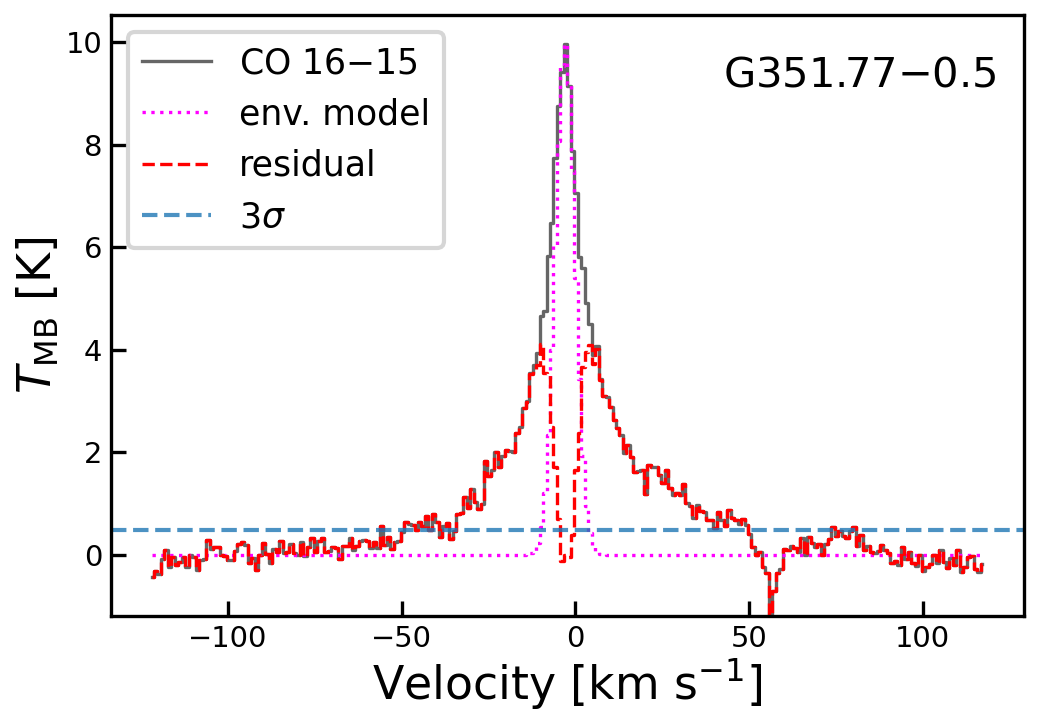}}
   \end{subfigure}%
    \begin{subfigure}{0.33\textwidth}
      \centering
      \resizebox{0.9\hsize}{!}{\includegraphics{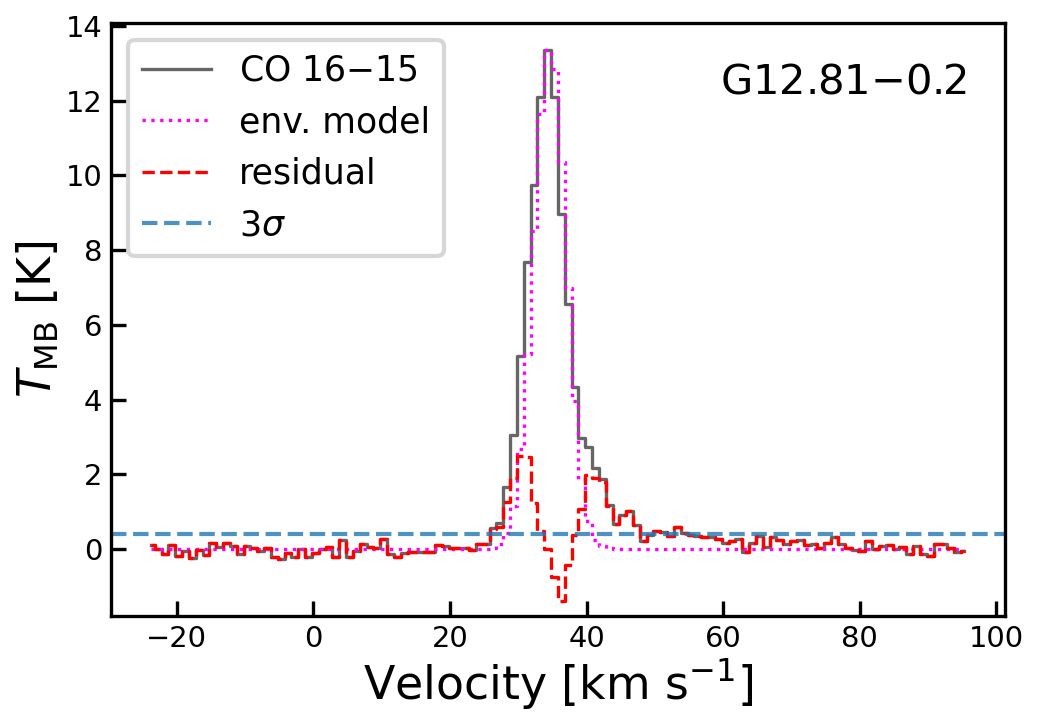}}
   \end{subfigure}
   \begin{subfigure}{0.33\textwidth}
      \centering
      \resizebox{0.9\hsize}{!}{\includegraphics{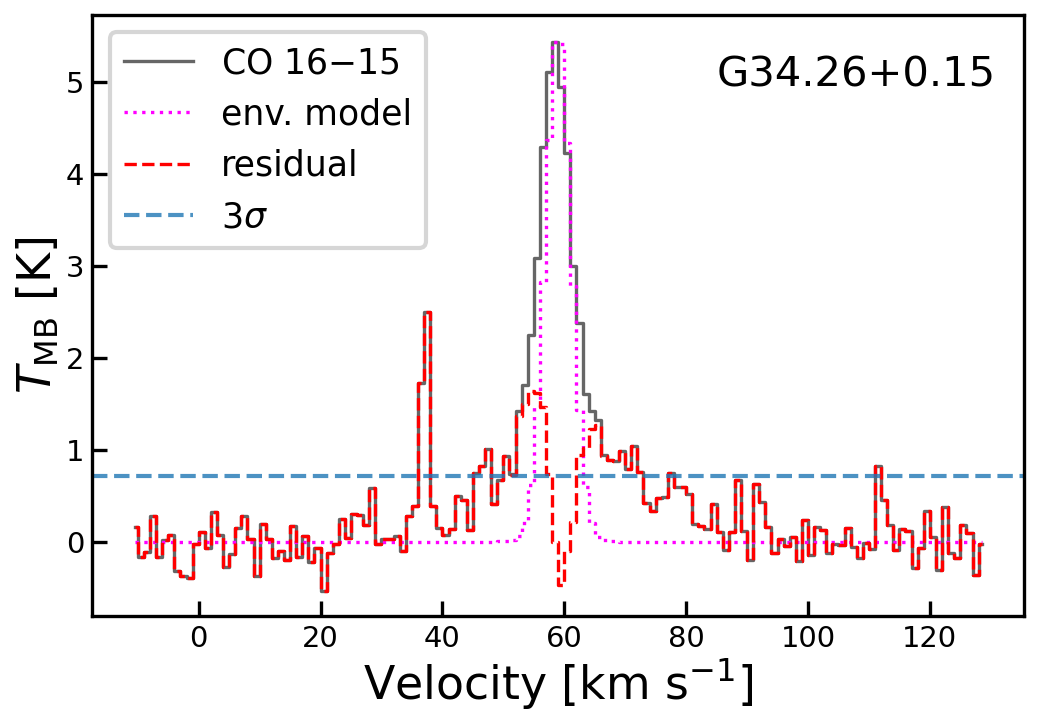}}
   \end{subfigure}%
   \begin{subfigure}{0.33\textwidth}
      \centering
      \resizebox{0.9\hsize}{!}{\includegraphics{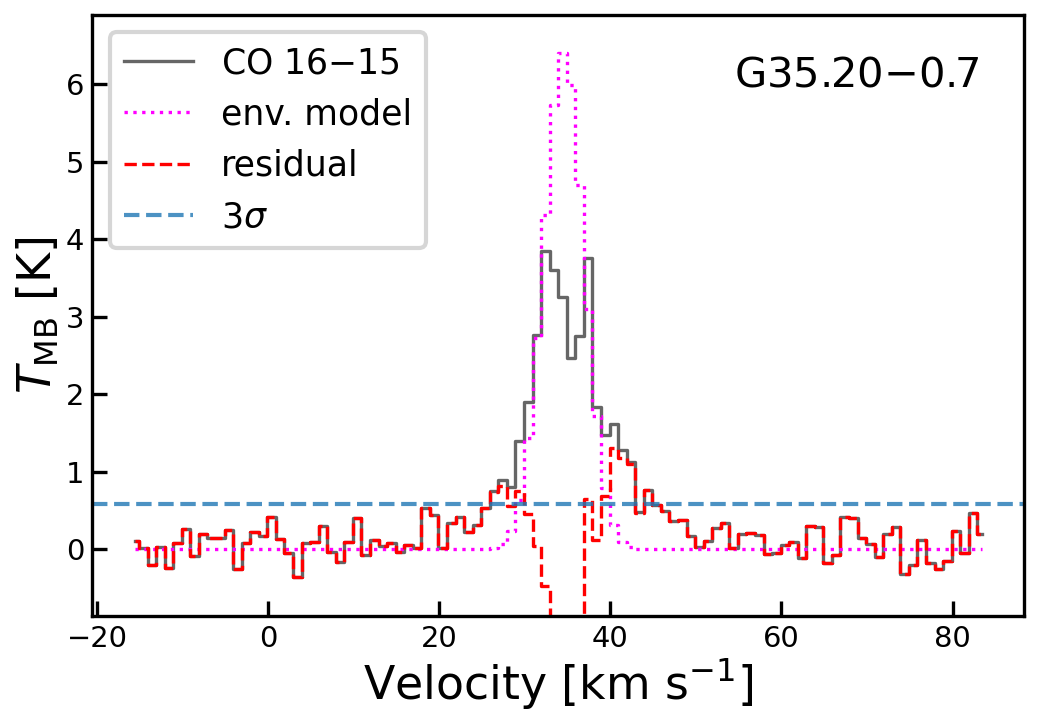}}
   \end{subfigure}
   \caption{\costft{} wing emission (dashed red profile). The color-coding and line styles are the same as in Fig.\,\ref{fig:all_11_wing}.}
   \label{fig:all_16_wing}
\end{figure*}

\begin{figure*}
\centering
   \begin{subfigure}{0.45\hsize}
      \centering
      \resizebox{0.9\hsize}{!}{\includegraphics{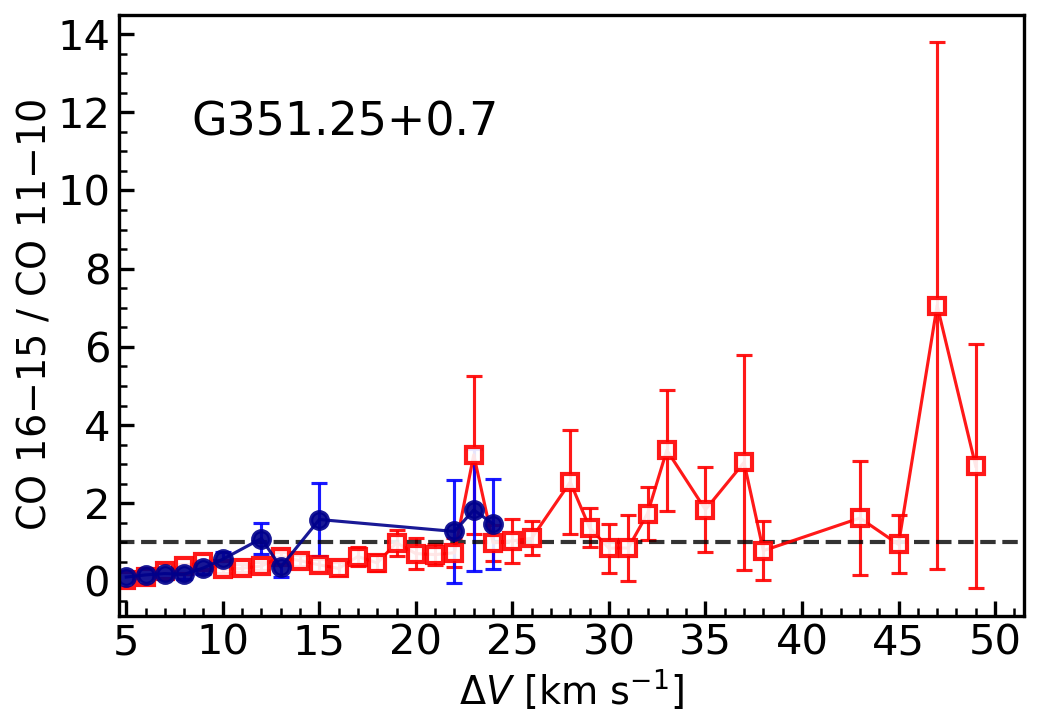}}
   \end{subfigure}%
   \begin{subfigure}{0.45\hsize}
      \centering
      \resizebox{0.9\hsize}{!}{\includegraphics{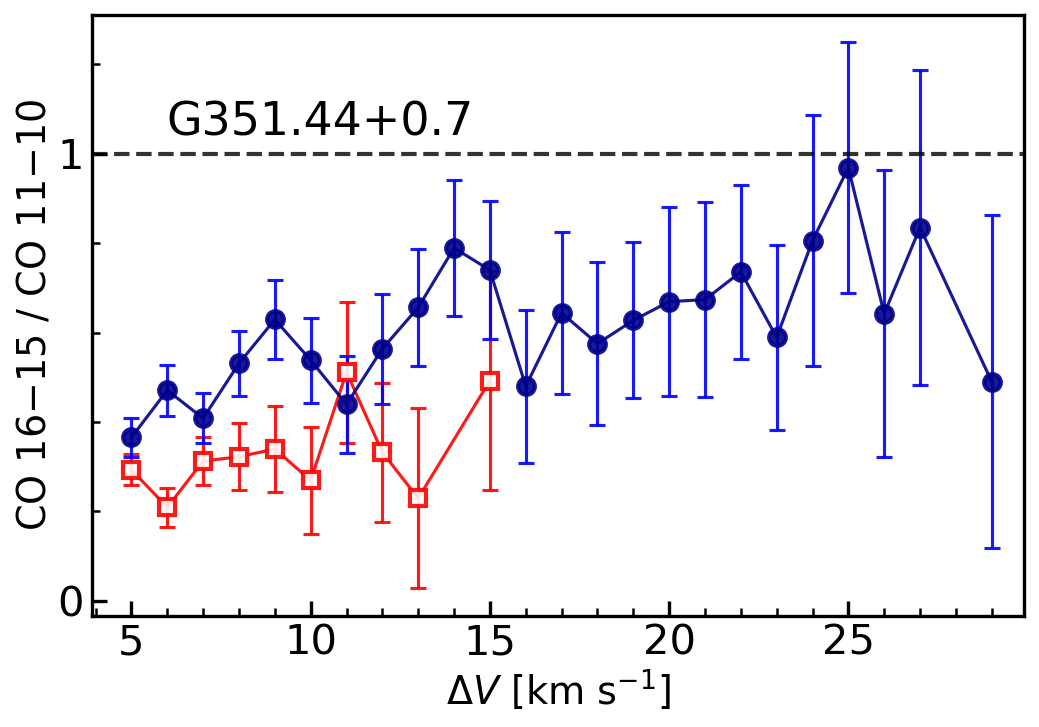}}
   \end{subfigure}
   \begin{subfigure}{0.45\hsize}
      \centering
      \resizebox{0.9\hsize}{!}{\includegraphics{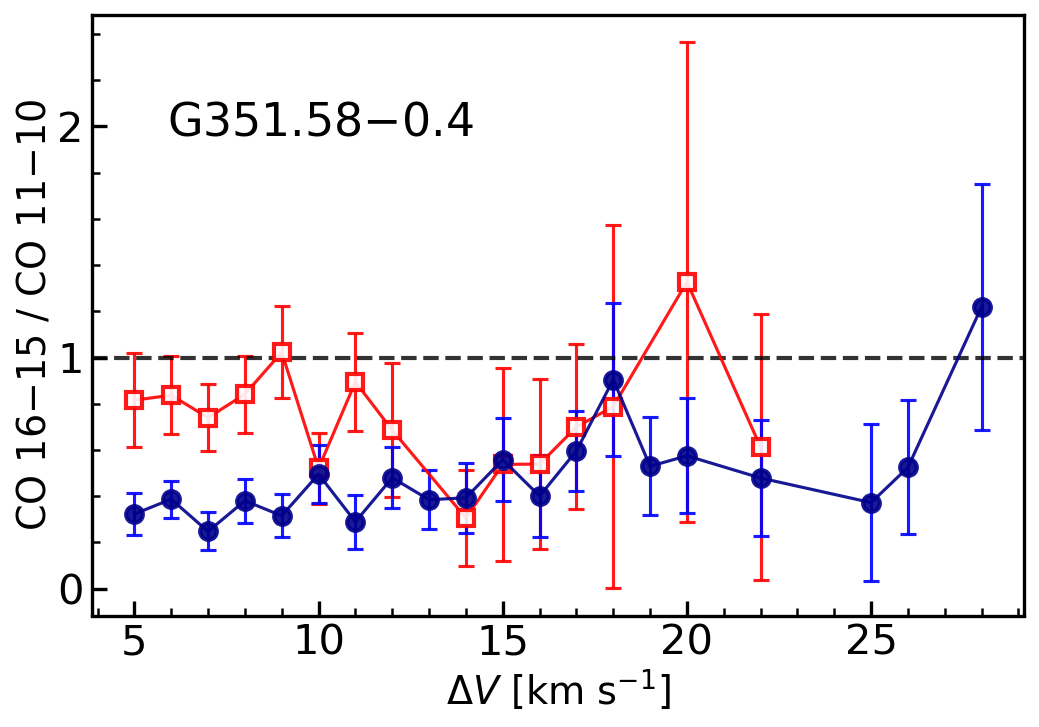}}
   \end{subfigure}%
   \begin{subfigure}{0.45\hsize}
      \centering
      \resizebox{0.9\hsize}{!}{\includegraphics{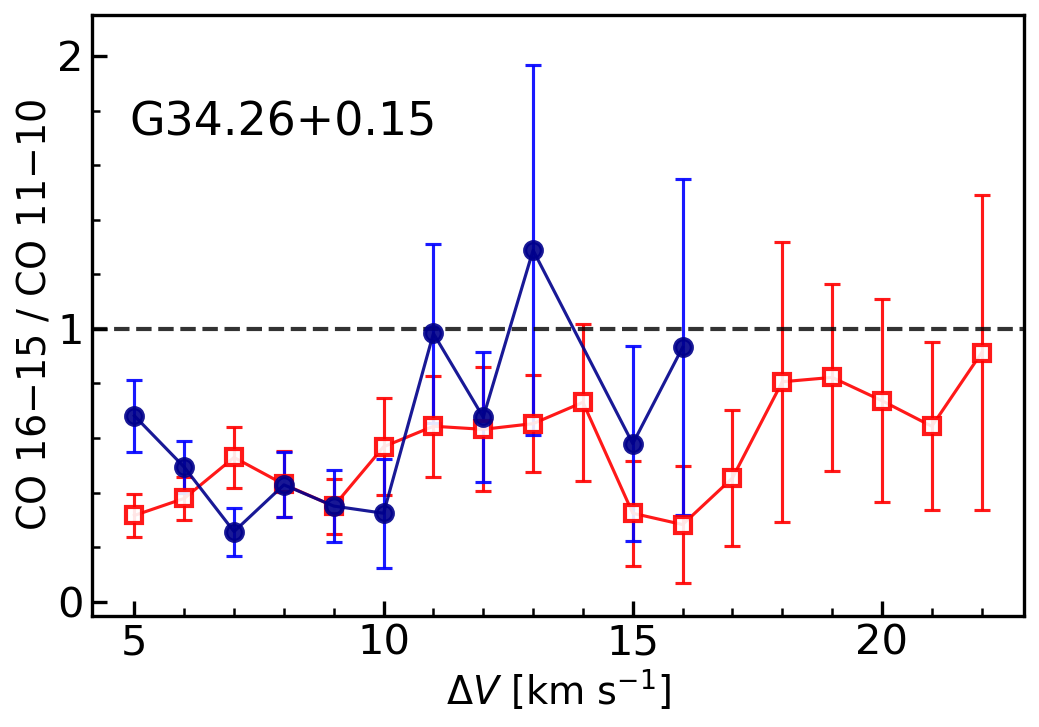}}
   \end{subfigure}
   \caption{The ratio of line wing emission in the \costft{} and 11--10 transitions versus absolute velocity offset from source velocity. The red-shifted emission is shown in red squares while the blue shifted emission is in blue circle. The dashed horizontal line presents the level above which \costft{} is greater than \coet{}.}
   \label{fig:line_wing_ratio_all}
\end{figure*}

\subsection{Wing emission in \cosf{}} \label{sec:wing-extract-co65}

\begin{figure*}[h]
   \begin{subfigure}{0.33\textwidth}
      \centering
      \resizebox{0.9\hsize}{!}{\includegraphics{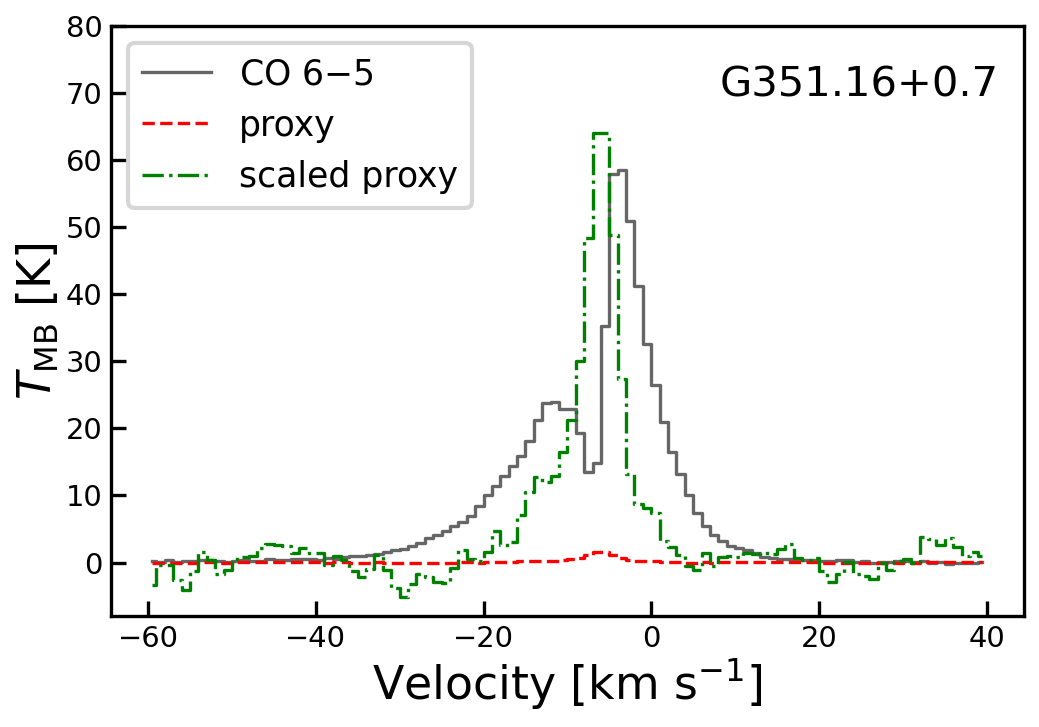}}
   \end{subfigure}%
   \begin{subfigure}{0.33\textwidth}
      \centering
      \resizebox{0.9\hsize}{!}{\includegraphics{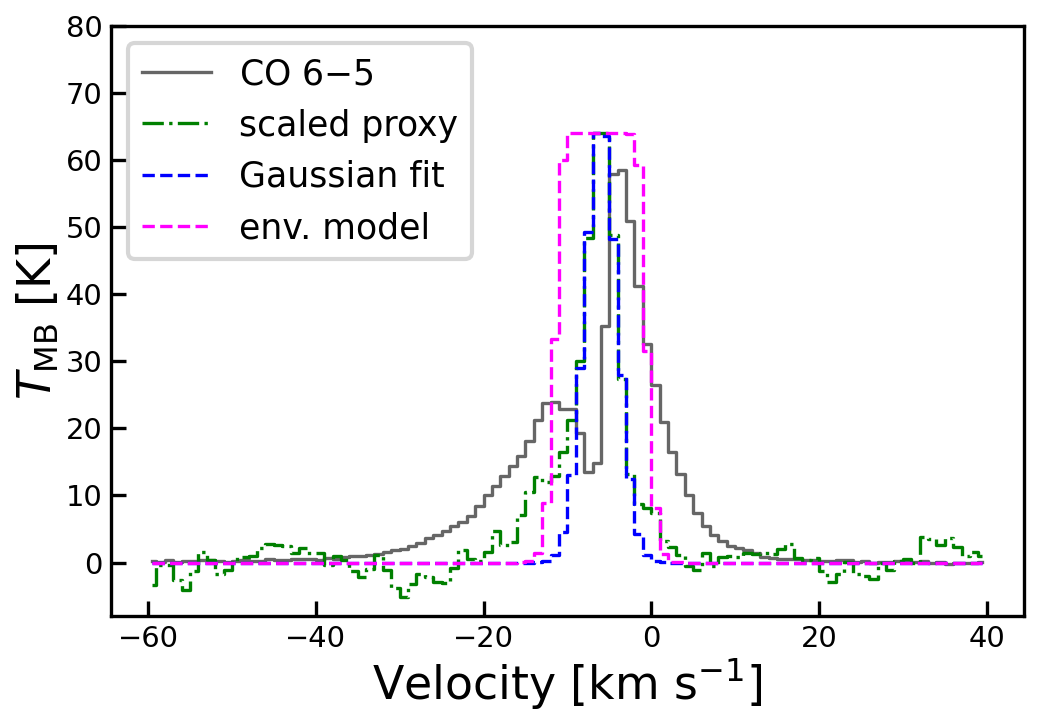}}
   \end{subfigure}
   \begin{subfigure}{0.33\textwidth}
      \centering
      \resizebox{0.9\hsize}{!}{\includegraphics{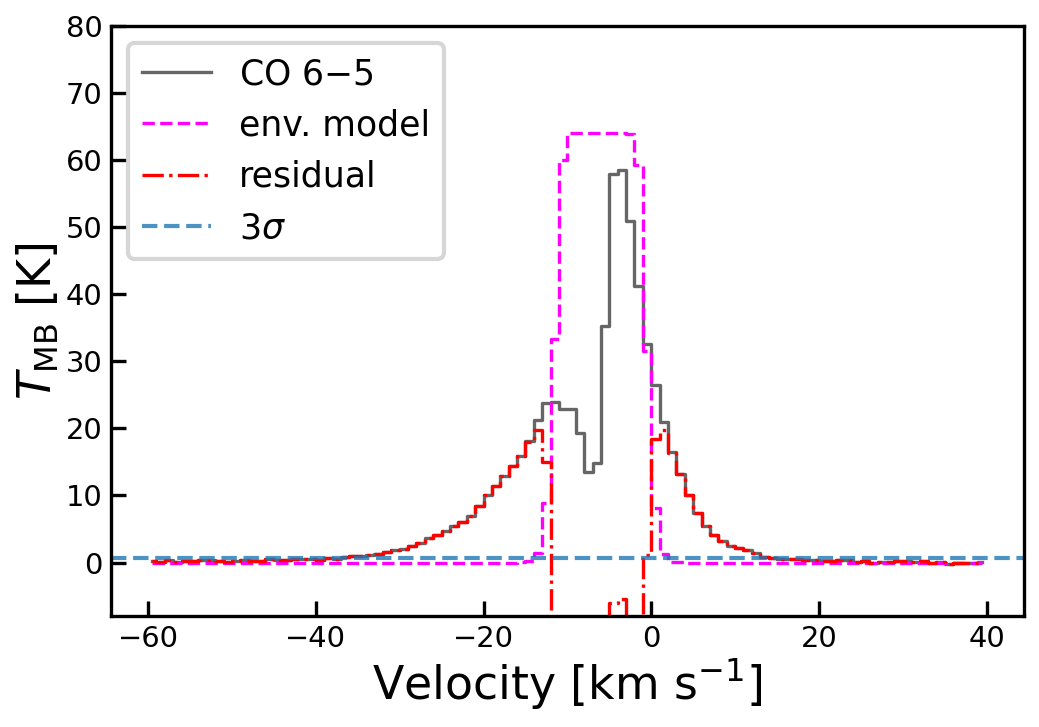}}
   \end{subfigure}
   \caption{Wing extraction steps for the \cosf{} line of G351.16+0.7 from left to right (see Appendix \ref{app:decomposition-method}).}
   \label{fig:wing_detect_midj}
\end{figure*}

\citet{navarete2019atlasgal} found broad components associated with outflows on \cosf{} lines of our sources. Here, we attempt to extract wing emission from the \cosf{} line of seven sources in our sample (Table\,\ref{tab:wing_properties}) by applying the decomposition method in \citep{yang22} (see Fig.\,\ref{fig:wing_detect_midj}). This method is similar to the decomposition method we use for the high-$J$ CO lines, but an additional step is added after step (2) to account for the opacity broadening effect of the optically thick \cosf{}.

To correct for the opacity broadening, one needs to derive opacity in each frequency channel:
\begin{equation}\label{eqn:tau_distribution}
        \tau_{\nu} = \tau_0 \times \text{exp}\left(-\frac{(\nu-\nu_0)^2}{2\sigma^2}\right),
\end{equation}
where $\tau_0$ and $\nu_0$ are central opacity and line frequency, $\sigma$ is the gas intrinsic velocity dispersion, namely, the FWHM of the Gaussian fit to the scaled proxy line. \cosf{} central opacity is estimated by multiplying the optical depth of \ttcosf{} (Dat et al., in preparation) with the abundance ratio of $^{12}$CO relative to $^{13}$CO, $X_{^{12}\mathrm{CO}/^{13}\mathrm{CO}}$, which is calculated using the galactic radius relation derived in \citet{giannetti2014atlasgal} and galactic distance. The broaden line, which can be used as model for the envelope emission, then, can be obtained by the radiative transfer equation:
\begin{equation}
    T_{\text{mb},\nu} = [F_{\nu}(T_{\text{ex}})-F_{\nu}(T_{\text{bg}})] \times [1-\text{exp}(-\tau_{\nu})],
\end{equation}
where $F_{\nu}(T) = \frac{h\nu/k}{\text{exp}(h\nu/kT)-1} $, $T_{\text{ex}}$ and $T_{\text{bg}}$ are excitation and background temperature, respectively. The term $[J_{\nu}(T_{\text{ex}})-$J$_{\nu}(T_{\text{bg}})]$ is directly obtained from the peak of the Gaussian fit to the scaled proxy line.

The very high optical depths of the \cosf{} line (up to 122) result in broad lines with flat peaks (see Fig.\,\ref{fig:wing_detect_midj}, middle). Such line profiles are not seen in observed spectra, which suggests that our envelope models are not perfect. However, the wing emission retained after subtracting such envelope models from the full line profiles can still serve as a lower limit for the outflow emission. In total, we could determine \cosf{} wing emission for all seven sources.

\begin{table*}[h!]
\centering
\caption{Line wing emission in the \coet{}, 16--15, 13--12, and 6--5 line.}
\label{tab:wing_properties}
\begin{tabular}{ll|lcc|lcc}
 \hline
 \hline
\multirow{3}{0.5cm}{No.} & \multirow{3}{2cm}{Source} & \multicolumn{3}{c|}{Blue wing} & \multicolumn{3}{c}{Red wing} \\ \cline{3-8}
                & & $V_{\text{range}}$ & $T_{\text{peak}}$ & $S_{\text{int}}$ & $V_{\text{range}}$ & $T_{\text{peak}}$ & $S_{\text{int}}$ \\
                & & (km s$^{-1}$) & (K) & (K\,km\,s$^{-1}$) & (km s$^{-1}$) & (K) & (K\,km\,s$^{-1}$) \\
 \hline
 \multicolumn{8}{c}{\coet{}} \\ 
 \hline
1 & G351.16+0.7 & [-32, -6] & 2.1 & 23.7  & [-5, 8] & 1.6 & 9.4 \\
2 & G351.25+0.7 & [-47, -5] & 12.1 & 44.9 & [0, 33] & 9.1 & 43.6\\
3 & G351.44+0.7 & [-43, -5] & 5.9 & 60.4 & [-3, 21] & 6.3 & 41.7 \\
4& G351.58$-$0.4 & [-137, -96] & 2.8 & 41.4  & [-93, -68] & 1.7 & 18.0 \\
5 & G351.77$-$0.5 & [-81, -3] & 9.1 & 171.4 & [0, 76] & 9.9 & 154.3 \\
\hline
6 & G12.81$-$0.2 & [-1, 34] & 11.1 & 59.1 & [36, 65] & 9.9 & 81.4\\
7 & G14.19$-$0.2 & [21, 38] & 0.2 & 1.4 & [40, 54] & 0.3 & 1.5\\
8 & G14.63$-$0.6 & [2, 18] & 0.6 & 3.9 & [20, 30] & 0.8 & 3.6\\
\hline
9 & G34.41+0.2  & [28, 58] & 0.9 & 11.2 & -- & -- & -- \\
10 & G34.26+0.15 & [41, 56] & 2.8 & 19.6 & [57, 98] & 4.4 & 53.1 \\
11 & G34.40$-$0.2  & [30, 57] & 1.9 & 14.5 & [58, 75] & 2.3 & 13.3\\
12 & G35.20$-$0.7  & [-8, 31] & 2.3 & 23.4 & [35, 58] & 9.0 & 49.5\\
  \hline
  \multicolumn{8}{c}{\costft{}} \\ 
  \hline
1 & G351.16+0.7 & [-31, -8] & 1.4 & 13.9 & [-7, 20] & 1.5 & 11.7\\
2 & G351.25+0.7 & [-28, -1] & 2.4 & 15.2 & [0, 47] & 1.3 & 18.0\\
3 & G351.44+0.7 & [-36, -7] & 1.7 & 23.7 & [-5, 14] & 2.2 & 15.9 \\
4 & G351.58$-$0.4 & [-119, -97] & 1.1 & 14.1 & [-93, -72] & 1.3 & 12.0 \\
5 & G351.77$-$0.5 & [-89, -4] & 4.1 & 77.1 & [-1, 93] & 4.1 & 89.0\\
\hline
6 & G12.81$-$0.2 & [23, 34] & 2.5 & 10.9 & [38, 81] & 2.0 & 19.0\\
7 & G34.26+0.15 & [40, 58] & 1.6 & 14.5 & [61, 88] & 1.3 & 15.9\\
8 & G35.20$-$0.7 & [20, 32] & 0.8 & 5.2 & [37, 50] & 1.3 & 7.8\\
 \hline

  \multicolumn{8}{c}{\cott{}} \\ 
  \hline
1 & G351.25+0.7 & [-11, -2] & 6.3 & 23.7 & [0, 12] & 2.0 & 8.0\\
2 & G12.81$-$0.2 & [20, 35] & 3.7 & 21.0 & [35, 57] & 3.4 & 29.7 \\
\hline
 \multicolumn{8}{c}{\cosf{}} \\ 
\hline
1 & G351.16+0.7 & [-68, -12] & 19.8 & 174.7 & [-1, 26] & 19.7 & 111.8\\
2 & G351.25+0.7 & [-53, -8] & 22.6 & 114.9 & [2, 47] & 16.6 & 143.8 \\
3 & G351.58-0.4 & [-150, -103] & 6.7 & 85.6 & [-88, -56] & 5.2 & 45.3 \\
4 & G351.77$-$0.5 & [-98, -12] & 16.9 & 371.2 & [6, 98] & 19.1 & 289.8 \\
\hline
5 & G12.81$-$0.2  & [9, 27] & 7.2 & 27.5 & [41, 68] & 25.4 & 170.3 \\
6 & G34.26+0.15 & [2, 51] & 7.0 & 54.1 & [65, 129] & 8.8 & 108.4 \\
7 & G35.20$-$0.7 & [-8, 27] & 11.6 & 91.0 & [40, 68] & 11.2 & 69.1 \\
\hline
\end{tabular}
\begin{flushleft}
    \tablefoot{$V_{\mathrm{range}}$ is the velocity range of line wing. $T_{\mathrm{peak}}$ and $S_{\mathrm{int}}$ are peak and integrated intensity of the line wing emission, respectively.}
\end{flushleft}
\end{table*}

\end{appendix}

\end{document}